\documentclass[lineo]{jfm}
\usepackage{amssymb,amsmath,mathrsfs,bm}
\usepackage{color,tikz,breakcites}
\usepackage{psfrag,overpic}
\usepackage{graphicx}
\usepackage{newtxtext}
\usepackage{newtxmath}
\usepackage{natbib}
\usepackage[breaklinks=true]{hyperref}
\usepackage[export]{adjustbox}
\usepackage{multirow}

\makeatletter
\renewcommand*{\p@section}{\S\,}
\renewcommand*{\p@subsection}{\S\S\,}
\makeatother

\definecolor{bblue}{RGB}{52,152,219}
\hypersetup{
    colorlinks = true,
    urlcolor   = black,
    citecolor  = black,
    linkcolor  = red,
}

\newcommand{\RomanNumeralCaps}[1]
\linenumbers

\shorttitle{Interactions between the near-wall turbulent structures and heavy particles}
\shortauthor{M. Yu, L.H. Zhao, Y.B. Du, X.X. Yuan, C.X. Xu}

\title{Interactions between the near-wall turbulent structures and heavy particles
in compressible turbulent boundary layers}
\author{Ming Yu \aff{1},
Lihao Zhao \aff{2},
Yibin Du \aff{1,3},
Xianxu Yuan \aff{1}\corresp{\email{yuanxianxu2023@163.com}},
Chunxiao Xu \aff{2}\corresp{\email{xucx@tsinghua.edu.cn}}}

\affiliation{
\aff{1}
State Key Laboratory of Aerodynamics, Mianyang 621000, China
\aff{2} Key Laboratory of Applied Mechanics, Ministry of Education, Institute of
Fluid Mechanics, Department of Engineering Mechanics, Tsinghua University, Beijing 100084, China
\aff{3} School of Aeronautic Science and Engineering, Beihang University, Beijing, 100191, China}

\begin{document}

\maketitle


\begin{abstract}
In the present study, we conduct direct numerical simulations to investigate the near-wall
dynamics of compressible turbulent boundary layers at the free-stream Mach number of 6 
laden with heavy particles.
By inspecting the instantaneous near-wall flow structures, Reynolds stresses and 
the impacts of particle forces on solenoidal and dilatational motions,
we observed that higher particle mass loadings lead to the less meandering 
yet almost equally intense velocity streaks, but the weakened wall-normal velocity fluctuations 
induced by vortices and near-wall dilatational motions organized as travelling wave packets.
The strong correlation between the particle force and dilatational velocities
indicates that particles are accelerated/decelerated while travelling through 
these travelling wave packets composed of expansive and compressive events, 
and in return, leading to the weakened dilatational motions of the fluid during this process.
This correlation further supports the elucidation by Yu {\it et al.} (J. Fluid. Mech., vol. 984, 
2024, pp. A44) that dilatational motions are generated by the vortices that 
induce strong bursting events, rather than the evolving perturbations beneath the velocity streaks.
Nevertheless, the variation of skin friction in the presently considered cases with 
moderate mass loadings, either increased or decreased by the presence of particles,
is found to be primarily attributed to the solenoidal Reynolds shear stress as 
in incompressible turbulence, suggesting the essentially unaltered nature of wall-bounded turbulence
populated by vortical and shear motions instead of gradually switching to the state dominated by
dilatational motions.
\end{abstract}

\section{Introduction} \label{sec:intro}

Compressible turbulent boundary layers laden with particles can be encountered in a wide range
of engineering applications, such as high-speed vehicles travelling through dusty environments
~\citep{ching2021sensitivity,cheng2012stochastic},
and the internal flows in solid-fueled combustion engines~\citep{merker2005simulating,
zhang2020grid,junlong2023numerical}.
The compressible fluid flow, characterized by shear, vortical, and dilatational motions, 
transports dispersed phase particles, leading to their motions and spatial organization 
in various patterns. 
The particles exert feedback forces, modulating turbulent motions at all scales
~\citep{crowe1996numerical,crowe2011multiphase,balachandar2010turbulent,brandt2022particle}.
Provided that the particles are smaller than the Kolmogorov scales in turbulence 
and the total volume fraction of the particle phase is low, 
they can be treated as points and tracked under Lagrangian coordinates when the continuous 
fluid phase is simulated under Eulerian coordinates~\citep{soldati2009physics,m2016point}. 
This approach, known as the Eulerian-Lagrangian point-particle method, 
has been extensively employed to study particle-laden turbulent flows,
as summarized by the related review papers, to name but a few,
~\citet{crowe1996numerical,balachandar2010turbulent,brandt2022particle}.

When the particle mass loadings (the mass ratio between particles and 
the fluid per unit volume) are low, the influences of particle feedback forces can be disregarded, 
so only the particle transport by turbulence should be considered~\citep{m2016point}. 
In such scenarios, the dynamics of the particles depend on their inertia, 
characterized by the particle Stokes number $St^+$, with the superscript $+$ denoting the
normalization under viscous scales in wall-bounded turbulence.
The $St^+$ is defined as the ratio between the particle response time $\tau_p$ and 
the characteristic time of the fluid flow $\tau_f$,
the latter of which is constructed by wall shear stress $\tau_w$, 
dynamic viscosity $\mu_w$ and fluid density $\rho_w$ 
(details of these notations will be given in section~\ref{sec:phys}). 
Particles with low $St^+$ tend to follow the motions of the fluid, distributing uniformly within 
turbulent channels or boundary layers. 
Those with finite $St^+$, on the other hand, act as filters, responding most effectively
to fluid motions with similar time scales~\citep{motoori2022role}. 
This behaviour leads to a non-uniform particle spatial distribution
~\citep{eaton1994preferential,marchioli2008statistics,bragg2014new,fong2019velocity}, 
with the tendency to accumulate near the wall and organized in the shape of streamwise elongated
patterns with the scales of such characteristic flow structures as velocity streaks
~\citep{marchioli2002mechanisms,
picciotto2005characterization,rouson2001preferential,soldati2009physics,sardina2012wall}, 
large-scale circulations and very large-scale motions at high Reynolds numbers
~\citep{jie2022existence,motoori2022role,berk2020transport,bernardini2013effect}. 
In compressible wall-bounded turbulence, similar phenomena are observed, 
but due to the non-uniform density distributions, 
the particles are found to cluster within the low-density regions in the outer region
~\citep{xiao2020eulerian}.
Additionally, variations in mean density and viscosity, affecting the particle response time, 
coupled with weaker streamwise vortices in compressible turbulent boundary layers and channels, 
mitigate the degree of near-wall preferential accumulation and small-scale clustering
with the increasing Mach number~\citep{yu2024transport,wang2024turbophoresis}.

When the particle mass loadings reach a level where the impact of particle feedback force 
becomes significant, turbulence will be modulated in the aspect of augmentation or reduction of
the skin friction~\citep{vreman2007turbulence,zhao2010turbulence,li2016direct,gualtieri2023effect},
the intensification or suppression of the turbulent fluctuations~\citep{dritselis2011numerical,
wang2019modulation,zhou2020non},
and the variation of the morphology or the intensity of the velocity streaks
and streamwise vortices~\citep{zhou2020non,gao2023direct,richter2014modification,
richter2015turbulence,lee2015modification}.
Specific conclusions are dependent on both the particle Stokes number $St^+$ and the mass loadings. 
For a given mass loading, particles with $St^+<1$ amplify the turbulent fluctuations, 
leading to more abundant vortical structures in turbulent channels, 
while the case is reversed for particles with $St^+>1$
~\citep{lee2015modification,mortimer2020density}.
On the other hand, at a given particle Stokes number, 
increasing mass loading gradually diminishes turbulent fluctuations
eventually inducing flow laminarization beyond a critical threshold
~\citep{muramulla2020disruption,zhou2020non,zhao2013interphasial}.
This critical value for flow laminarization is also dependent on the particle Stokes number
$St^+$, beyond which near-wall turbulent motions are governed by particle forces rather than 
production, particularly in terms of turbulent kinetic energy transport
~\citep{vreman2007turbulence,muramulla2020disruption,capecelatro2018transition}. 
From the perspective of particle motions, the two-way coupling of particles and fluid 
results in a reduced degree of near-wall particle accumulation~\citep{vreman2007turbulence,
zhao2013interphasial}.

In compressible turbulent boundary layers, related studies are relatively scanty.
\citet{chen2022two} conducted direct numerical simulations to explore the interactions of
boundary layer turbulence, combustion and particle dynamics.
Their findings revealed that reacting flows with heavy particles and high mass loadings 
lead to reduced turbulent intensities and wall heat flux. 
Higher mass loadings result in weaker sweep and ejection events that are associated with
near-wall particle accumulation and hence a more uniform particle distribution. 
These observations are generally in accordance with those in incompressible turbulence
~\citep{brandt2022particle} due to the comparatively low Mach numbers considered,
and the essentially the similar turbulent structures in these flows.
\citet{yu2024momentum} further explored the effects of particle Stokes number and mass loadings
on the momentum and kinetic energy transport in compressible turbulent boundary layers at 
the free-stream Mach number of 2.0.
Besides the abatement of the skin friction, Reynolds stress and turbulent kinetic energy production,
it is also found that the presence of particles reduces the near-wall turbulent heat flux,
inhibiting the heat transfer from transporting towards and free-stream and 
thus enhancing the temperature of the fluid.
However, in higher Mach number turbulent boundary layers over cold walls, 
the dilatational motions progressively dominate the turbulent fluctuations in the 
viscous sublayer and buffer region, particularly for the wall-normal velocity
~\citep{yu2019genuine,yu2021compressibility}. 
These dilatational motions, manifested as positive-negative-alternating patterns and 
organized as travelling wave packets, significantly impact wall pressure, 
wall shear stress, and wall heat flux fluctuations~\citep{yu2020compressibility,
yu2022wall1,yu2022wall2,zhang2022wall,zhang2023conditional,xu2021effect}. 
They are dynamically associated with velocity streaks and quasi-streamwise vortices 
constituting the near-wall regeneration cycle~\citep{yu2024generation}. 
Considering the particles are mostly clustered near the wall where the dilatational
travelling wave packets reside, it is highly possible that they are closely dynamically coupled,
which has not been studied so far, to the best of our knowledge.
This serves as the motivation for the present study.
To investigate the particle modulations of near-wall vortical and dilatational motions 
in high Mach number turbulent boundary layers, we perform
direct numerical simulations (DNS) of two-way force coupling particle-laden 
compressible turbulent boundary layers at the free-stream Mach number of 6.0 over adiabatic 
and cooling walls. The primary focus lies on the examination of the variations in solenoidal 
and dilatational motions under different mass loadings and wall temperatures, 
and the influences of particle forces on these motions.

The remainder of this paper is structured as follows.
In section~\ref{sec:phys}, we present the governing equations for fluid flow and particles, 
along with the numerical simulation methods.
Section~\ref{sec:result} gives the results and discussions, 
including the mean and fluctuating velocity and temperature (Section~\ref{subsec:mean}), 
comparisons of instantaneous velocity distributions, particle feedback force, and 
particle distributions (Section~\ref{subsec:inst}), modulation of solenoidal and 
dilatational motions by particles (Section~\ref{subsec:mod}), 
the statistics and work of particle feedback forces and their spatial correlation with the 
near-wall dilatational motions (Section~\ref{subsec:force}),
and the modulation of the turbulent dynamics in the aspect of skin friction integral identity 
(Section~\ref{subsec:skin}).
Section~\ref{sec:con} summarizes the conclusions of this study.

\section{Physical model and numerical methods} \label{sec:phys}

In this study, we consider the two-way force coupling particle-laden 
compressible turbulent boundary layers over flat plates at the free-stream Mach number of $6$.
The simulations are conducted using the classical Eulerian-Lagrangian point particle method, 
treating the compressible fluid flow as a continuous phase under Eulerian coordinates, 
while tracking the motions of the particles independently as a dispersed phase under 
Lagrangian coordinates~\citep{m2016point}. 
The particles are assumed to be in thermal equilibrium with the fluid, 
allowing us to disregard the interphasial heat transfer effects.

The compressible fluid flow is governed by the following Navier-Stokes equation for Newtonian
perfect gases,
\begin{equation}
\frac{\partial \rho}{\partial t} + \frac{\partial \rho u_j}{\partial x_j} =0
\label{eqn:gov-1}
\end{equation}
\begin{equation}
\frac{\partial \rho u_i}{\partial t} + \frac{\partial \rho u_i u_j}{\partial x_j} = 
- \frac{\partial p}{\partial x_i} + \frac{\partial \tau_{ij}}{\partial x_j} + F_{pi}
\label{eqn:gov-2}
\end{equation}
\begin{equation}
\frac{\partial \rho E}{\partial t} + \frac{\partial \rho E u_j}{\partial x_j} = 
- \frac{\partial p u_j}{\partial x_j} + \frac{\partial \tau_{ij} u_i}{\partial x_j} 
- \frac{\partial q_j}{\partial x_j} + F_{pi} u_i
\label{eqn:gov-3}
\end{equation}
where $u_i$ is the velocity in the $x_i$ direction ($i$=1,2,3, also denoted by $x$, $y$, $z$,
representing the streamwise, wall-normal and spanwise directions), $\rho$, $p$ and $E$ are
density, pressure and total energy of the fluid, related with the temperature $T$ by
the state equation for perfect gases
\begin{equation}
E = \frac{1}{2} u_i u_i + C_V T,~~p=\rho R T,
\end{equation}
with $C_V$ being the constant volume specific heat and $R$ the gas constant.
The viscous stress tensor $\tau_{ij}$ is determined by the constitutive equations of 
the Newtonian fluid, and the molecular heat conduction $q_j$ by Fourier's law
\begin{equation}
\tau_{ij} = \mu \left( \frac{\partial u_i}{\partial x_j} + \frac{\partial u_j}{\partial x_i} \right)
- \frac{2}{3} \mu \frac{\partial u_k}{\partial x_k} \delta_{ij},~~
q_j = - \kappa \frac{\partial T}{\partial x_j},
\end{equation}
where $\mu$ is the dynamic viscosity, calculated by Sutherland's law, and $\kappa$ 
is the molecular heat conductivity, related with the dynamic viscosity as $\kappa = \mu C_p/Pr$
($C_p$ is constant pressure specific heat and $Pr=0.71$ is the Prandtl number).
$F_{pi}$ is the particle feedback force, the approximation of which will be given subsequently.

The simulations are conducted in the computational domains in the shape of rectangular boxes. 
Turbulent inflow at the inlet is generated using the synthetic digital filtering method
~\citep{klein2003digital}. 
At the lower wall, no-slip and no-penetration conditions are given for velocity, 
and the isothermal condition is given for temperature. 
The no-reflection condition is imposed at the upper boundary in the $y$ direction
and outlet boundary in the $x$ direction~\citep{pirozzoli2013generalized}, 
and periodic conditions are applied in the spanwise direction.

Some other notations are introduced to simplify subsequent discussions.
The ensemble average of a generic flow quantity $\varphi$ is denoted by 
$\bar \varphi$, and the corresponding fluctuation by $\varphi'$. 
The density-weighted average is represented by $\tilde \varphi$, and its fluctuation by $\varphi''$. 
Free-stream flow quantities are identified by the subscript $\infty$. 
The superscript $+$ represents the normalization by characteristic flow quantities at the wall, 
namely the wall shear stress $\tau_w$, wall fluid density $\rho_w$, and wall viscosity $\mu_w$. 
The friction velocity $u_\tau$, the viscous length scale $\delta_\nu$ and the friction Reynolds
number $Re_\tau$ are defined accordingly as
\begin{equation}
u_\tau = \sqrt{\frac{\tau_w}{\rho_w}},~~\delta_\nu = \frac{\mu_w}{\rho_w u_\tau},~~
Re_\tau = \frac{\rho_w u_\tau \delta}{\mu_w},
\end{equation}
where $\delta$ represents the nominal boundary layer thickness at a specific streamwise station.

The flow parameters are listed in Table~\ref{tab:param}. 
The free-stream Mach number $M_\infty = U_\infty /a_\infty$ is set as 6.0, 
where $a_\infty = \sqrt{\gamma R T_\infty}$ ($\gamma=1.4$ is the specific heat ratio) 
is the free-stream sound speed. 
The free-stream temperature is set as $55 {\rm K}$.
Two groups of cases with different wall temperatures are considered, 
one with the recovery temperature $T_w=T_r$ and the other with $T_w=0.25 T_r$,
corresponding to scenarios of a quasi-adiabatic wall (with trivial mean heat flux) and a cold wall, 
denoted as cases M6 and M6C, respectively. 
The friction Reynolds number at the turbulent inlet, $Re_{\tau,in}$, is set as 200 for all cases,
according to which the free-stream Reynolds number
$Re_\infty = \rho_\infty U_\infty \delta_{in}/\mu_\infty$ is estimated ($\delta_{in}$ is the
nominal boundary layer thickness at the turbulent inlet).

\begin{table}
\centering
\begin{tabular*}{0.7\textwidth}{@{\extracolsep{\fill}}cccccccc}
Case &  $Re_\infty$ & $M_\infty$ & $T_w/T_r$ & $St^+$ & $N_p$ & $\phi_m$ & $Re_\tau$  \\ 
M6-0 &  96960 & 6.0 & 1.0  & -   & 0      & 0.0    & 288 \\
M6-1 &  96960 & 6.0 & 1.0  & 82  & 1.8E7  & 0.188  & 281 \\
M6-2 &  96960 & 6.0 & 1.0  & 76  & 4.2E7  & 0.440  & 265 \\
M6-3 &  96960 & 6.0 & 1.0  & 67  & 6.0E7  & 0.628  & 245 \\
M6C-0 & 13548 & 6.0 & 0.25 & -   & 0      & 0.0    & 323 \\
M6C-1 & 13548 & 6.0 & 0.25 & 20  & 1.8E7  & 0.188  & 340 \\
M6C-2 & 13548 & 6.0 & 0.25 & 18  & 4.2E7  & 0.440  & 298 \\
M6C-3 & 13548 & 6.0 & 0.25 & 19  & 6.0E7  & 0.628  & 314 \\
\end{tabular*}
\caption{Flow parameters. Here, $St^+$ is the particle Stokes number, 
$N_p$ is the number of particles and $\phi_m$ is the particle mass loading. 
$Re_\tau$ is the averaged friction Reynolds number within $x=(55 \sim 70) \delta_{in}$ .}
\label{tab:param}
\end{table}

The governing Equations~\eqref{eqn:gov-1}-\eqref{eqn:gov-3} 
are directly solved using a finite difference method with 
the open-source code developed by~\citet{bernardini2021streams}, which has been validated 
extensively~\citep{cogo2022direct,bernardini2023unsteadiness,ceci2022numerical}. 
The convective terms are approximated using the sixth-order kinetic energy preserving scheme
~\citep{kennedy2008reduced,pirozzoli2010generalized} in smooth regions, 
and the fifth-order weighted essentially non-oscillatory (WENO) scheme~\citep{shu1988efficient} 
in discontinuity regions identified by the refined Ducro's shock sensor
~\citep{ducros1999large,pirozzoli2011numerical}. 
Viscous terms are expanded into Laplacian form and then approximated using a sixth-order viscous 
scheme. Time advancement is achieved by the third-order low-storage Runge-Kutta scheme
~\citep{spalart1991spectral}.

For all the cases considered, the computational domain sizes in the streamwise, wall-normal, 
and spanwise directions are set as $L_x = 80 \delta_{in}$, $L_y = 9 \delta_{in}$, 
and $L_z = 8 \delta_{in}$, respectively, discretized by grid points of $2400 \times 280 \times 280$. 
The mesh is uniformly distributed in the streamwise and spanwise directions, 
resulting in grid intervals of $\Delta x^+ \approx 6.68$ and $\Delta z^+ \approx 5.71$
under the viscous scales at the turbulent inlet. 
In the wall-normal direction, grid stretching is implemented using the hyperbolic sine function, 
with the first off-wall grid point at $\Delta y^+_w = 0.5$ and a grid interval of approximately 
$\Delta y^+ \approx 5.37$ at the edge of the boundary layer. 
These grid settings are deemed adequate for capturing small-scale fluctuations in DNS 
under the low-dissipative numerical schemes employed in this study~\citep{pirozzoli2011turbulence}.
An initial assessment of skin friction coefficients, boundary layer shape factors, 
mean velocity, and temperature suggests that turbulence reaches quasi-equilibrium states 
downstream of $x=50\delta_{in}$. 
Therefore, subsequent discussions primarily focus on the domain within the range 
$x= (55 \sim 70) \delta_{in}$, where the averaged nominal boundary layer thickness is
denoted by $\delta$.

The trajectories and velocities of the particles are calculated by solving the following equations
for each particle individually,
\begin{equation}
\frac{{\rm d} r_{pi}}{{\rm d} t} = v_{i},~~
\frac{{\rm d} v_i}{{\rm d} t} = \frac{f_D}{\tau_p} (u_i - v_i).
\label{eqn:parm}
\end{equation}
Here, $r_{pi}$ and $v_i$ represent the particle location and velocity, respectively. 
The particle relaxation time, denoted as $\tau_p$, is defined as $\tau_p =\rho_p d^2_p/(18 \mu)$, 
where $\rho_p$ stands for the particle density and $d_p$ the particle diameter. 
The modified coefficient $f_D$ is estimated by the formula proposed by~\citet{loth2021supersonic},
incorporating the high Reynolds and Mach number effects
\begin{equation}
f_D = (1+0.15 Re^{0.687}_p) H_M,
\end{equation}
\begin{equation}
\begin{aligned}
H_M =
\begin{cases}
      0.0239 M^3_p + 0.212 M^2_p - 0.074 M_p + 1,~~~ & M_p \le 1 \\
      0.93 + \frac{1}{3.5 + M^5_p}, ~~~ & M_p >1   
\end{cases}
\end{aligned}
\label{eqn:hm}
\end{equation}
with $Re_p = \rho |u_i - v_i| d_p / \mu$ and $M_p = |u_i - v_i| / \sqrt{\gamma R T}$ representing 
the particle Reynolds number and particle Mach number, respectively. 
The ordinary differential equations described in Equation~\eqref{eqn:parm} are solved using 
a third-order Runge-Kutta scheme. 
The fluid flow quantities at the particle positions are obtained by trilinear interpolation.
Initially, the particles are randomly distributed in the streamwise and spanwise directions within 
$y = \delta_{in}$. 
Periodic conditions are applied in the spanwise direction, and elastic collisions are assumed 
at the wall. 
As the particles penetrate the flow outlet and upper boundaries, 
they are reintroduced at random locations within the boundary layer thickness at the flow inlet.

The particle feedback force is estimated using the particle-in-cell method
~\citep{zhao2013interphasial,gao2023direct}. 
This method calculates the volume-averaged forces exerted by the particles within a cell,
\begin{equation}
F_{pi} = - \frac{1}{V_{cell}}\sum^{n_p}_{j=1} m_p \frac{{\rm d} v_i}{{\rm d} t}
\end{equation}
where $m_p$ represents the mass of a single particle,
and $n_p$ denotes the number of particles within a mesh volume $V_{cell}$.
To ensure numerical stability, the particle feedback force is not incorporated until 
downstream of $x=10\delta_{in}$.
The DNS solver for one-way and two-way coupling in particle-laden wall-bounded turbulence
has been validated in the prior studies ~\citep{yu2024transport,yu2024momentum}
by comparing the fluid and particle statistics in low Mach number turbulent channel flows 
with those in incompressible turbulence~\citep{zhao2013interphasial}.

For all cases under consideration, the particle density is set as $\rho_p = 1600 \rho_\infty$.
The particle diameter is $d_p = 0.002 \delta_{in}$, which is $d^+_p = 0.4$ under viscous scales
at the turbulent inlet and approximately 0.3 times the Kolmogorov length scales at the wall,
satisfying the requirement of the point-particle simulation approach.
The particle Stokes numbers, evaluated based on wall flow quantities $St^+$ as reported in 
Table~\ref{tab:param}, vary across these cases.
The differences amongst the cases with the same wall temperature are caused by the disparities 
in wall shear stress, while those with different wall temperatures by the disparities in the 
wall viscosity and wall density.
Specifically, the $St^+$ values in the quasi-adiabatic wall cases (M6) are approximately 
three times higher than those in the cold wall cases (M6C).
The mass loading $\varphi_m = N_p m_p / (L_x L_z \delta)$ ranges from 0.0 
(the particle-free flow) to a moderate value of 0.628, remaining below a critical threshold
where laminarization of flows occurs~\citep{vreman2007turbulence,yu2024momentum}.
The friction Reynolds number $Re_\tau$ is lower in the cases laden with particles,
suggesting the reduction of the skin friction.
Details of the turbulent statistics will be discussed in the next section.

\section{Results and discussion} \label{sec:result}

\subsection{Mean and fluctuating velocity and temperature} \label{subsec:mean}

\begin{figure}
\centering
\begin{overpic}[width=0.5\textwidth]{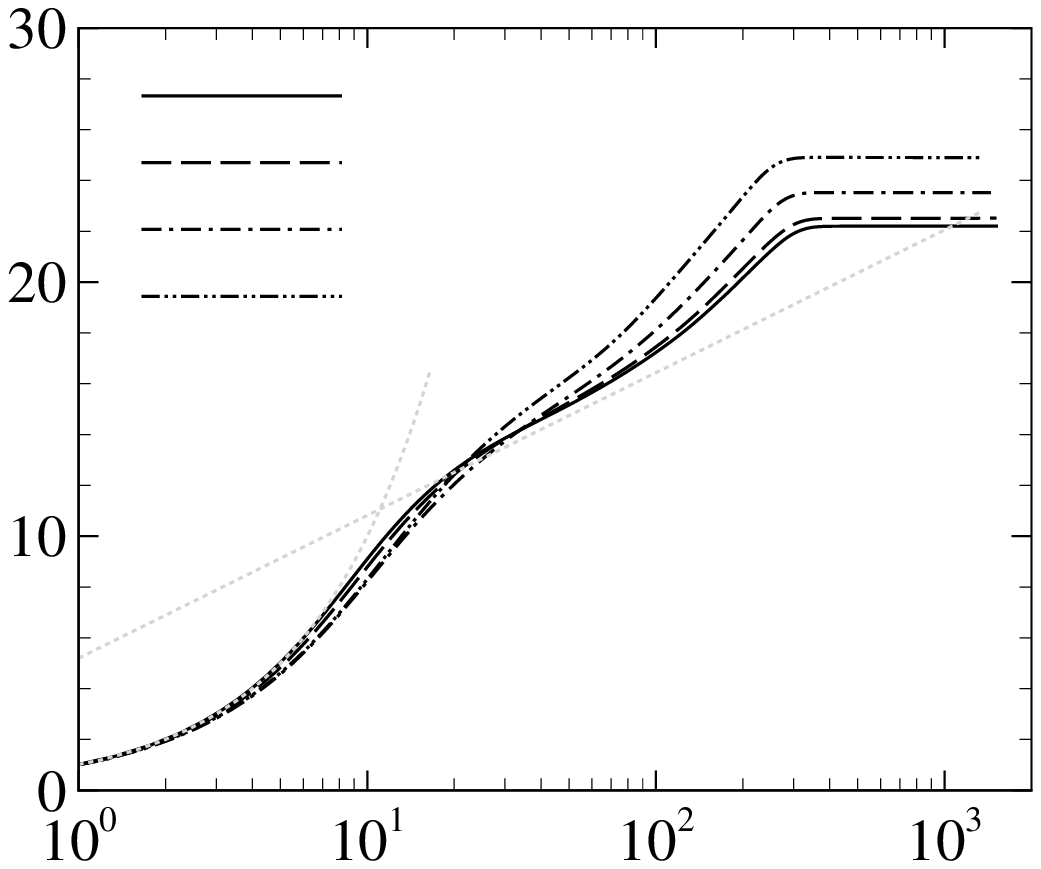}
\put(0,70){(a)}
\put(48,0){$y^+$}
\put(0,35){\rotatebox{90}{$u^+_{VD}$}}
\put(37,65){\small M6-0}
\put(37,60){\small M6-1}
\put(37,54.5){\small M6-2}
\put(37,49){\small M6-3}
\end{overpic}~
\begin{overpic}[width=0.5\textwidth]{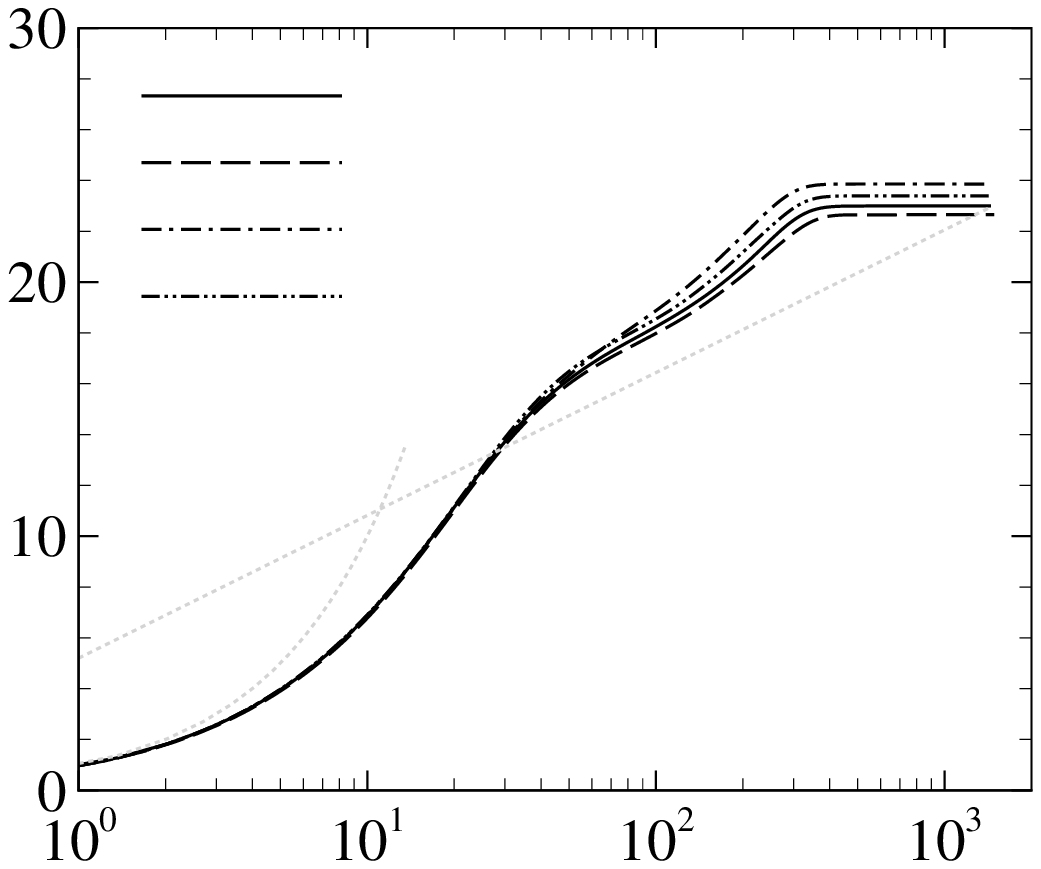}
\put(0,70){(b)}
\put(48,0){$y^+$}
\put(0,35){\rotatebox{90}{$u^+_{VD}$}}
\put(37,65){\small M6C-0}
\put(37,60){\small M6C-1}
\put(37,54.5){\small M6C-2}
\put(37,49){\small M6C-3}
\end{overpic}\\[1.0ex]
\begin{overpic}[width=0.5\textwidth]{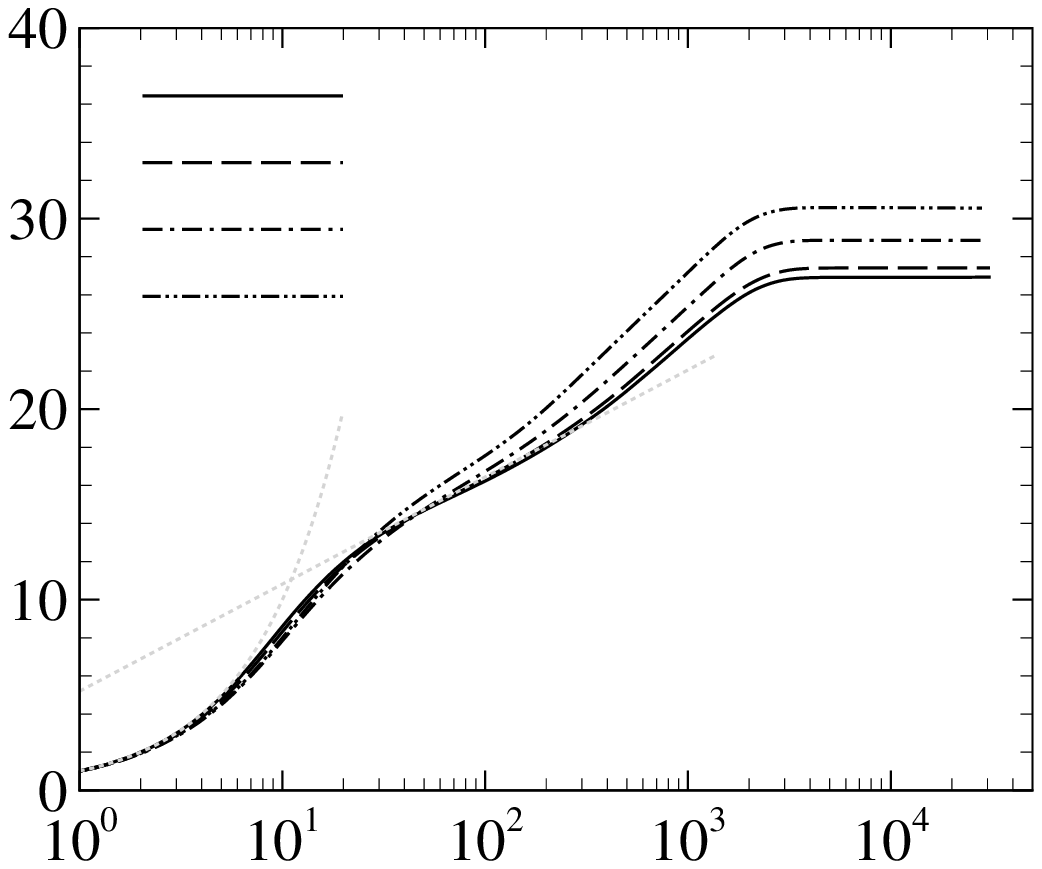}
\put(0,70){(c)}
\put(48,0){$y^+_{VP}$}
\put(0,35){\rotatebox{90}{$u^+_{VP}$}}
\put(37,65){\small M6-0}
\put(37,60){\small M6-1}
\put(37,54.5){\small M6-2}
\put(37,49){\small M6-3}
\end{overpic}~
\begin{overpic}[width=0.5\textwidth]{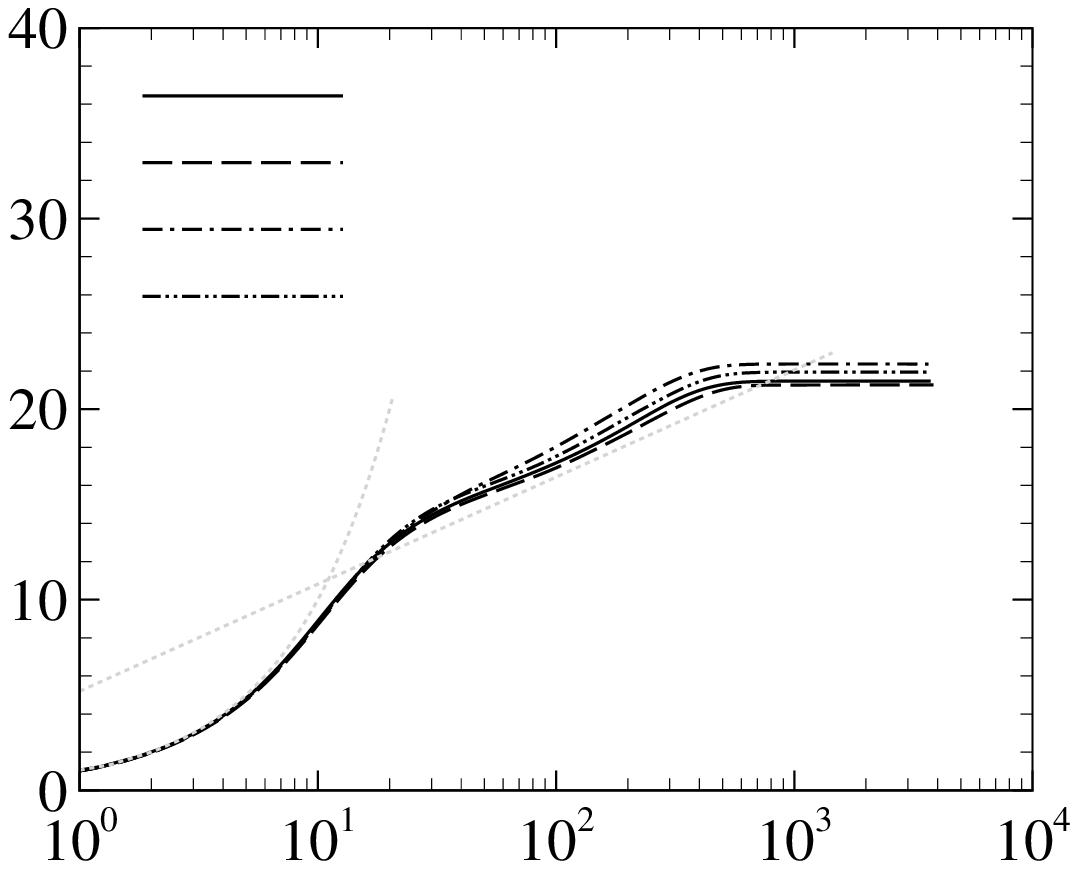}
\put(0,70){(d)}
\put(48,0){$y^+_{VP}$}
\put(0,35){\rotatebox{90}{$u^+_{VP}$}}
\put(37,65){\small M6C-0}
\put(37,60){\small M6C-1}
\put(37,54.5){\small M6C-2}
\put(37,49){\small M6C-3}
\end{overpic}\\
\caption{Mean velocity distribution under the integral transformation of 
(a,b) Equation \eqref{eqn:vd} and (c,d) Equation \eqref{eqn:vp} in
(a,c) cases M6, (b,d) cases M6C.}
\label{fig:meanvelo}
\end{figure}

We first consider the impact of particles on the mean velocity of the fluid.
In compressible turbulent boundary layers, it is customary to perform integrations
on the wall-normal coordinate and mean velocity to collapse the profiles with the scaling laws 
of the incompressible turbulence with the general form of~\citep{modesti2016reynolds}
\begin{equation}
y^+_I = \frac{1}{\delta_\nu} \int^y_0 f_I {\rm d} y,~~
u^+_I = \frac{1}{u_\tau} \int^{\tilde u}_0 g_I {\rm d} \tilde u.
\end{equation}
In the present study, we consider the classical van Driest transformation~\citep{van1951turbulent} 
\begin{equation}
f_{VD} = 1.0,~~g_{VD} = \sqrt{\frac{\bar \rho}{\bar \rho_w}},
\label{eqn:vd}
\end{equation}
and the one proposed by \citet{volpiani2020data}
\begin{equation}
f_{VP} = \sqrt{\frac{\bar \rho}{\bar \rho_w} \left( \frac{\bar \mu_w}{\bar \mu} \right)^3},~~
g_{VP} = \sqrt{\frac{\bar \rho}{\bar \rho_w} \left( \frac{\bar \mu_w}{\bar \mu} \right)}.
\label{eqn:vp}
\end{equation}
For the particle-free case M6-0 over a quasi-adiabatic wall, 
both of these transformations yield the profiles that obey the linear law
in the viscous sublayer and the log law in the log region 
(if any at the current low Reynolds numbers).
The presence of particles (cases M6-1, M6-2, and M6-3) tends to reduce the mean velocities
in the viscous sublayer while increasing them from the log region to the free stream
under both transformations.
In the log region, both the slopes and the intercepts of the log law are altered compared with
case M6-0, the degree of which depends on the particle mass loading.
In the cold wall case M6C-0, the van Driest transformation of mean velocity is lower than 
the linear law in the viscous sublayer but higher than the classical log law for 
incompressible flows within $y^+ \approx 30 \sim 100$,
which are amended by the transformation~\eqref{eqn:vp} that incorporates the variation of
the mean viscosity.
In these cold wall cases, particle feedback forces only weakly influence 
the transformed mean velocities in the viscous sublayer but elevate them in the log region 
and above.
Under these two integral transformations, the deviations of the mean velocity profiles in
the two-way coupling cases from the particle-free cases are almost identical, suggesting that
modulation of the mean velocity by the particle feedback force cannot be ameliorated by
the incorporation of the variations of the fluid density and viscosity.
By comparison, the particle modulation of the mean velocity is weaker
in turbulent boundary layers over cold walls than over adiabatic walls,
despite the identical particle populations and mass loadings.
Such differences can be attributed to the different Stokes numbers
in these cases, the influences of which have been proven
to be significant in incompressible turbulent flows~\citep{gualtieri2023effect}.

\begin{figure}
\centering
\begin{overpic}[width=0.5\textwidth]{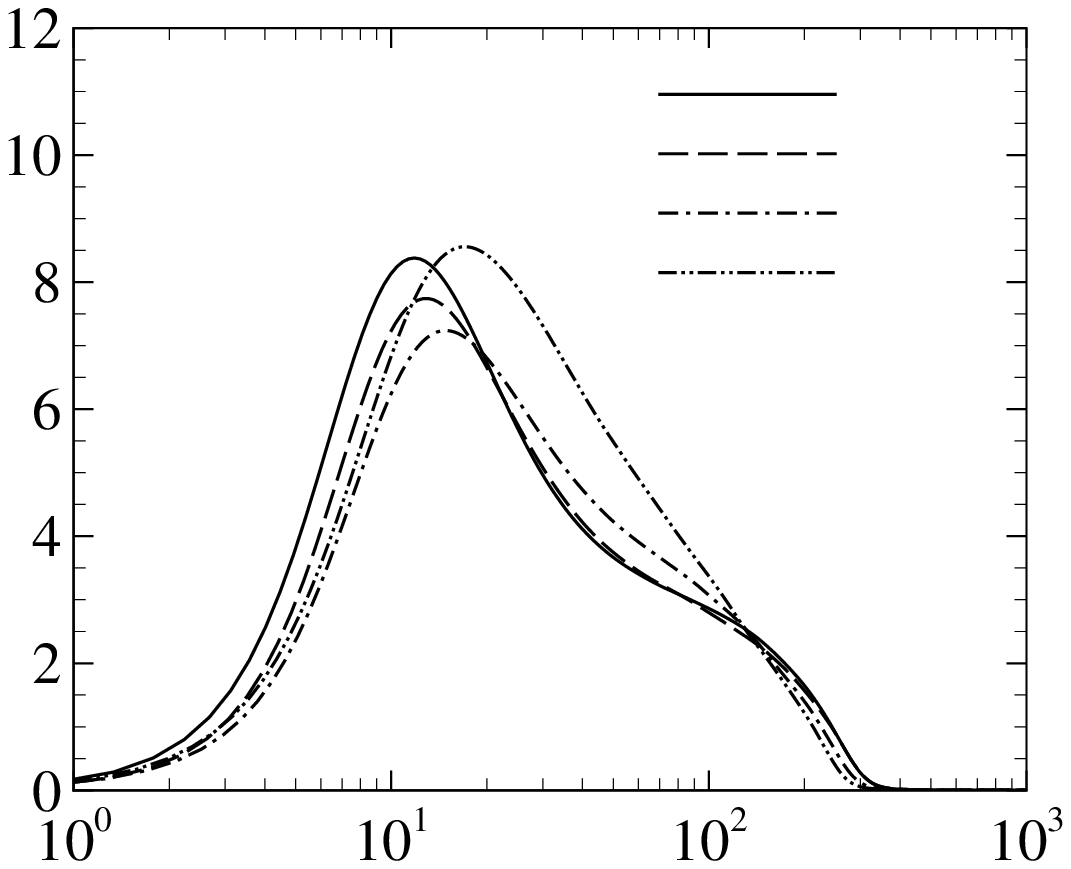}
\put(-2,70){(a)}
\put(48,0){$y^+$}
\put(-2,35){\rotatebox{90}{$R^+_{11}$}}
\put(75,65.5){\small M6-0}
\put(75,60.5){\small M6-1}
\put(75,55.5){\small M6-2}
\put(75,50.5){\small M6-3}
\end{overpic}~
\begin{overpic}[width=0.5\textwidth]{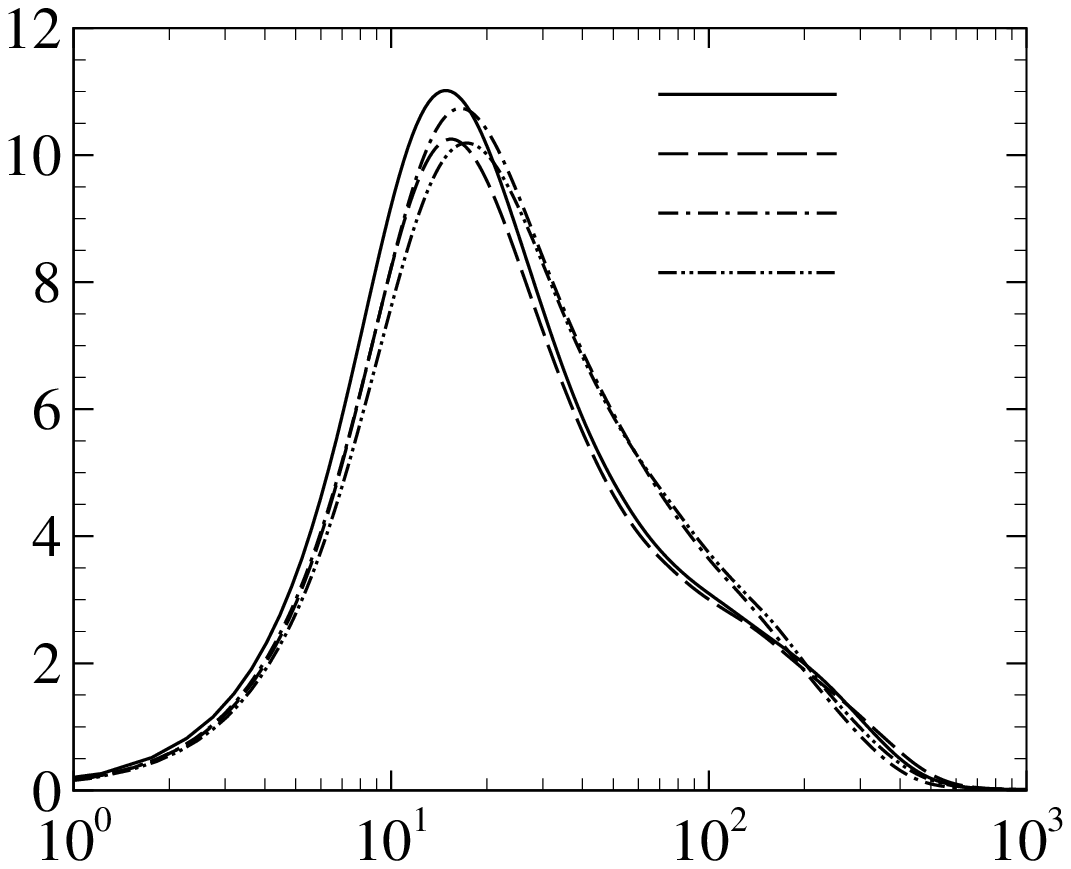}
\put(-2,70){(b)}
\put(48,0){$y^+_{VP}$}
\put(-2,35){\rotatebox{90}{$R^+_{11}$}}
\put(74,65.5){\small M6C-0}
\put(74,60.5){\small M6C-1}
\put(74,55.5){\small M6C-2}
\put(74,50.5){\small M6C-3}
\end{overpic}\\[1.0ex]
\begin{overpic}[width=0.5\textwidth]{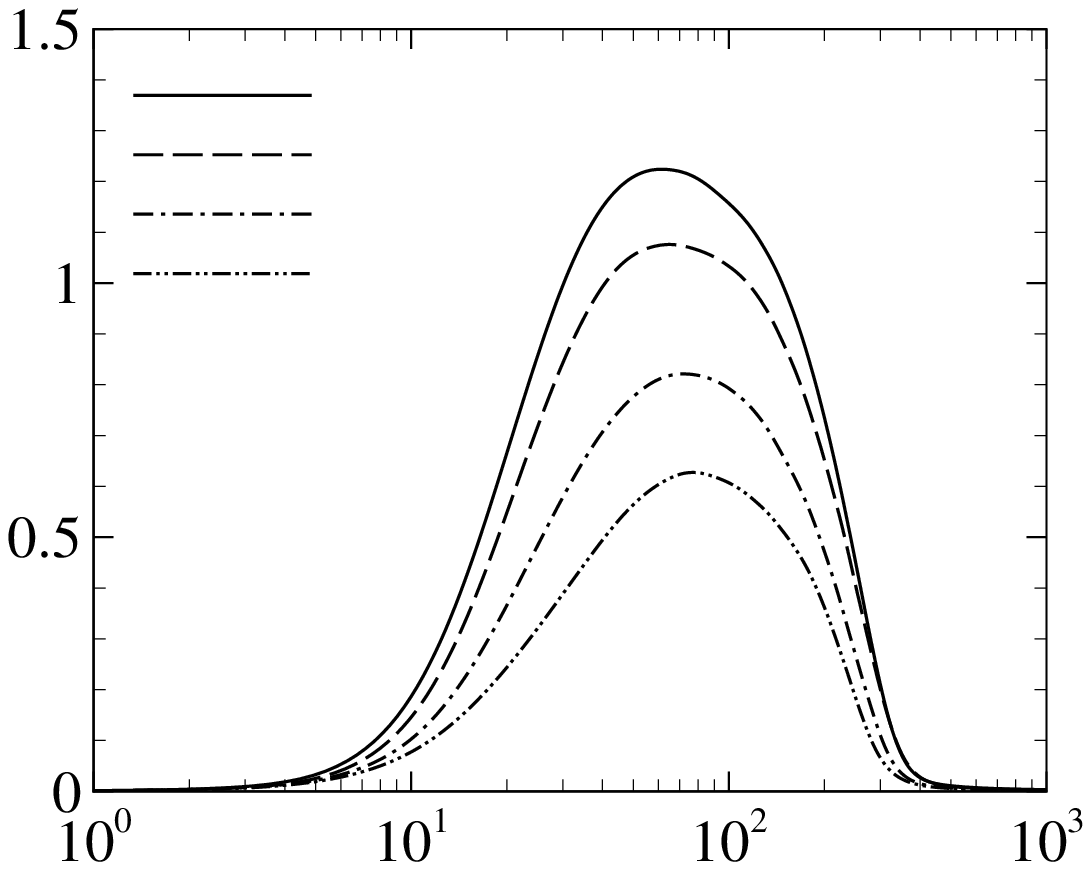}
\put(-2,70){(c)}
\put(48,0){$y^+$}
\put(-2,35){\rotatebox{90}{$R^+_{22}$}}
\put(32,65.5){\small M6-0}
\put(32,60.5){\small M6-1}
\put(32,55.5){\small M6-2}
\put(32,50.5){\small M6-3}
\end{overpic}~
\begin{overpic}[width=0.5\textwidth]{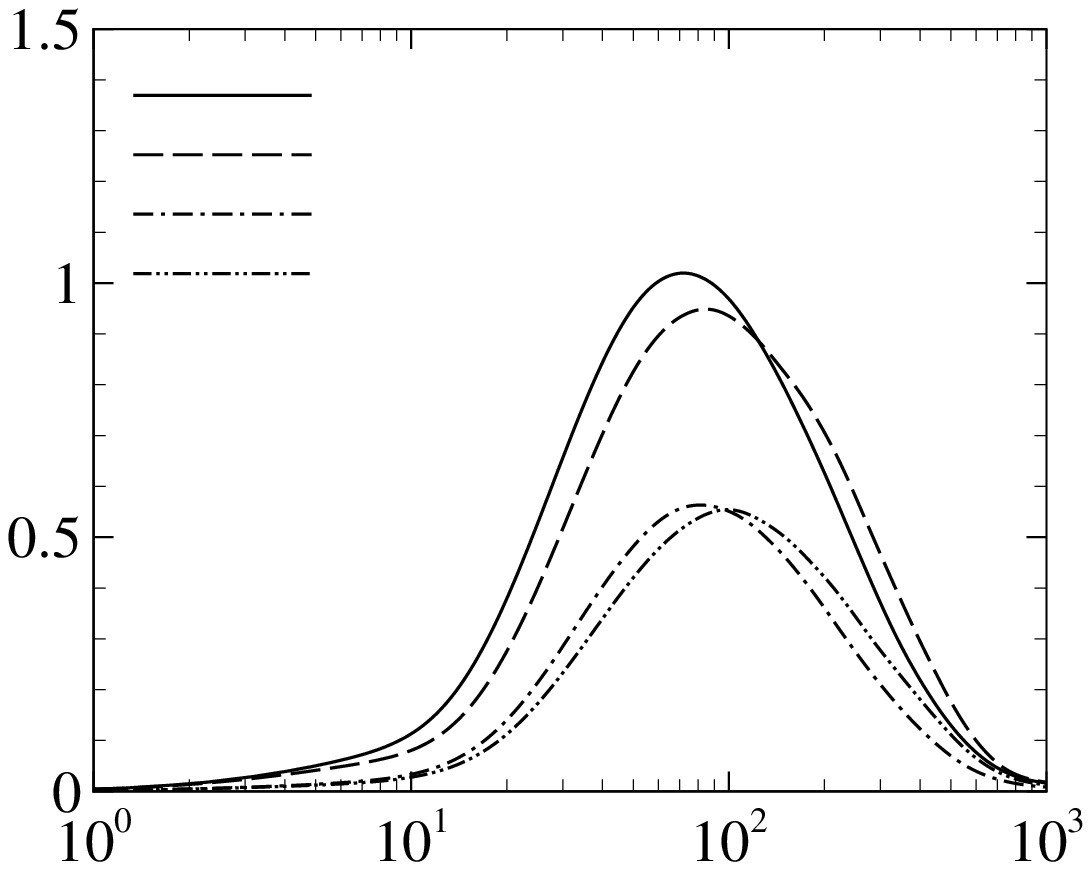}
\put(-2,70){(d)}
\put(48,0){$y^+_{VP}$}
\put(-2,35){\rotatebox{90}{$R^+_{22}$}}
\put(32,65.5){\small M6C-0}
\put(32,60.5){\small M6C-1}
\put(32,55.5){\small M6C-2}
\put(32,50.5){\small M6C-3}
\end{overpic}\\[1.0ex]
\begin{overpic}[width=0.5\textwidth]{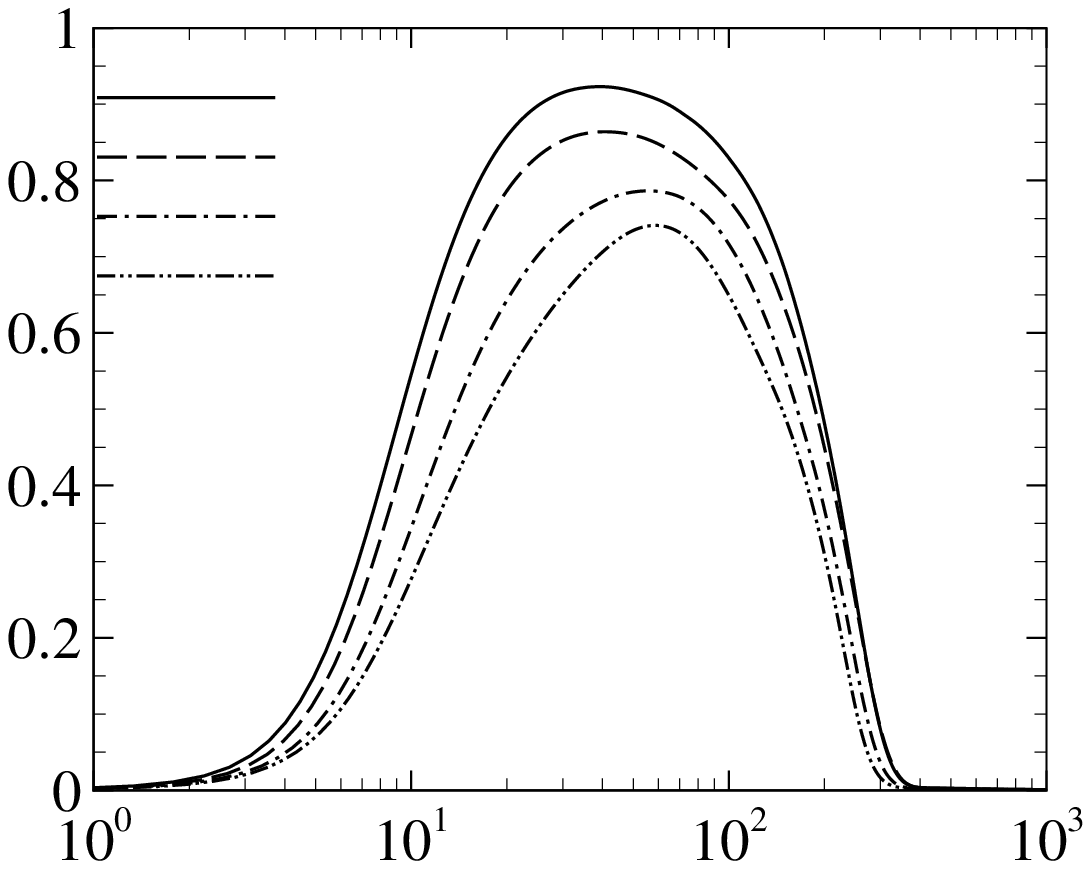}
\put(-2,70){(e)}
\put(48,0){$y^+$}
\put(-2,35){\rotatebox{90}{$-R^+_{12}$}}
\put(28,65.5){\small M6-0}
\put(28,60.5){\small M6-1}
\put(28,55.5){\small M6-2}
\put(28,50.5){\small M6-3}
\end{overpic}~
\begin{overpic}[width=0.5\textwidth]{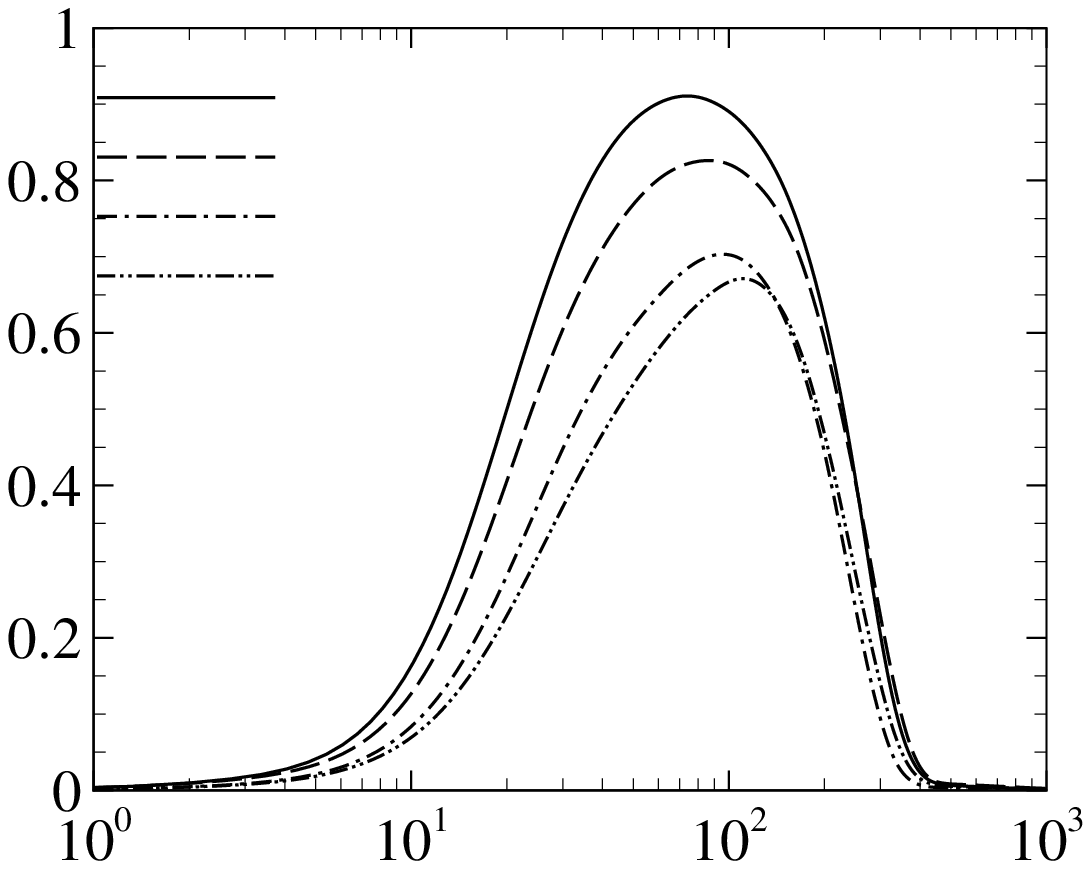}
\put(-2,70){(f)}
\put(48,0){$y^+_{VP}$}
\put(-2,35){\rotatebox{90}{$-R^+_{12}$}}
\put(28,65.5){\small M6C-0}
\put(28,60.5){\small M6C-1}
\put(28,55.5){\small M6C-2}
\put(28,50.5){\small M6C-3}
\end{overpic}\\[1.0ex]
\caption{Wall-normal distribution of the Reynolds stresses, (a,b) $R^+_{11}$, (c,d) $R^+_{22}$,
(e,f) $-R^+_{12}$, (a,c,e) cases M6, (b,d,f) cases M6C.}
\label{fig:rey}
\end{figure}

In Figure~\ref{fig:rey}, we present the wall-normal distributions of the Reynolds stress 
components, defined as $R_{ij} = \overline{\rho u''_i u''_j}$.
Considering their better consistency with the scaling laws observed in incompressible turbulence 
as depicted in Figure~\ref{fig:meanvelo}, 
we display the results of cases M6 against $y^+$ and those of cases M6C against $y^+_{VP}$. 
The magnitudes of the peaks of $R^+_{11}$ are relatively unaffected by particle feedback forces
(Figures~\ref{fig:rey}(a,b)). 
In cases M6, the peaks of $R^+_{11}$ exhibit a slight shift towards higher off-wall positions with 
the increasing particle mass loading, although the magnitudes display a non-monotonic variation,
first decreasing for cases M6-1 and M6-2 before increasing for case M6-3. 
This non-monotonic behaviour in the streamwise Reynolds stress component with the mass loading 
has also been observed by~\citet{zhou2020non} in incompressible turbulent channel flows, which was 
found to be consistent with the variation of the addition of the production term and the work of
the particle feedback force in the transport of the Reynolds stress components.
The peaks of $R^+_{11}$ in cold wall cases M6C manifest a similar trend of variation 
with $\varphi_m$, albeit with weaker influences of the particles in comparison.
From the perspective of the turbulent structures, we can infer that the intensity of 
near-wall velocity streaks, as indicated by the $R^+_{11}$ component, remains almost unaffected, 
although their morphology may exhibit significant differences, 
which will be elaborated in subsequent discussions. 
The other Reynolds stress components, $R^+_{22}$ and $-R^+_{12}$ (Figures~\ref{fig:rey}(c-f)), 
decrease with increasing mass loading, implying reduced cross-stream velocity fluctuations 
and momentum associated with the cross-stream motions.
These observations are qualitatively in agreement with those in incompressible 
wall-bounded turbulence~\citep{zhao2013interphasial,zhou2020non,muramulla2020disruption}.

\begin{figure}
\centering
\begin{overpic}[width=0.5\textwidth]{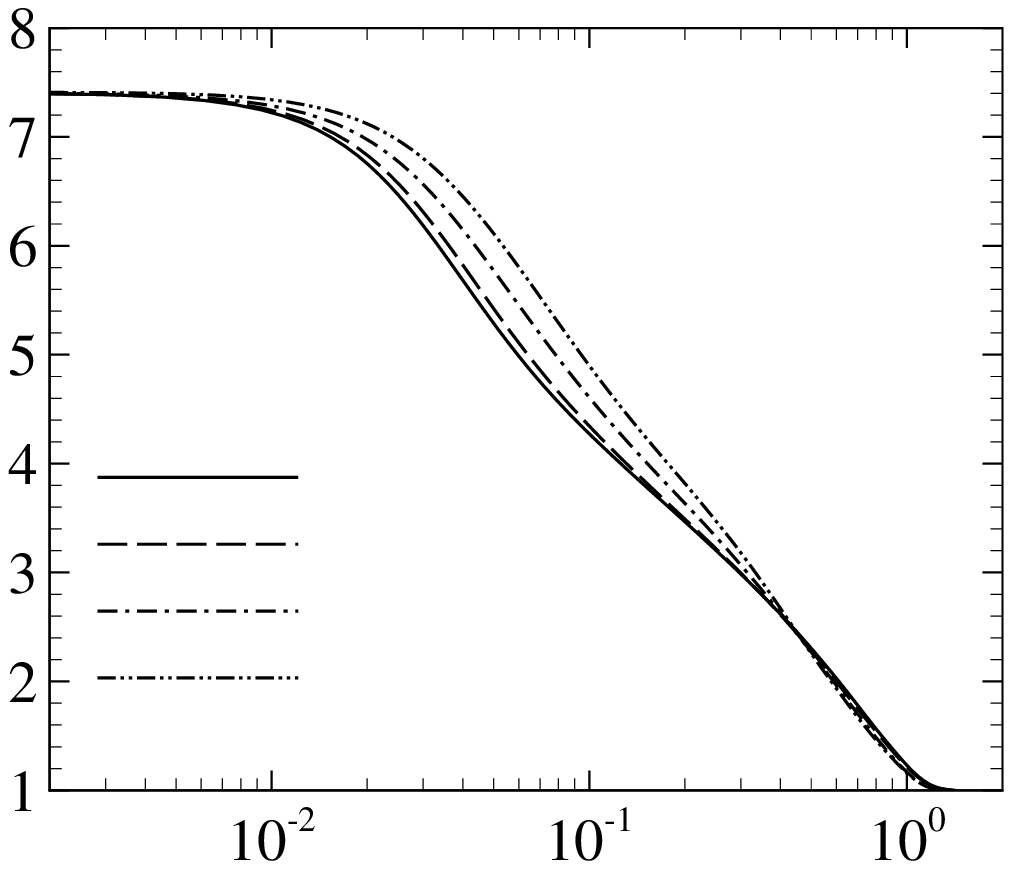}
\put(0,70){(a)}
\put(48,0){$y/\delta$}
\put(0,35){\rotatebox{90}{$\overline T/T_0$}}
\put(35,34){\small M6-0}
\put(35,29){\small M6-1}
\put(35,23.5){\small M6-2}
\put(35,18){\small M6-3}
\end{overpic}~
\begin{overpic}[width=0.5\textwidth]{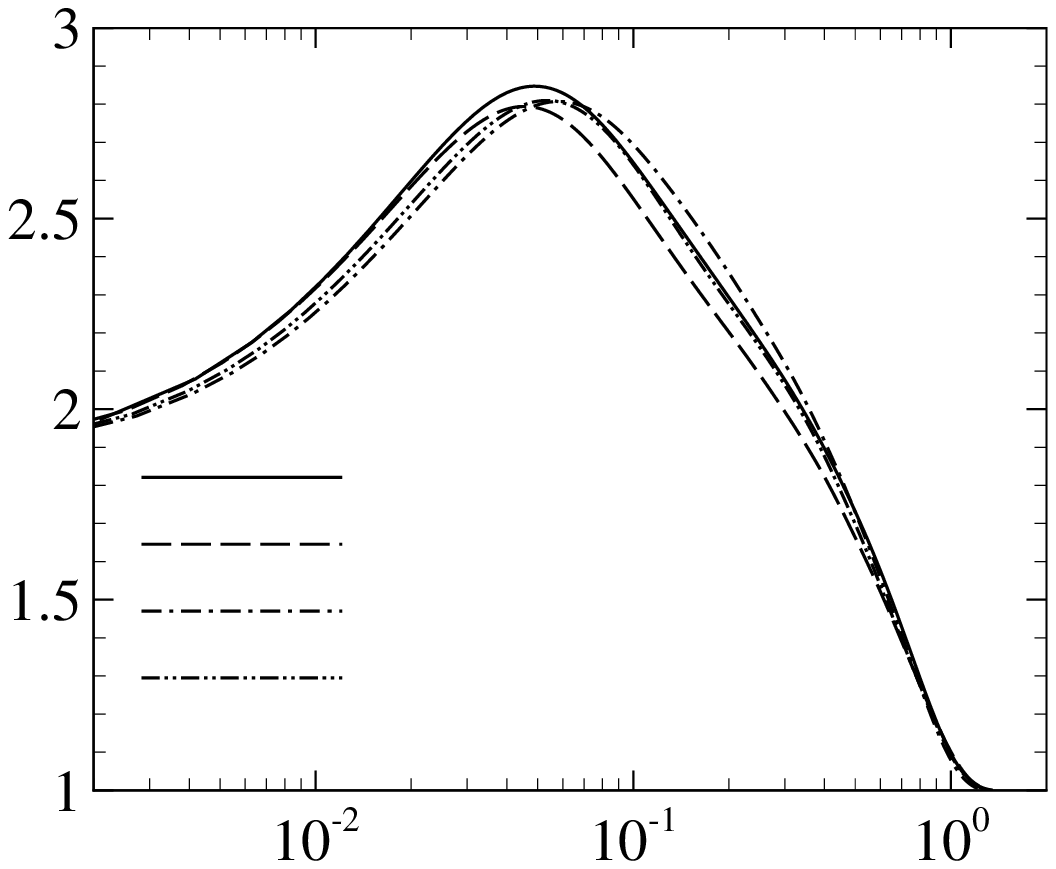}
\put(0,70){(b)}
\put(48,0){$y/\delta$}
\put(0,35){\rotatebox{90}{$\overline T/T_0$}}
\put(35,34){\small M6C-0}
\put(35,29){\small M6C-1}
\put(35,23.5){\small M6C-2}
\put(35,18){\small M6C-3}
\end{overpic}\\[1.0ex]
\begin{overpic}[width=0.5\textwidth]{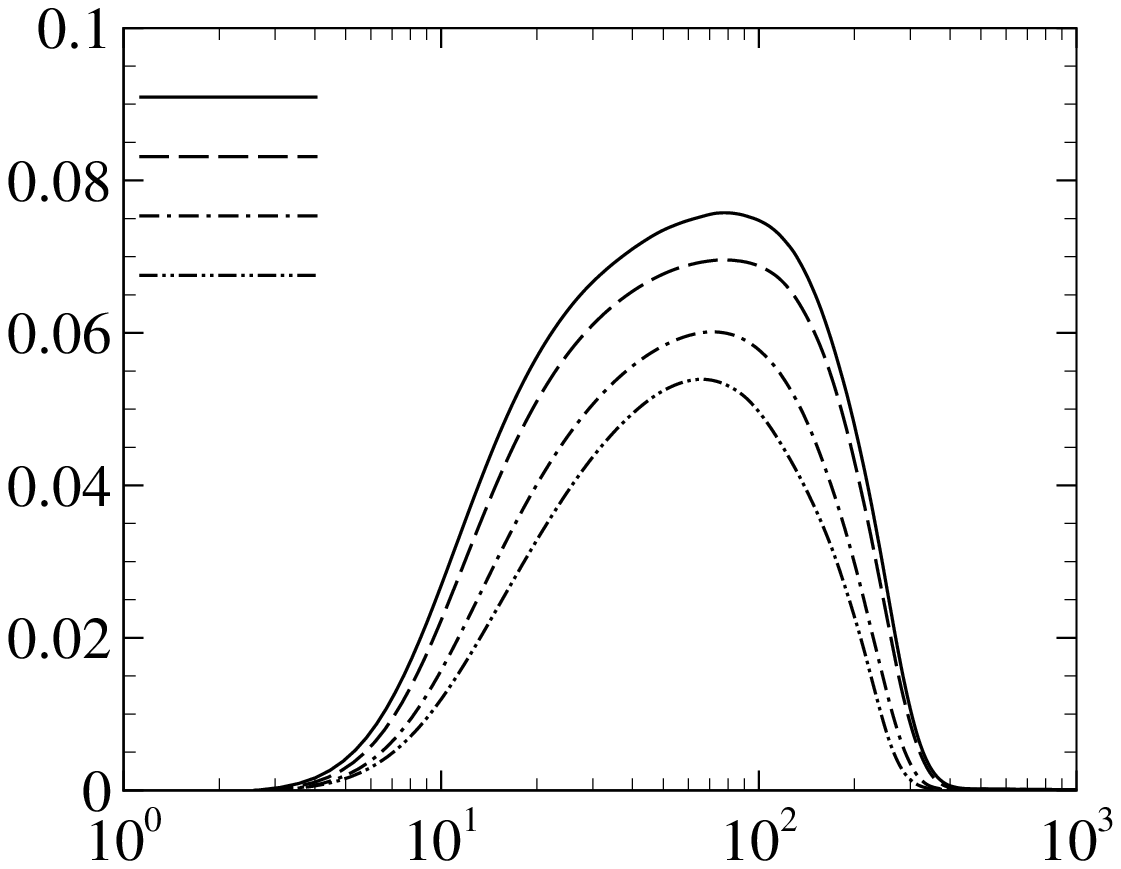}
\put(-2,70){(c)}
\put(48,0){$y^+$}
\put(-4,35){\rotatebox{90}{$\overline{\rho u''_2 T'}^+$}}
\put(32,65.5){\small M6-0}
\put(32,60.5){\small M6-1}
\put(32,55.5){\small M6-2}
\put(32,50.5){\small M6-3}
\end{overpic}~
\begin{overpic}[width=0.5\textwidth]{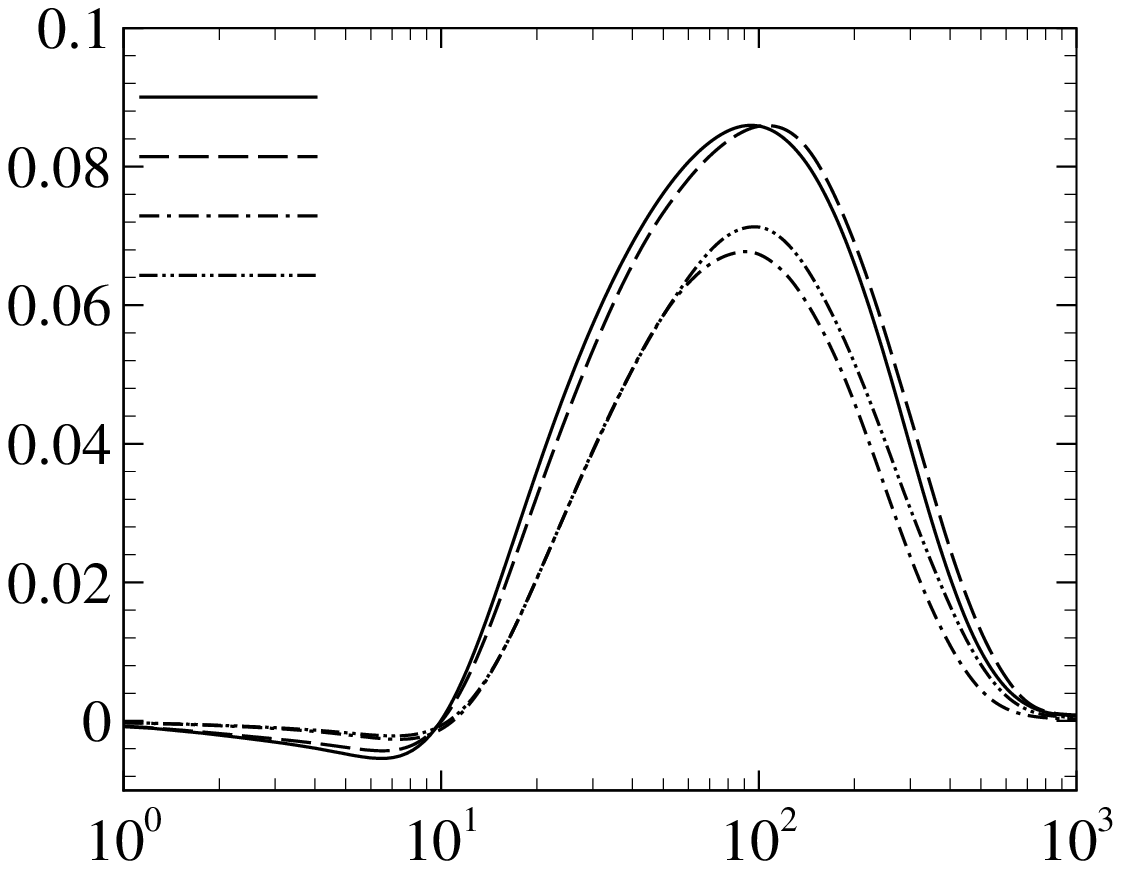}
\put(-2,70){(d)}
\put(48,0){$y^+_{VP}$}
\put(-4,35){\rotatebox{90}{$\overline{\rho u''_2 T'}^+$}}
\put(32,65.5){\small M6C-0}
\put(32,60.5){\small M6C-1}
\put(32,55.5){\small M6C-2}
\put(32,50.5){\small M6C-3}
\end{overpic}\\
\caption{Wall-normal distribution of (a,b) the mean temperature and (c,d) wall-normal turbulent
heat flux, (a,c) cases M6, (b,d) cases M6C.}
\label{fig:meantemp}
\end{figure}

The variation in flow dynamics induced by particles influences not only the mean velocity 
and Reynolds stresses but also the mean temperature and turbulent heat flux, 
in spite of the absence of direct heat coupling between the fluid and particles. 
As shown in Figure~\ref{fig:meantemp}(a), for the adiabatic wall cases M6, 
the mean temperatures below $y=0.3\delta$ exhibit significant enhancement in particle-laden flows, 
displaying a consistent increasing trend with mass loading. 
The temperature gradients at the wall remain small, indicating the trivial wall heat flux 
and maintenance of quasi-adiabatic thermal conditions. 
In the cold wall cases M6C (Figure~\ref{fig:meantemp}(b)), 
the wall heat flux reduces slightly in cases M6C-2 and M6C-3.
The highest mean temperatures are slightly reduced by the presence of particles, 
but the degree of the variation is much smaller compared with cases M6 over adiabatic walls.
The variations in mean temperature can be attributed to reduced turbulent heat flux that transports
the near-wall high temperatures either to the free-stream or to the wall, 
as shown in Figures~\ref{fig:meantemp}(c,d). 
However, the relationship between mean temperature and mean velocity can still be characterized by
the generalized Reynolds analogy~\citep{zhang2014generalized}.
The turbulent Prandtl number $Pr_T$ that associates the turbulent heat flux and 
the Reynolds shear stress is close to unity (not shown here for brevity),
suggesting that the energy-containing temperature fluctuations can still be regarded 
as passive scalars, probably due to the neglection of the interphasial heat transfer effects or
the retainment of the flows being in their turbulent states~\citep{yu2024momentum}.


\subsection{Instantaneous flow structures} \label{subsec:inst}

\begin{figure}
\centering
\begin{overpic}[width=0.5\textwidth,trim={0.2cm 0.2cm 0.2cm 0.2cm},clip]{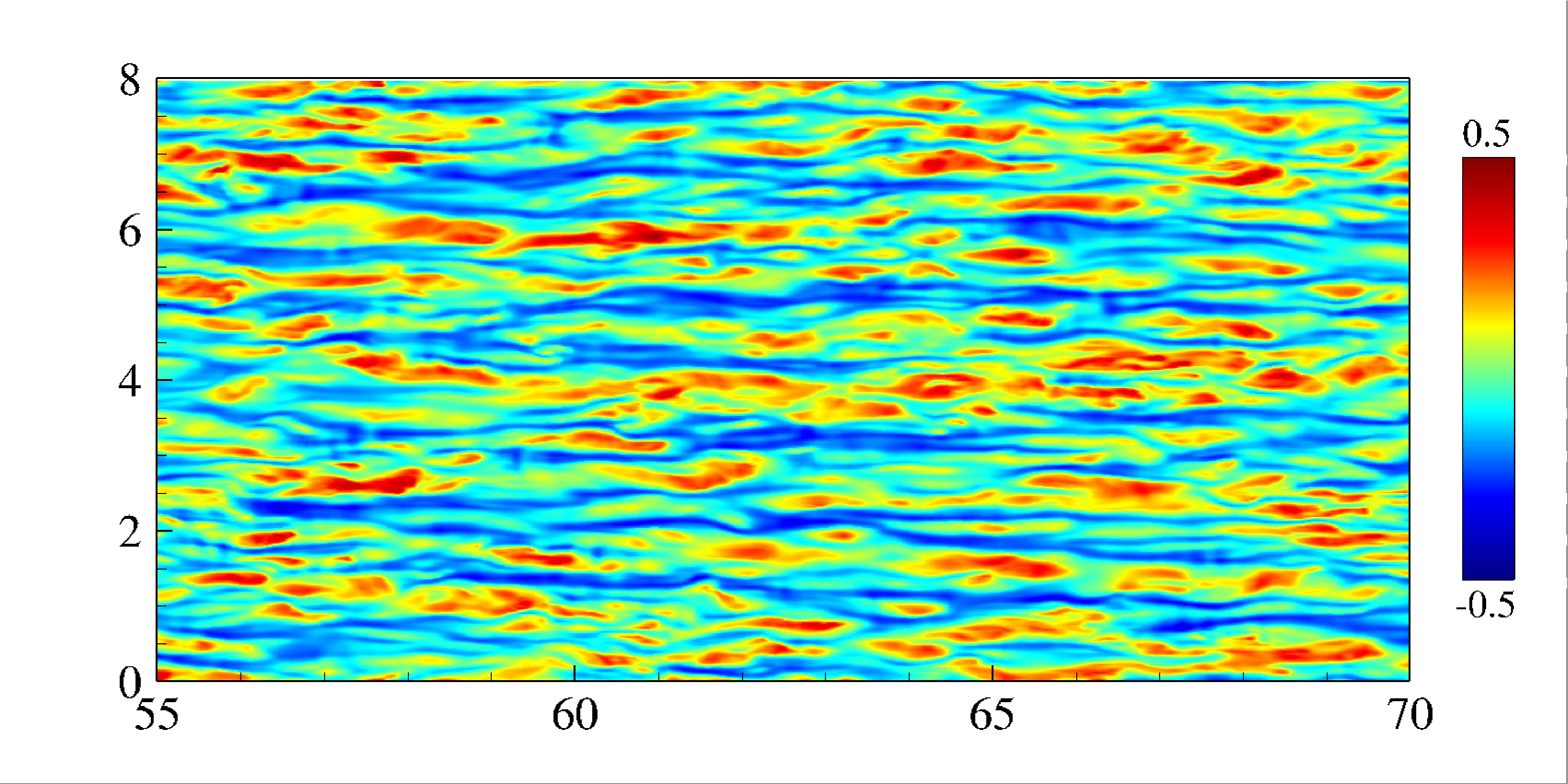}
\put(0,42){(a)}
\put(48,0){$x/\delta_{in}$}
\put(0,20){\rotatebox{90}{$z/\delta_{in}$}}
\end{overpic}~
\begin{overpic}[width=0.5\textwidth,trim={0.2cm 0.2cm 0.2cm 0.2cm},clip]{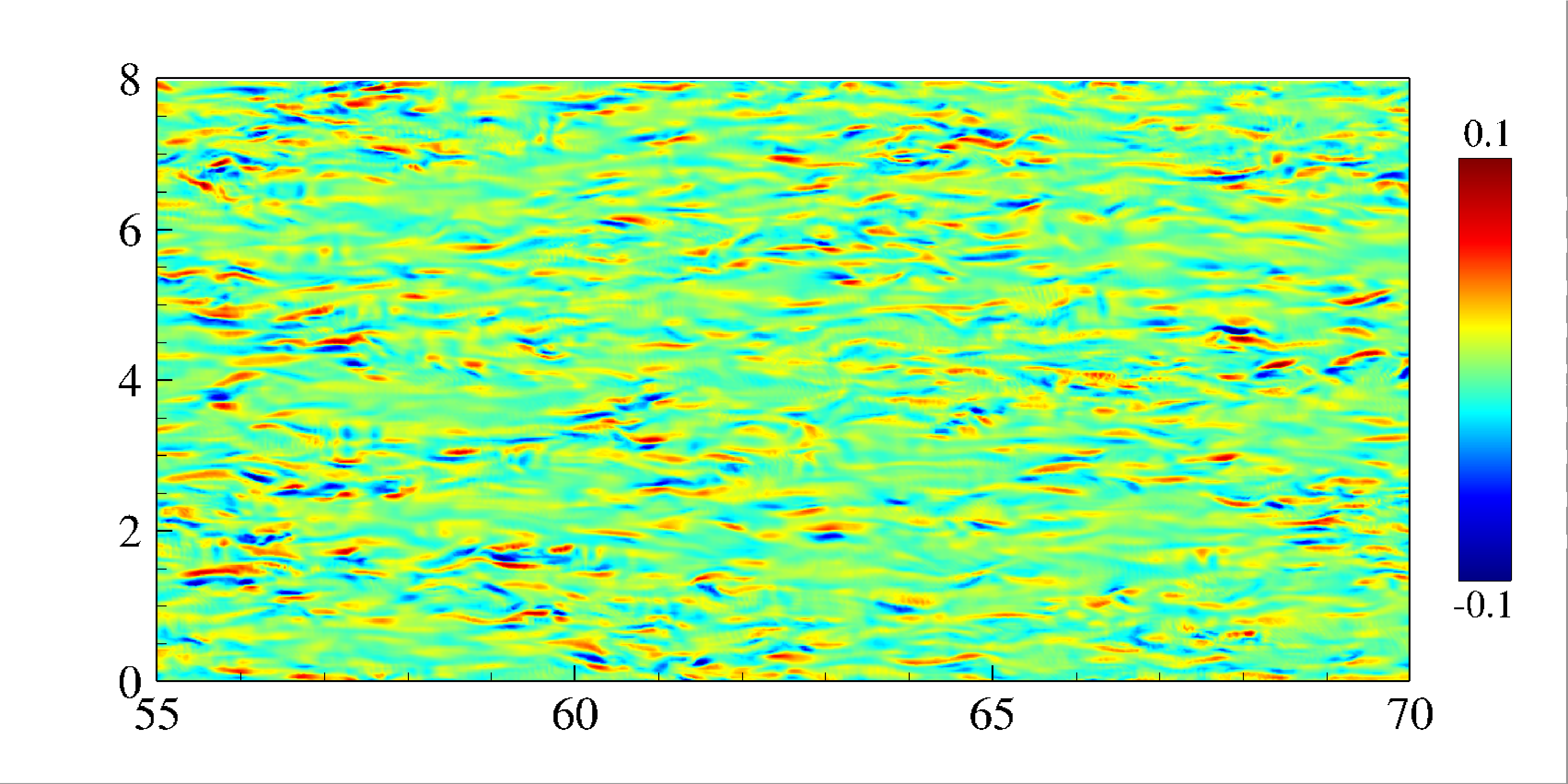}
\put(0,42){(b)}
\put(48,0){$x/\delta_{in}$}
\put(0,20){\rotatebox{90}{$z/\delta_{in}$}}
\end{overpic}\\[0.5ex]
\begin{overpic}[width=0.5\textwidth,trim={0.2cm 0.2cm 0.2cm 0.2cm},clip]{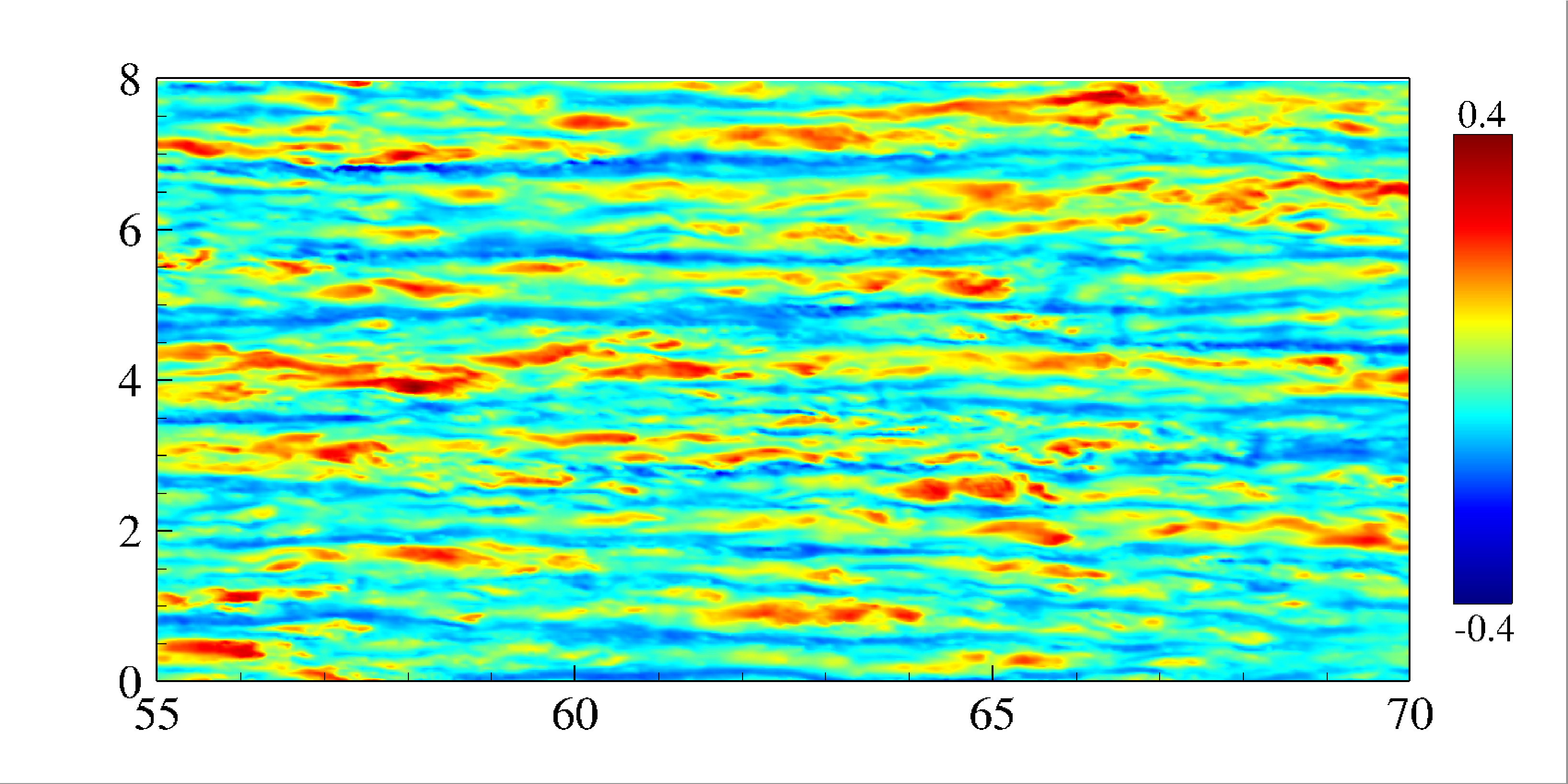}
\put(0,42){(c)}
\put(48,0){$x/\delta_{in}$}
\put(0,20){\rotatebox{90}{$z/\delta_{in}$}}
\end{overpic}~
\begin{overpic}[width=0.5\textwidth,trim={0.2cm 0.2cm 0.2cm 0.2cm},clip]{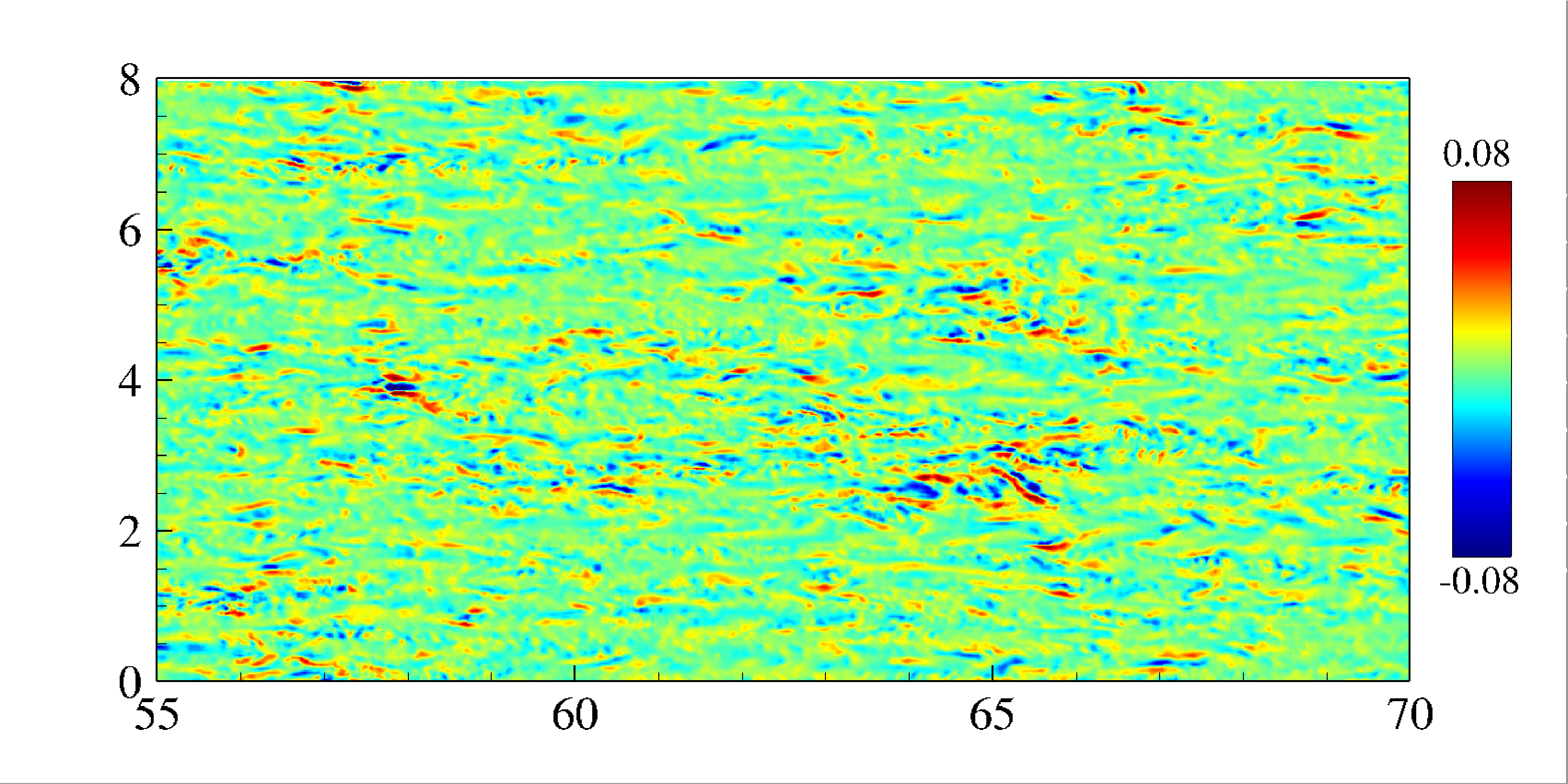}
\put(0,42){(d)}
\put(48,0){$x/\delta_{in}$}
\put(0,20){\rotatebox{90}{$z/\delta_{in}$}}
\end{overpic}\\[0.5ex]
\begin{overpic}[width=0.5\textwidth,trim={0.2cm 0.2cm 0.2cm 0.2cm},clip]{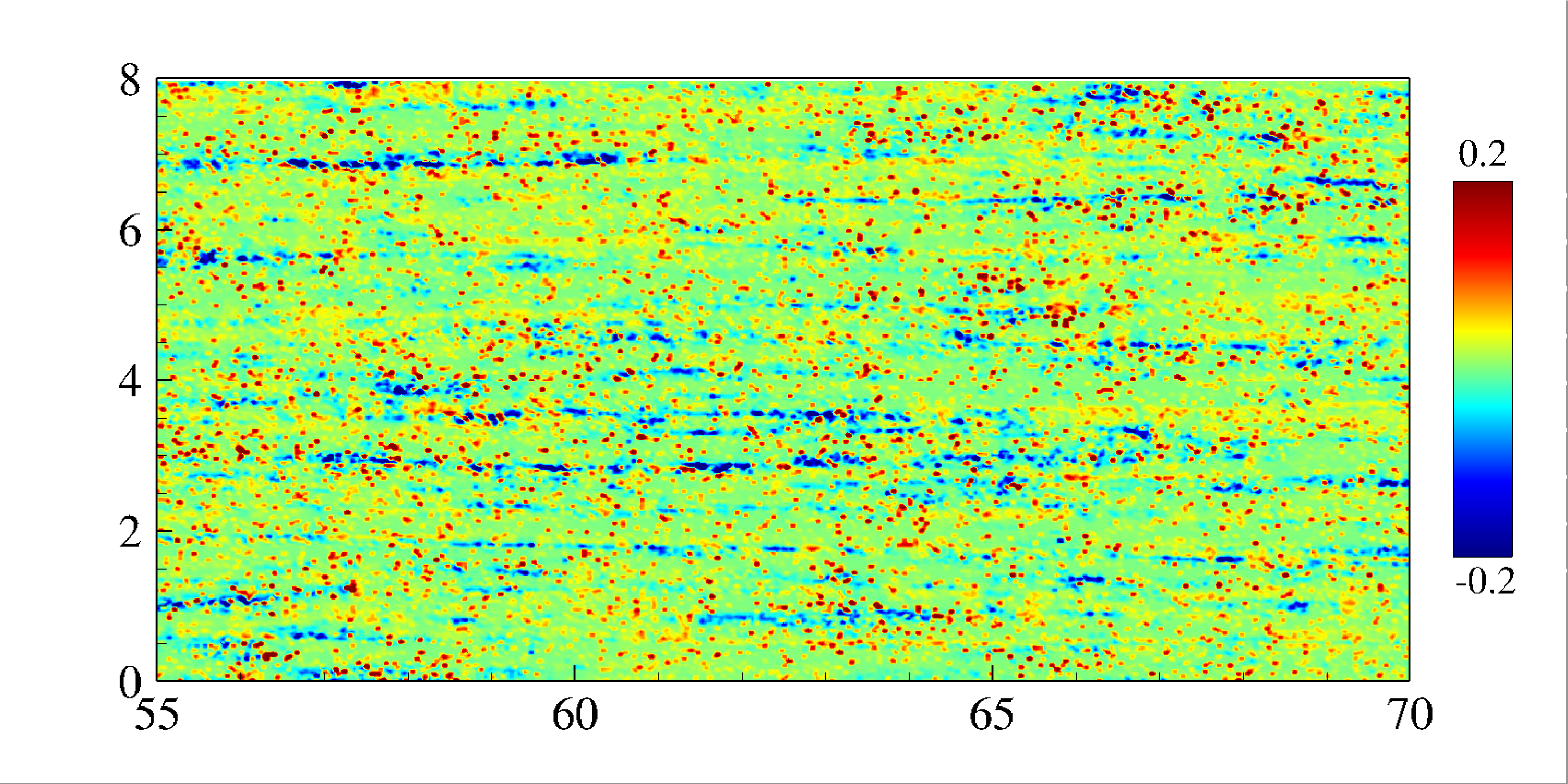}
\put(0,42){(e)}
\put(48,0){$x/\delta_{in}$}
\put(0,20){\rotatebox{90}{$z/\delta_{in}$}}
\end{overpic}~
\begin{overpic}[width=0.5\textwidth,trim={0.2cm 0.2cm 0.2cm 0.2cm},clip]{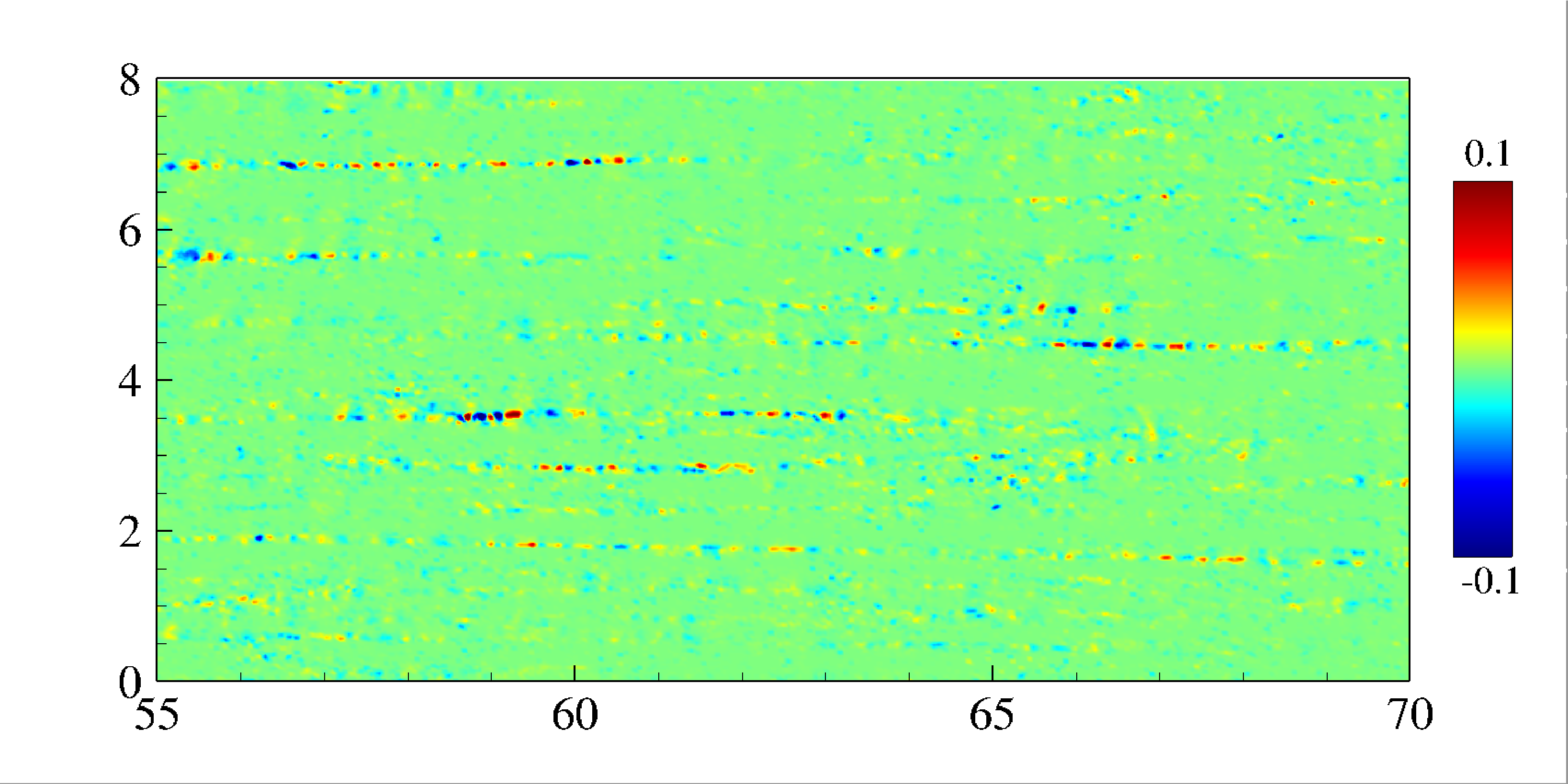}
\put(0,42){(f)}
\put(48,0){$x/\delta_{in}$}
\put(0,20){\rotatebox{90}{$z/\delta_{in}$}}
\end{overpic}\\[0.5ex]
\begin{overpic}[width=0.7\textwidth,trim={0.2cm 0.2cm 0.2cm 0.2cm},clip]{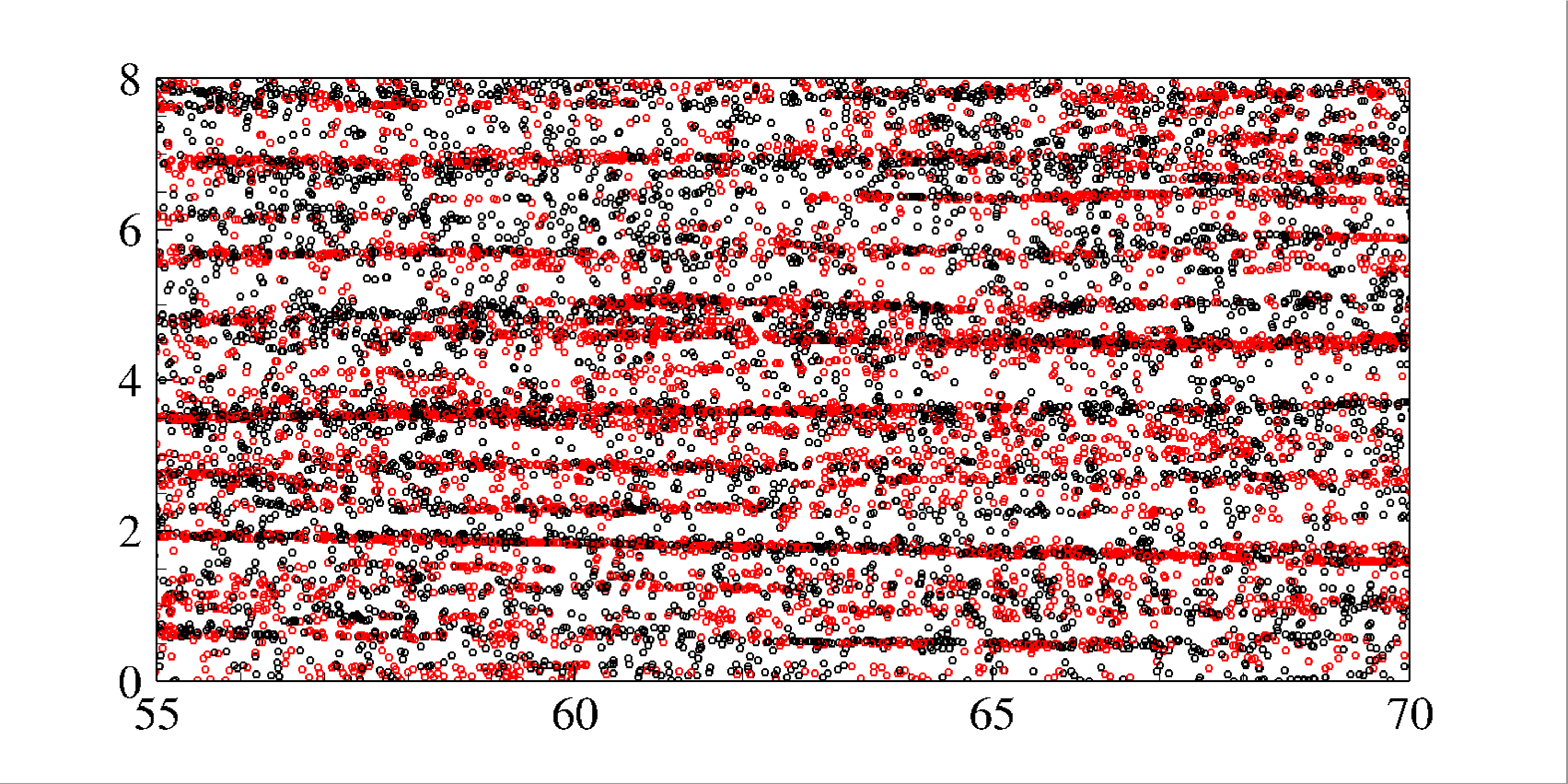}
\put(0,42){(g)}
\put(48,0){$x/\delta_{in}$}
\put(0,20){\rotatebox{90}{$z/\delta_{in}$}}
\put(95,25){\scriptsize {\color{red}$\circ$}~$a_x>0$ }
\put(95,20){\scriptsize $\circ$~$a_x<0$ }
\end{overpic}
\caption{Instantaneous distributions of fluid at $y^+=10$ and particles within $y^+=3 \sim 10$,
(a,c) $u''_1$ and  (b,d) $u''_2$ in cases (a,b) M6-0 and (c,d) M6-2,
(e) $F_{p1}$, (f) $F_{p2}$ and (g) particle distributions in case M6-2.}
\label{fig:inst-m6}
\end{figure}

In this subsection, we consider the instantaneous near-wall distributions of fluid velocity, 
particles, and their feedback forces to examine the effects of two-way coupling 
on flow structures and particle clustering. 

Figures~\ref{fig:inst-m6}(a-d) present streamwise and wall-normal fluid velocity fluctuations 
at $y^+=10$ in cases M6-0 and M6-2 over adiabatic walls. 
In case M6-0, the flow structures resemble those reported in previous studies
~\citep{duan2011direct,duan2010direct}, 
with negative streamwise velocity fluctuations $u''_1<0$ forming streamwise elongated streaks 
and the spanwise paired-up wall-normal velocity fluctuations $u''_2$ induced by the 
near-wall quasi-streamwise vortices. 
In case M6-2, the low-speed streaks are weaker and less meandering, 
while the $u''_2$ fluctuations from the quasi-streamwise vortices are less intense, 
consistent with the lower Reynolds stress at $y^+=10$ (recall Figure~\ref{fig:rey}). 
The presence of particles appears to disrupt near-wall turbulent dynamics at a moderate level,
rather than completely eradicate the self-regeneration cycle~\citep{jimenez2018coherent,
wang2019modulation}.
Additionally, both $u''_1$ and $u''_2$ structures exhibit sporadic spots induced by 
the particle feedback force $F_{p1}$, as evidenced in Figure~\ref{fig:inst-m6}(e). 
Discontinuous as it is, the $F_{p1}$ component generally manifests strong negative values beneath 
the high-speed regions and positive values beneath the low-speed regions of the fluid flow, 
indicating the negative work of particle feedback forces exerted on the fluid. 
In contrast, the wall-normal particle force $F_{p2}$ (Figure~\ref{fig:inst-m6}(f)) 
is notably smaller than $F_{p1}$, except in the regions aligned with low-speed fluid streaks 
where particles congregate (Figure~\ref{fig:inst-m6}(g)). 
Intriguingly, this component shows streamwise positive-negative alternating structures 
reminiscent of dilatational motions observed in canonical compressible turbulent 
boundary layers~\citep{yu2019genuine,yu2021compressibility}.

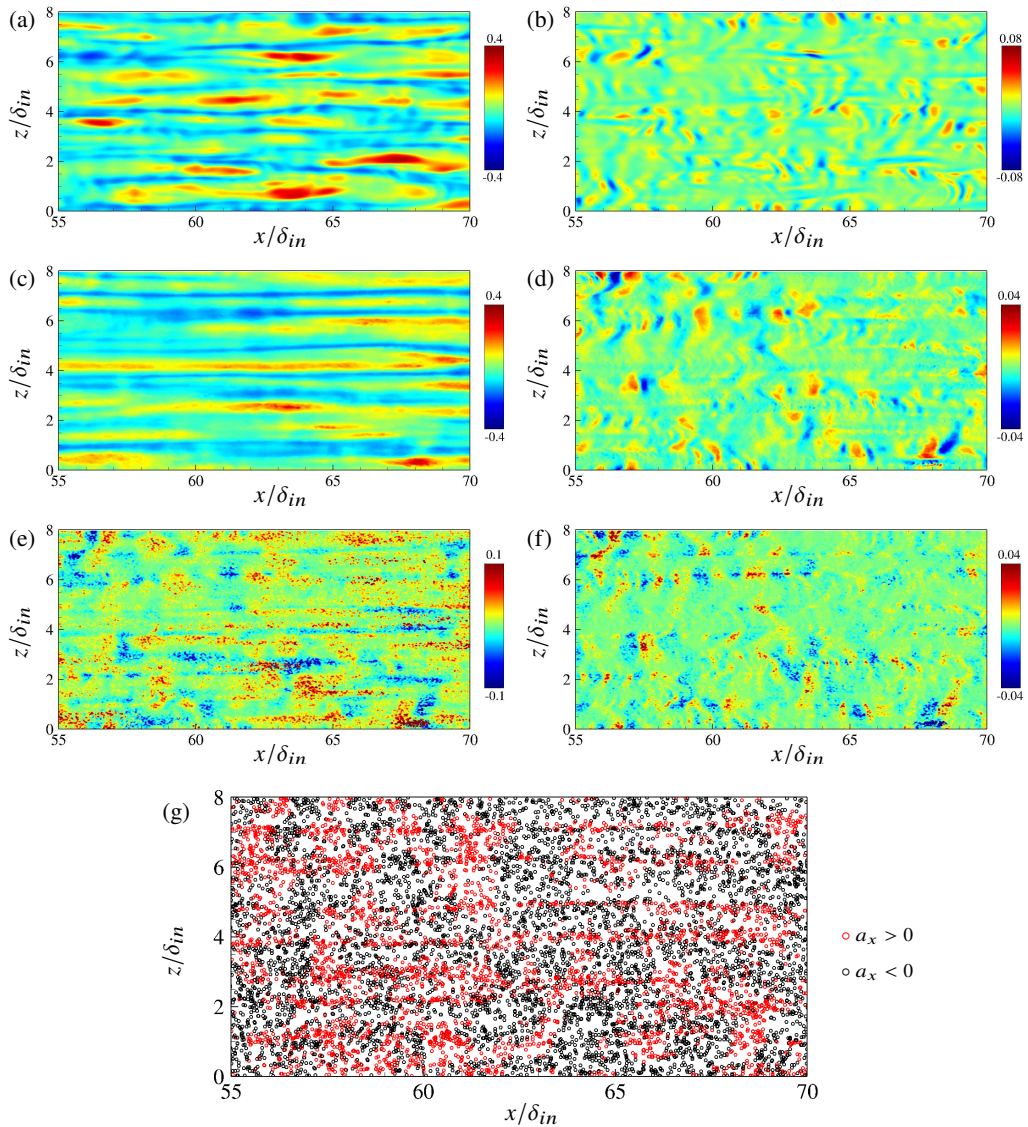
\begin{figure}
\centering
\begin{overpic}[width=0.5\textwidth,trim={0.2cm 0.2cm 0.2cm 0.2cm},clip]{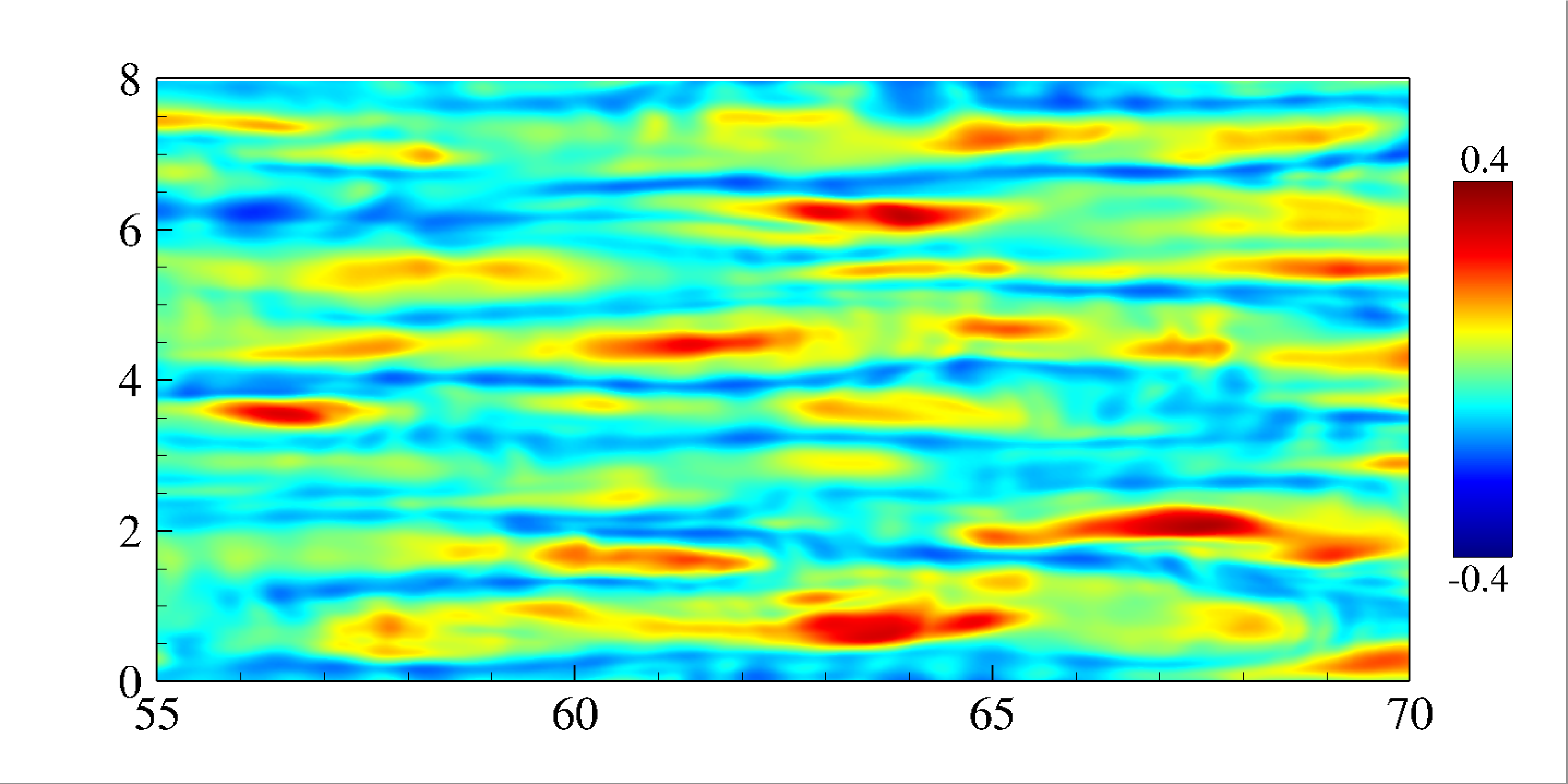}
\put(0,42){(a)}
\put(48,0){$x/\delta_{in}$}
\put(0,20){\rotatebox{90}{$z/\delta_{in}$}}
\end{overpic}~
\begin{overpic}[width=0.5\textwidth,trim={0.2cm 0.2cm 0.2cm 0.2cm},clip]{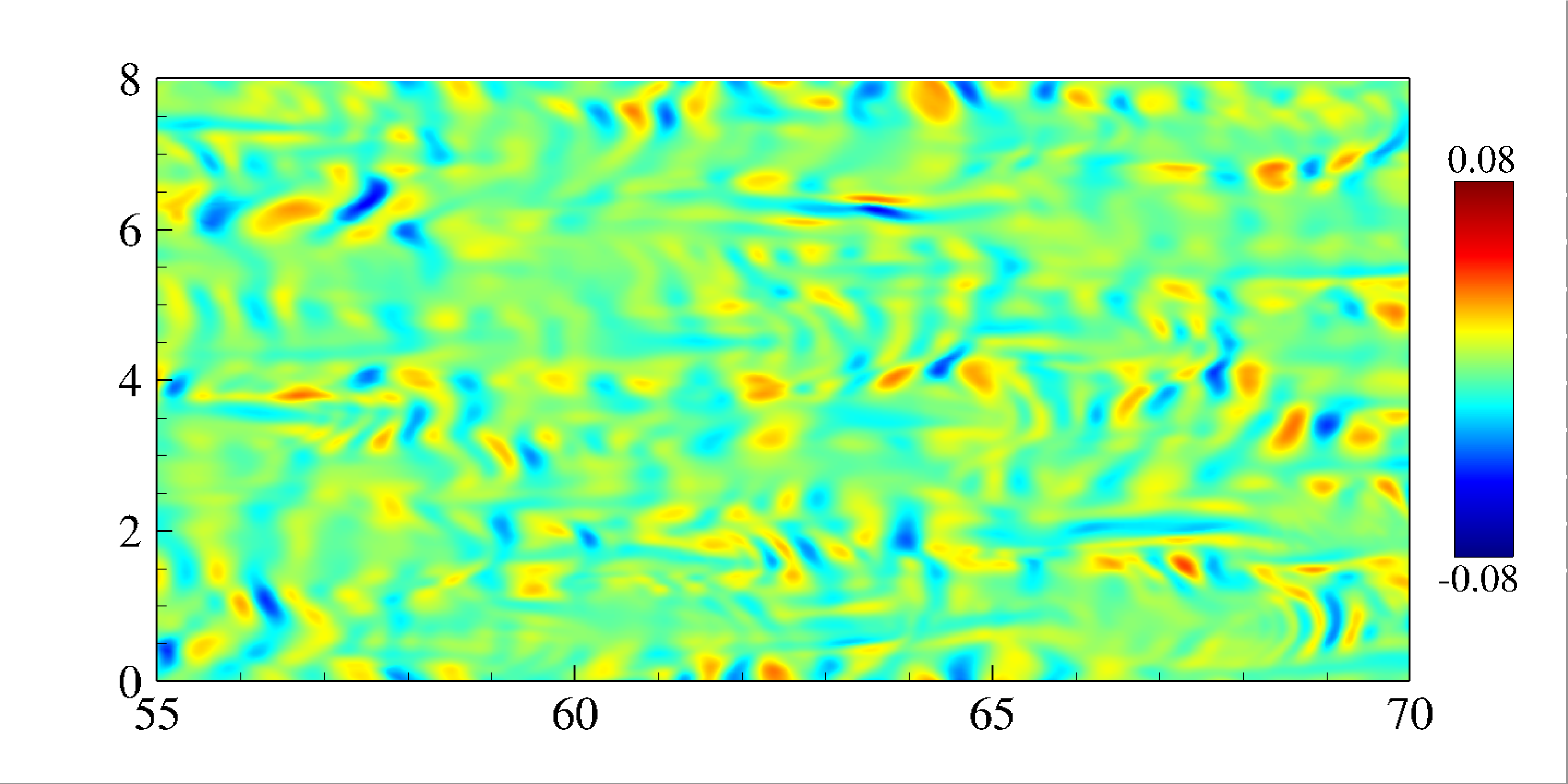}
\put(0,42){(b)}
\put(48,0){$x/\delta_{in}$}
\put(0,20){\rotatebox{90}{$z/\delta_{in}$}}
\end{overpic}\\[0.5ex]
\begin{overpic}[width=0.5\textwidth,trim={0.2cm 0.2cm 0.2cm 0.2cm},clip]{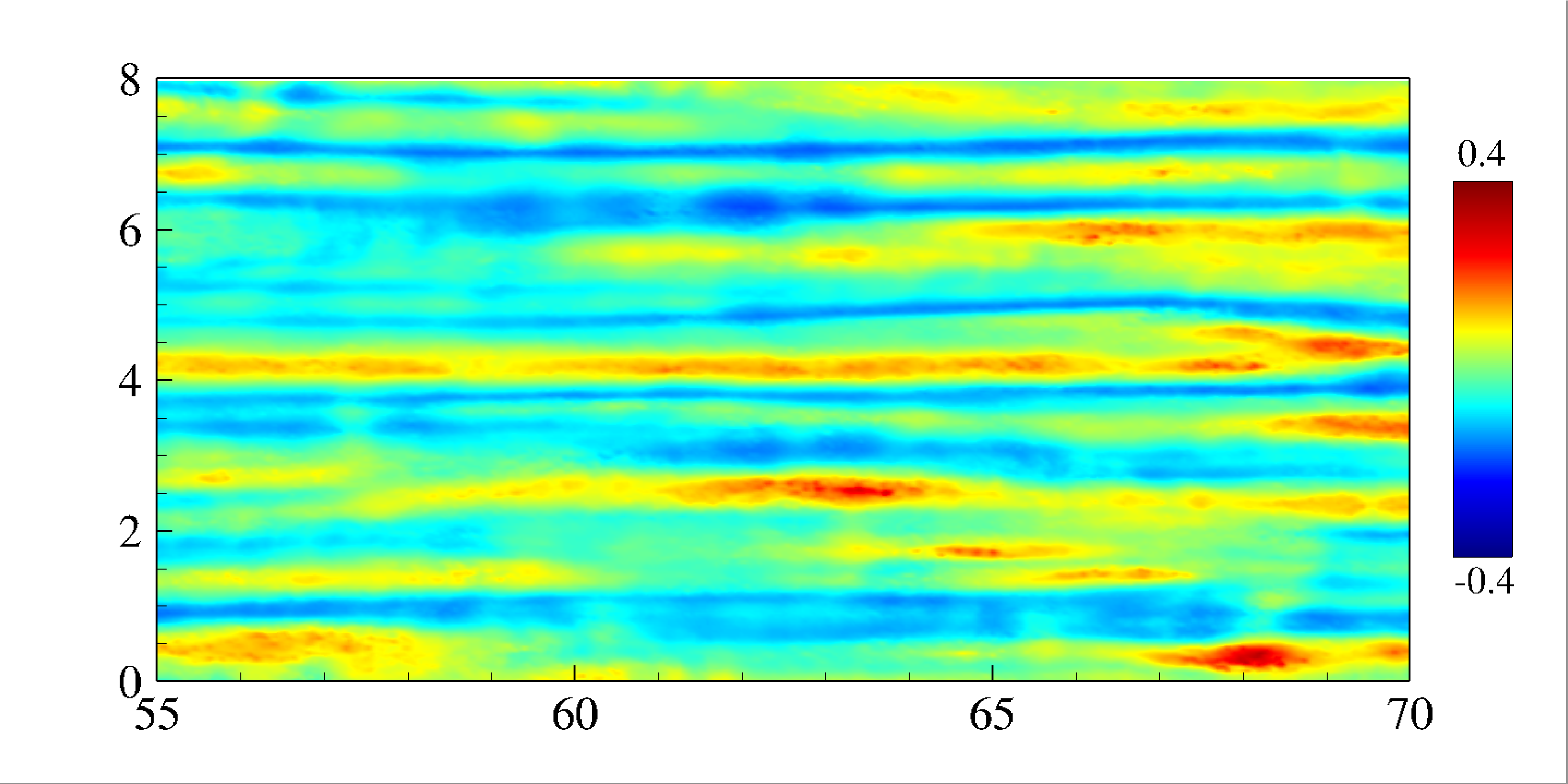}
\put(0,42){(c)}
\put(48,0){$x/\delta_{in}$}
\put(0,20){\rotatebox{90}{$z/\delta_{in}$}}
\end{overpic}~
\begin{overpic}[width=0.5\textwidth,trim={0.2cm 0.2cm 0.2cm 0.2cm},clip]{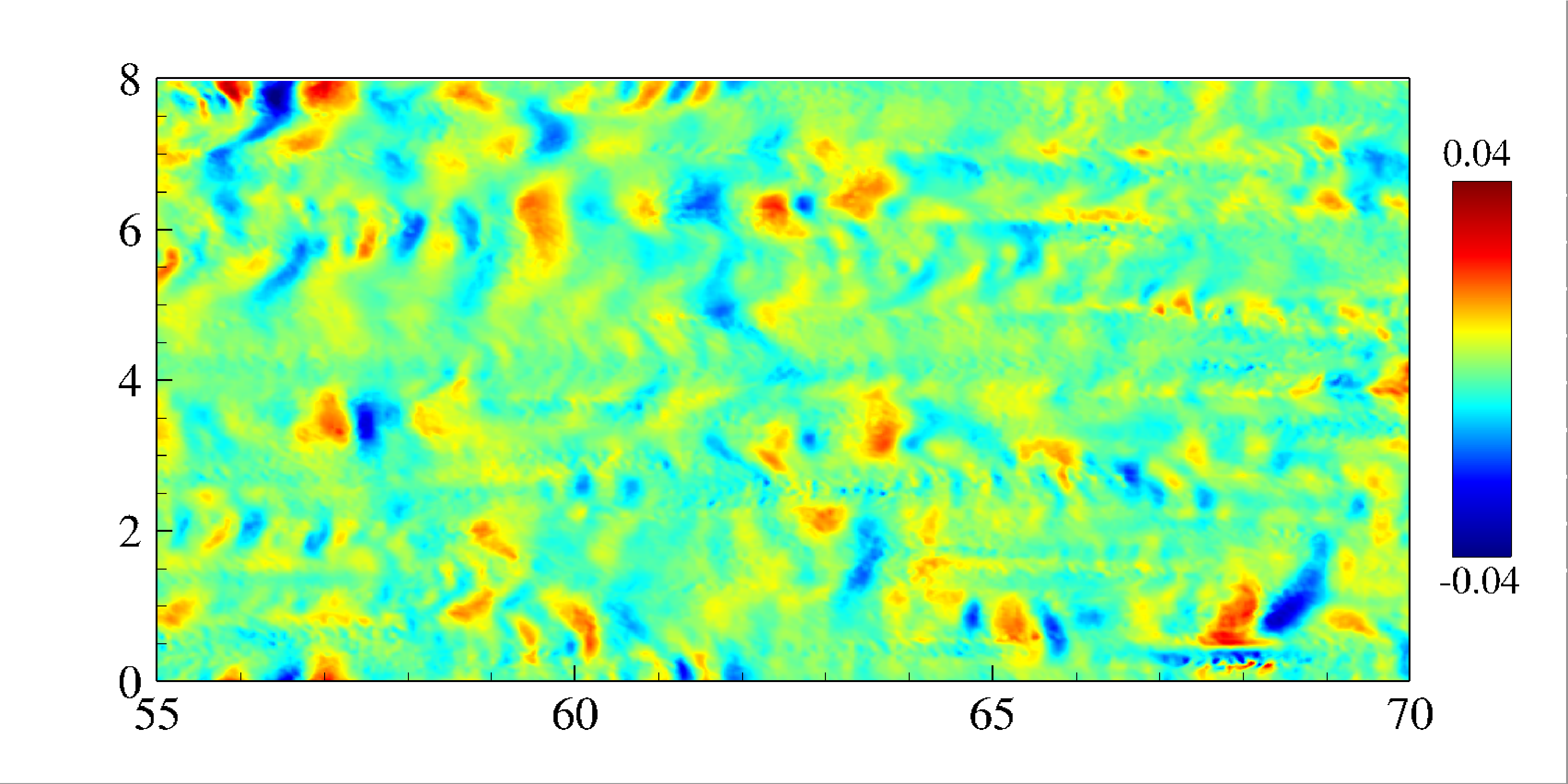}
\put(0,42){(d)}
\put(48,0){$x/\delta_{in}$}
\put(0,20){\rotatebox{90}{$z/\delta_{in}$}}
\end{overpic}\\[0.5ex]
\begin{overpic}[width=0.5\textwidth,trim={0.2cm 0.2cm 0.2cm 0.2cm},clip]{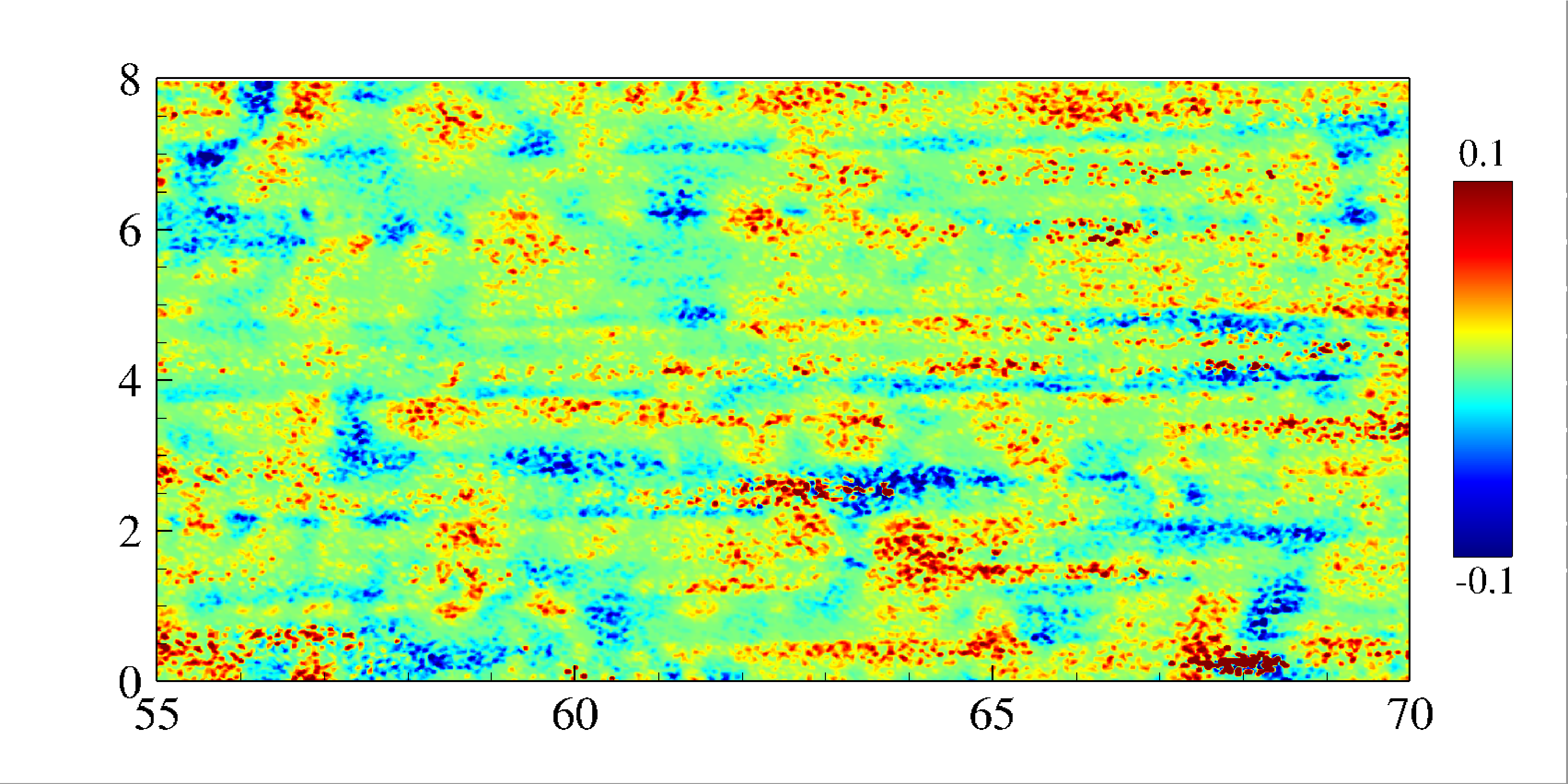}
\put(0,42){(e)}
\put(48,0){$x/\delta_{in}$}
\put(0,20){\rotatebox{90}{$z/\delta_{in}$}}
\end{overpic}~
\begin{overpic}[width=0.5\textwidth,trim={0.2cm 0.2cm 0.2cm 0.2cm},clip]{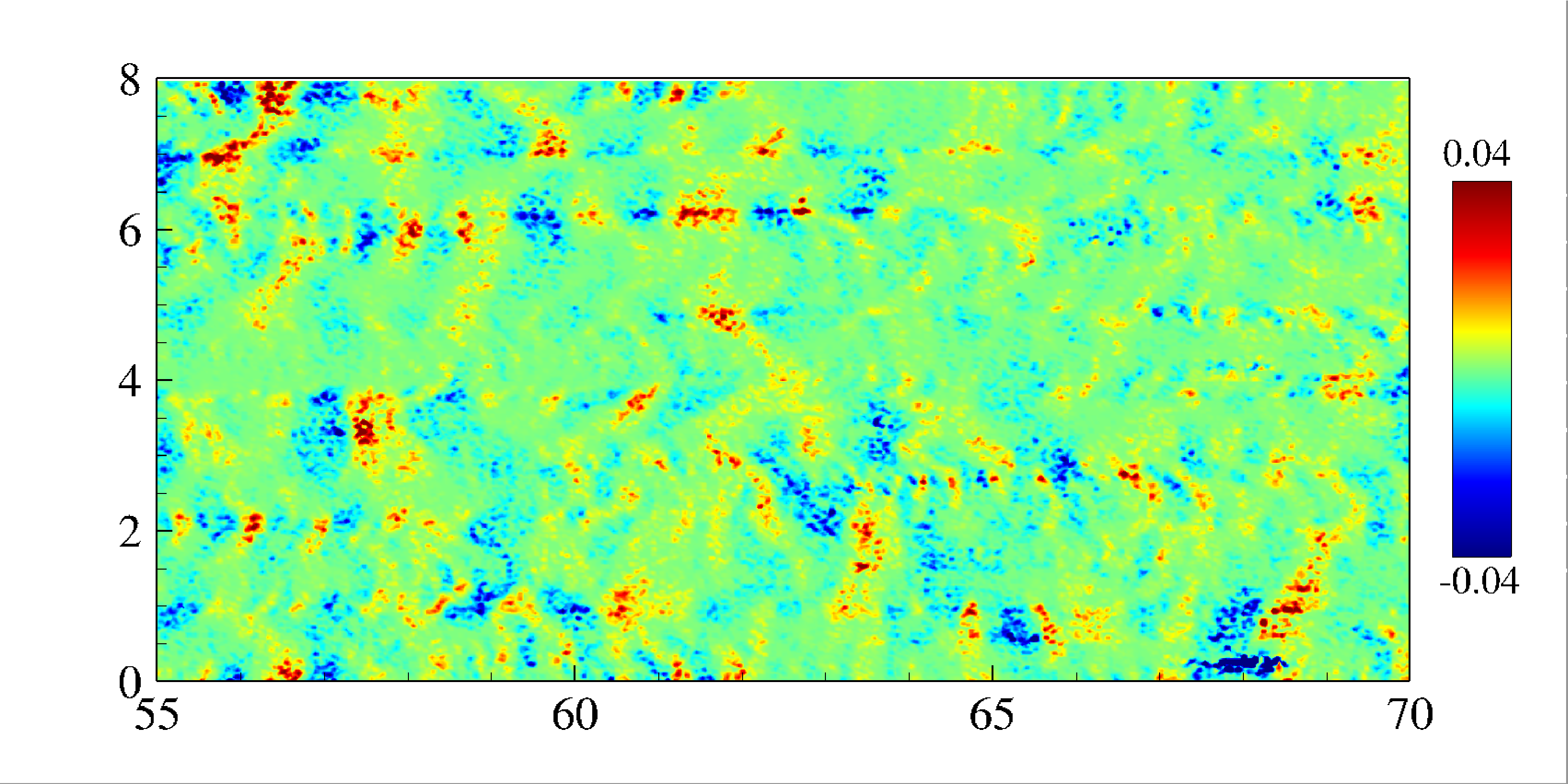}
\put(0,42){(f)}
\put(48,0){$x/\delta_{in}$}
\put(0,20){\rotatebox{90}{$z/\delta_{in}$}}
\end{overpic}\\[0.5ex]
\begin{overpic}[width=0.7\textwidth,trim={0.2cm 0.2cm 0.2cm 0.2cm},clip]{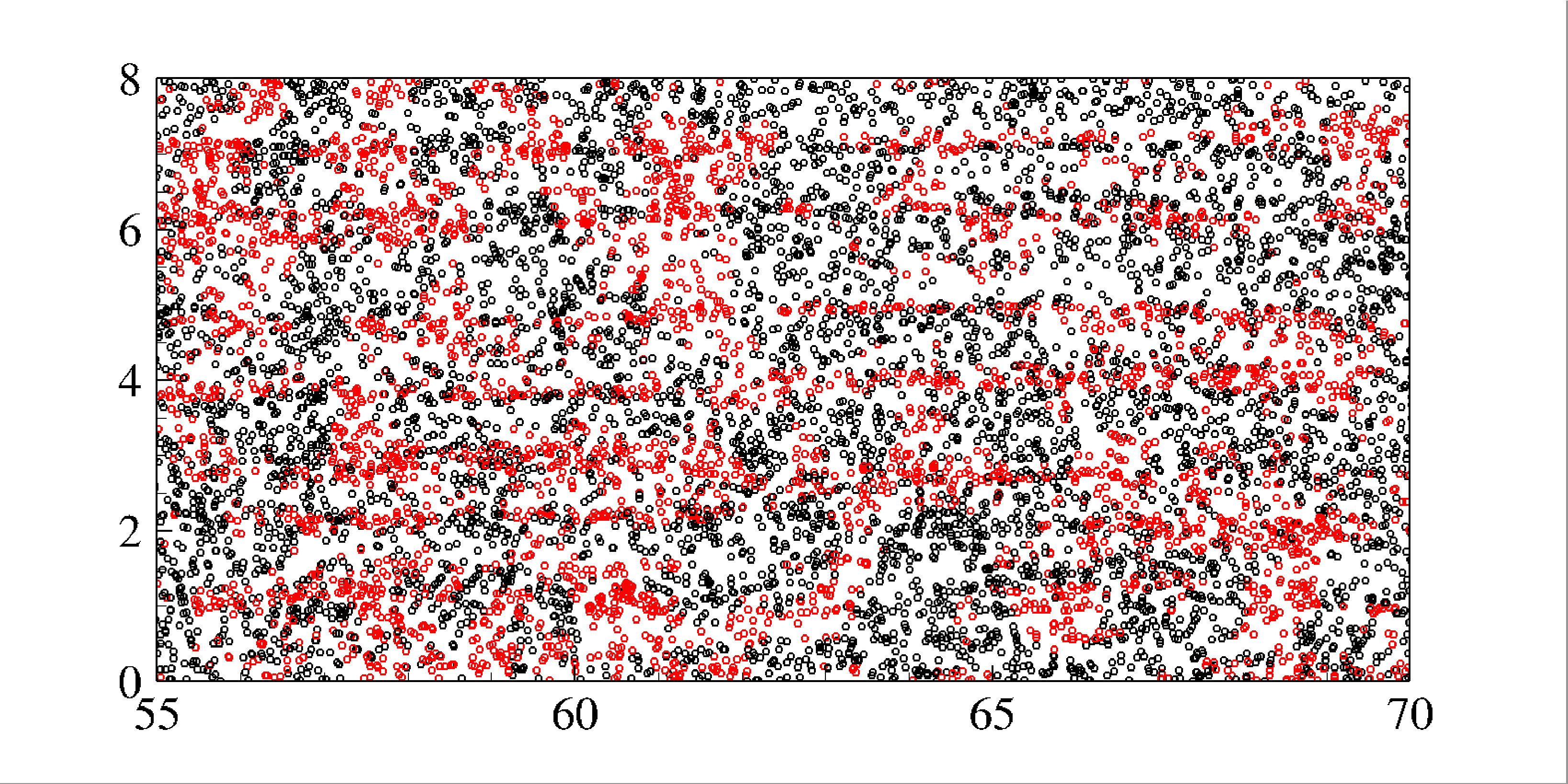}
\put(0,42){(g)}
\put(48,0){$x/\delta_{in}$}
\put(0,20){\rotatebox{90}{$z/\delta_{in}$}}
\put(95,25){\scriptsize {\color{red}$\circ$}~$a_x>0$ }
\put(95,20){\scriptsize $\circ$~$a_x<0$ }
\end{overpic}
\caption{Instantaneous distributions of fluid at $y^+=10$ and particles within $y^+=3 \sim 10$,
(a,c) $u''_1$ and  (b,d) $u''_2$ in cases (a,b) M6C-0 and (c,d) M6C-2,
(e) $F_{p1}$, (f) $F_{p2}$ and (g) particle distributions in case M6C-2.}
\label{fig:inst-m6c}
\end{figure}

Figure~\ref{fig:inst-m6c} displays the near-wall instantaneous velocity, 
particle feedback forces, and particle distributions in the cold wall cases M6C-0 and M6C-2. 
In case M6C-0, the coherence of low-speed streaks (Figure~\ref{fig:inst-m6c}(a))
is stronger compared to turbulence over adiabatic walls~\citep{duan2010direct,
hadjadj2015effect}, 
and the manifestation of streamwise positive-negative alternating structures, 
known as travelling wave packets~\citep{yu2021compressibility,zhang2022wall}, 
is more prominent in both $u''_1$ and $u''_2$ velocity components of the fluid
(Figure~\ref{fig:inst-m6c}(b)). 
These are the typical characteristics of cold wall turbulence, particularly in
high Mach number flows~\citep{yu2021compressibility,xu2021effect,zhang2022wall}. 
In the particle-laden case M6C-2, the velocity streaks (Figure~\ref{fig:inst-m6c}(c)) 
are less meandering and slightly weaker than those in M6C-0. 
The high-speed regions in the particle-laden case M6C-2 exhibit longer streaky structures,
a phenomenon that is absent in the particle-free case M6C-0.
The travelling wave packets are also weaker, as indicated by lower magnitudes of 
the wall-normal velocity fluctuations (Figure~\ref{fig:inst-m6c}(d)). 
In M6C-2, the tendency of particles to cluster in low-speed streaks is less apparent
(Figure~\ref{fig:inst-m6c}(g)), 
showing a more uniform distribution in the wall-parallel plane. 
However, the accelerated and decelerated particles appear to organize in patterns similar to 
travelling wave packets, leading to similar distributions of particle feedback forces $F_{p1}$ 
and $F_{p2}$ (Figure~\ref{fig:inst-m6c}(e-f)). 
Although these particle feedback forces exhibit spatial discontinuities, 
the velocity fluctuations in Figures~\ref{fig:inst-m6c}(c,d) are devoid of 
these patterns, probably due to the higher fluid viscosity that rapidly 
dissipates the high wavenumber fluctuations.

The resemblance observed in the instantaneous distributions between near-wall dilatational motions 
and particle feedback forces implies that particles probably influence the
near-wall dilatational motions of the fluid flow, reciprocally affecting particle dynamics. 
Further elaboration on this topic will be discussed in the subsequent subsection.

\subsection{Solenoidal and dilatational motions modulated by particles} \label{subsec:mod}

The particle feedback force distributions shown in the subsection above indicate 
the potential dynamic modulation of both solenoidal (near-wall velocity streaks and vortices) 
and dilatational (travelling wave packets) motions. 
In order to evaluate the influences of the particle feedback forces on the solenoidal and 
dilatational motions, we perform Helmholtz decomposition
~\citep{hirasaki1970boundary,pirozzoli2010dynamical} to split the velocity fluctuations 
into a vortical, solenoidal portion and a potential, dilatational portion.
By solving the following Poisson equations
for the velocity potential function $\varphi$ and vorticity vector potential function $A_i$
\begin{equation}
\frac{\partial^2 A_{i}}{\partial x_j \partial x_j} = -\omega'_i,~~
{\rm at}~y=0:~A_1 = A_3 = 0, \frac{\partial A_2}{\partial y}=0,
\end{equation}
\begin{equation}
\frac{\partial^2 \varphi}{\partial x_j \partial x_j} = \theta',~~
{\rm at}~y=0:~\frac{\partial \varphi}{\partial y}=0,
\end{equation}
with $\omega_i$ and $\theta$ the vorticity vector and velocity divergence,
the solenoidal and dilatational velocity components can be obtained as
\begin{equation}
u^s_i = -\epsilon_{ijk} \frac{\partial A_k}{\partial x_j},~~
u^d_i = \frac{\partial \theta}{\partial x_i},
\end{equation}
hereinafter denoted by the superscript $s$ and $d$, respectively.

\begin{figure}
\centering
\begin{overpic}[width=0.5\textwidth]{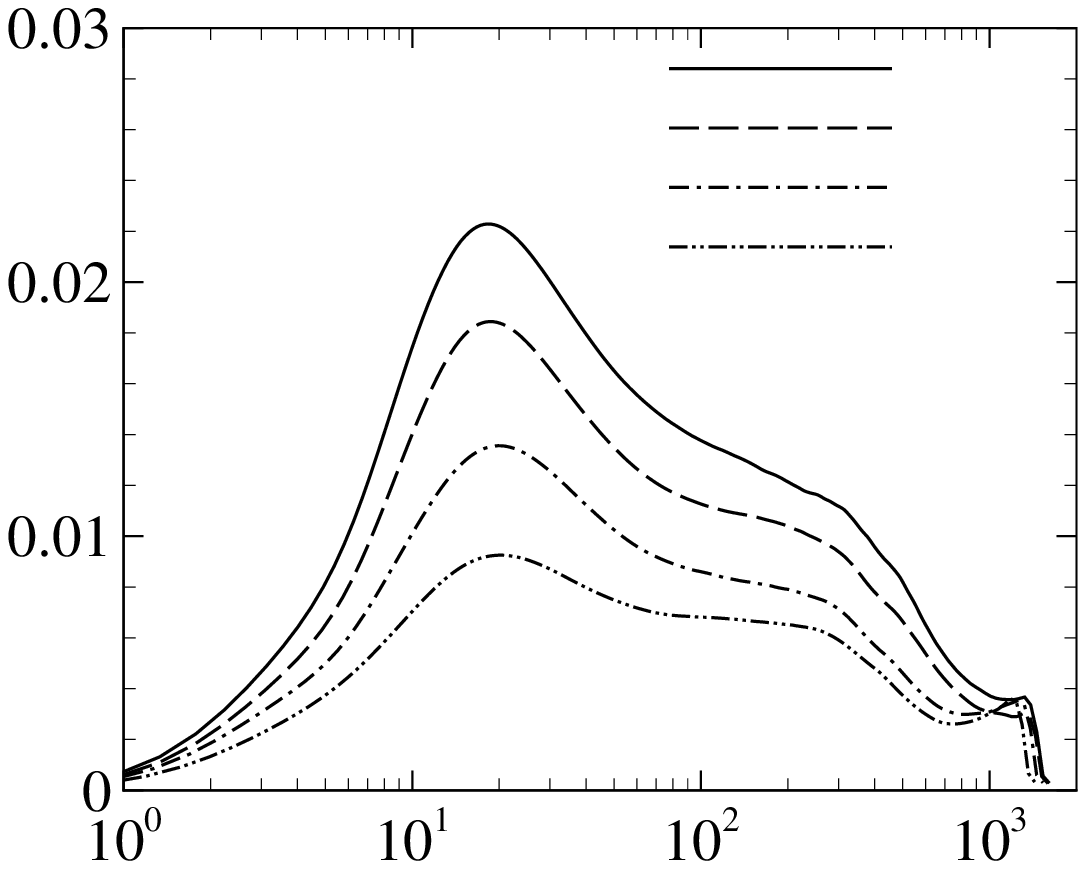}
\put(0,70){(a)}
\put(48,0){$y^+$}
\put(-2,35){\rotatebox{90}{$R^{d+}_{22}$}}
\put(75,67){\small M6-0}
\put(75,62){\small M6-1}
\put(75,57.5){\small M6-2}
\put(75,52.5){\small M6-3}
\end{overpic}~
\begin{overpic}[width=0.5\textwidth]{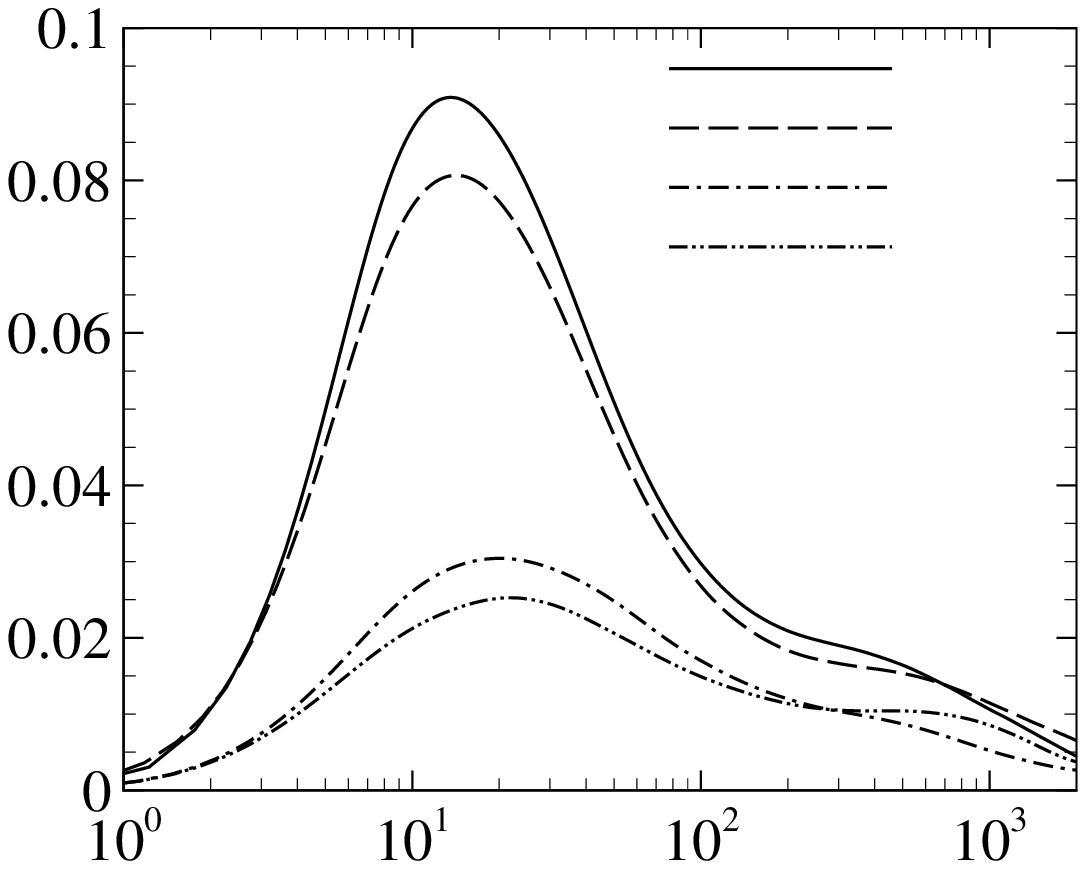}
\put(0,70){(b)}
\put(48,0){$y^+$}
\put(-2,35){\rotatebox{90}{$R^{d+}_{22}$}}
\put(75,67){\small M6C-0}
\put(75,62){\small M6C-1}
\put(75,57.5){\small M6C-2}
\put(75,52.5){\small M6C-3}
\end{overpic}\\[2.0ex]
\begin{overpic}[width=0.5\textwidth]{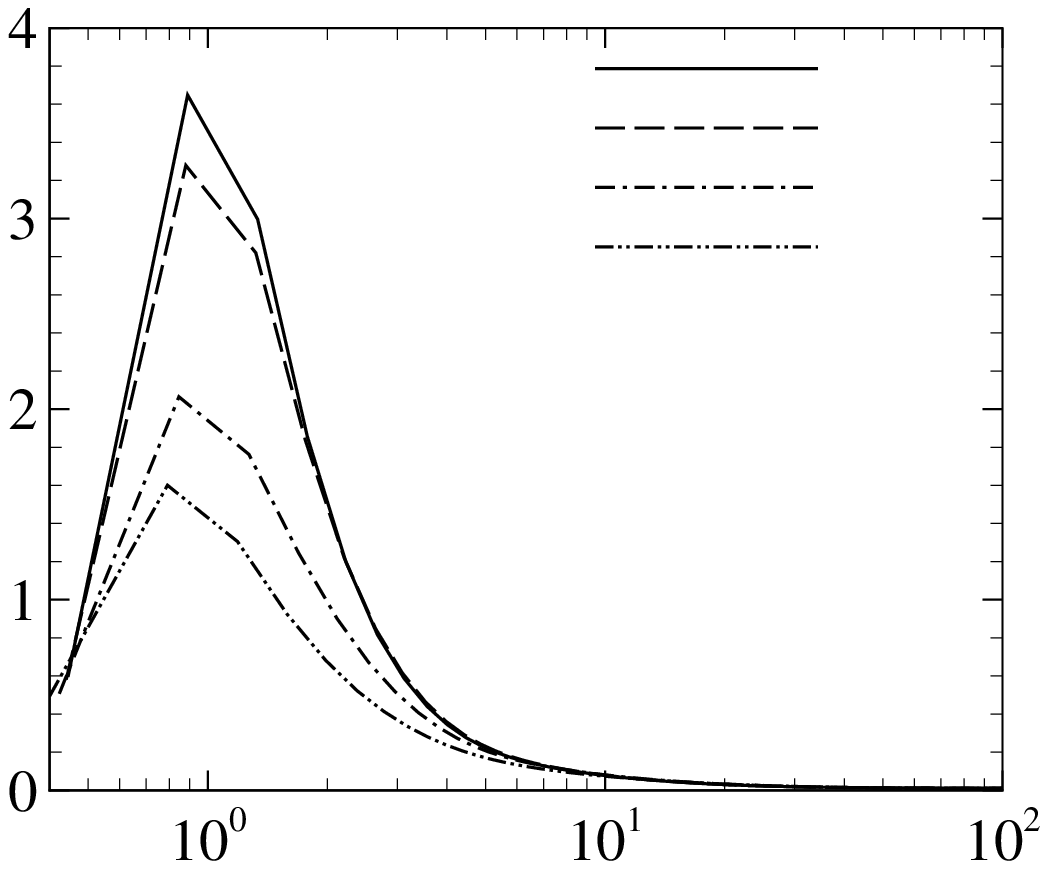}
\put(0,70){(c)}
\put(48,0){$y^+$}
\put(-2,30){\rotatebox{90}{$R^{d+}_{22}/R^{s+}_{22}$}}
\put(75,67){\small M6-0}
\put(75,62){\small M6-1}
\put(75,57.5){\small M6-2}
\put(75,52.5){\small M6-3}
\end{overpic}~
\begin{overpic}[width=0.5\textwidth]{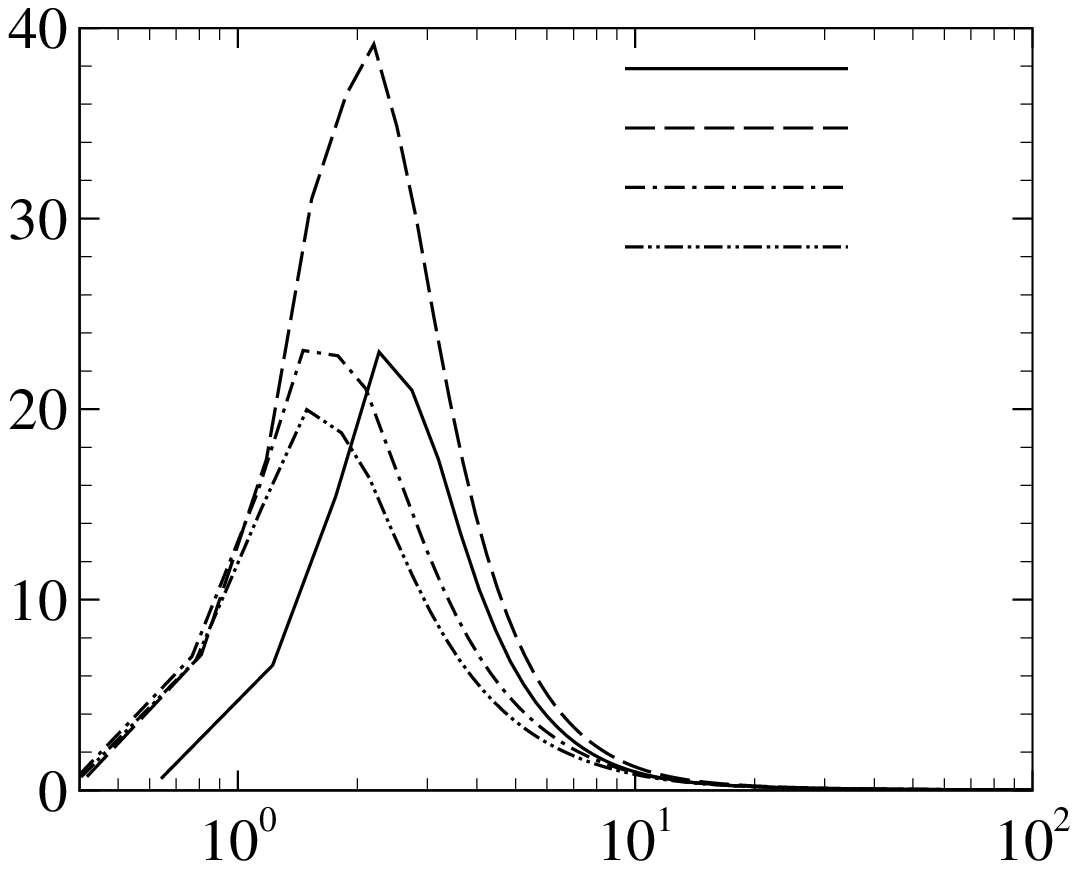}
\put(0,70){(d)}
\put(48,0){$y^+$}
\put(-2,30){\rotatebox{90}{$R^{d+}_{22}/R^{s+}_{22}$}}
\put(75,67){\small M6C-0}
\put(75,62){\small M6C-1}
\put(75,57.5){\small M6C-2}
\put(75,52.5){\small M6C-3}
\end{overpic}
\caption{Wall-normal distribution of (a,b) dilatational wall-normal Reynolds stress $R^{d+}_{22}$
and (c,d) its ratio with the solenoidal component $R^{d+}_{22}/R^{s+}_{22}$.}
\label{fig:reyhd}
\end{figure}

In Figures~\ref{fig:reyhd}(a,b), we present the dilatational portion of 
the wall-normal Reynolds stress component $R^{d}_{22} = \overline{\rho u''^d_2 u''^d_2}$,
normalized by the mean wall shear stress. 
For both the adiabatic (M6) and cold wall (M6C) cases, $R^{d+}_{22}$ consistently decreases 
with increasing mass loading, with a more pronounced abatement observed in the latter ones.
Figures~\ref{fig:reyhd}(c,d) present the ratio between $R^{d+}_{22}$ and $R^{s+}_{22}$.
The highest ratio values are observed within the viscous sublayer.
In adiabatic wall cases, this ratio peaks for case M6-0 at approximately $y^+ = 4.0$
and decreases with mass loading. 
The declining solenoidal component coupled with lower ratios in these cases 
suggests that particle feedback forces suppress dilatational motions, and the degree of such
suppression is higher than that of solenoidal velocities.
Conversely, in cold wall cases, the ratios remain at high levels without a clear decreasing trend,
and are notably higher than those in adiabatic wall cases.
This is consistent with the previous findings that the dilatational motions are stronger
in turbulence over cold walls~\citep{yu2021compressibility}.


\begin{figure}
\centering
\begin{overpic}[width=0.5\textwidth,trim={0.2cm 0.2cm 0.2cm 0.2cm},clip]{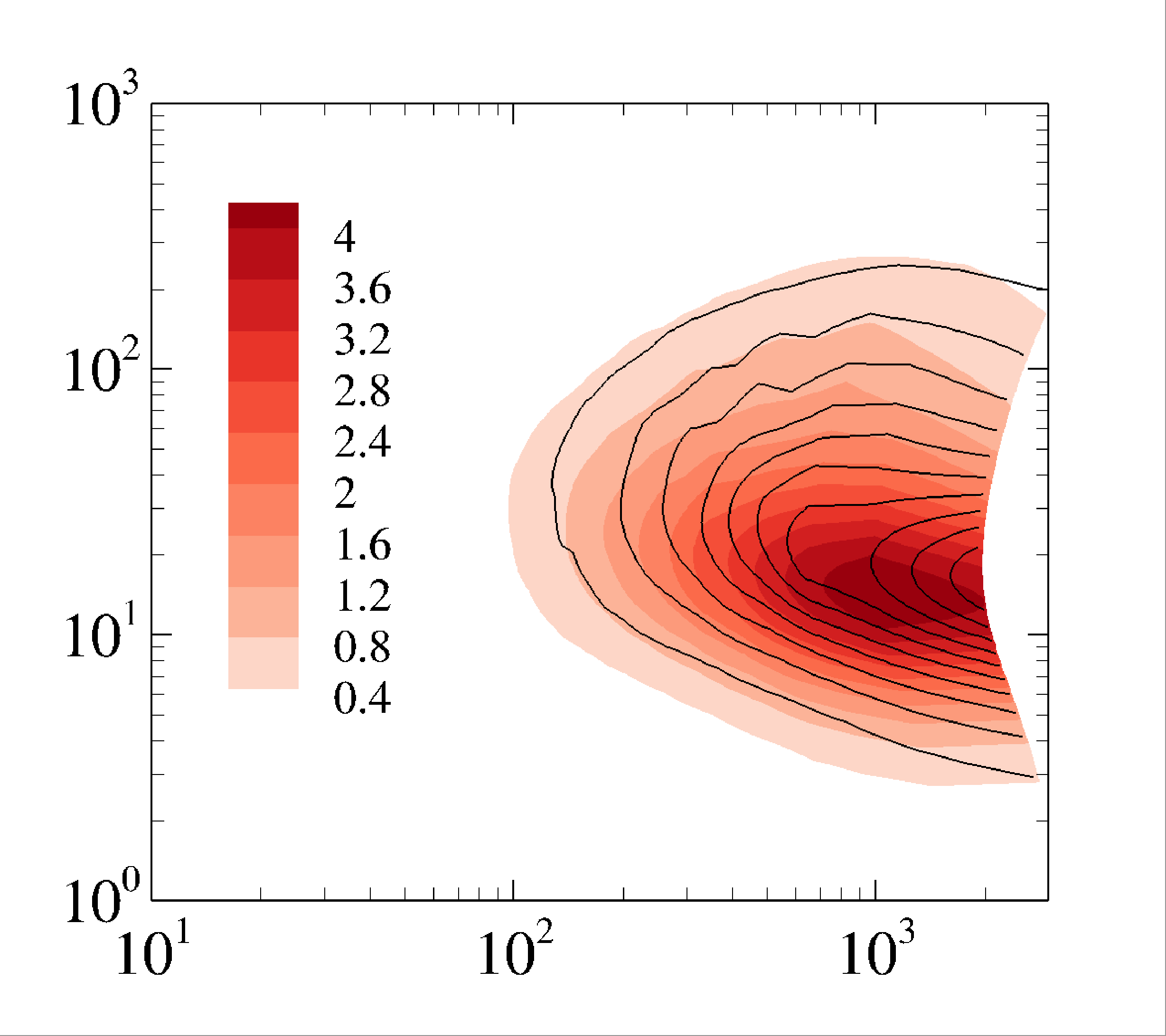}
\put(-5,80){(a)}
\put(48,0){$\lambda^+_{x,VP}$}
\put(-4,40){\rotatebox{90}{$y^+_{VP}$}}
\end{overpic}~
\begin{overpic}[width=0.5\textwidth,trim={0.2cm 0.2cm 0.2cm 0.2cm},clip]{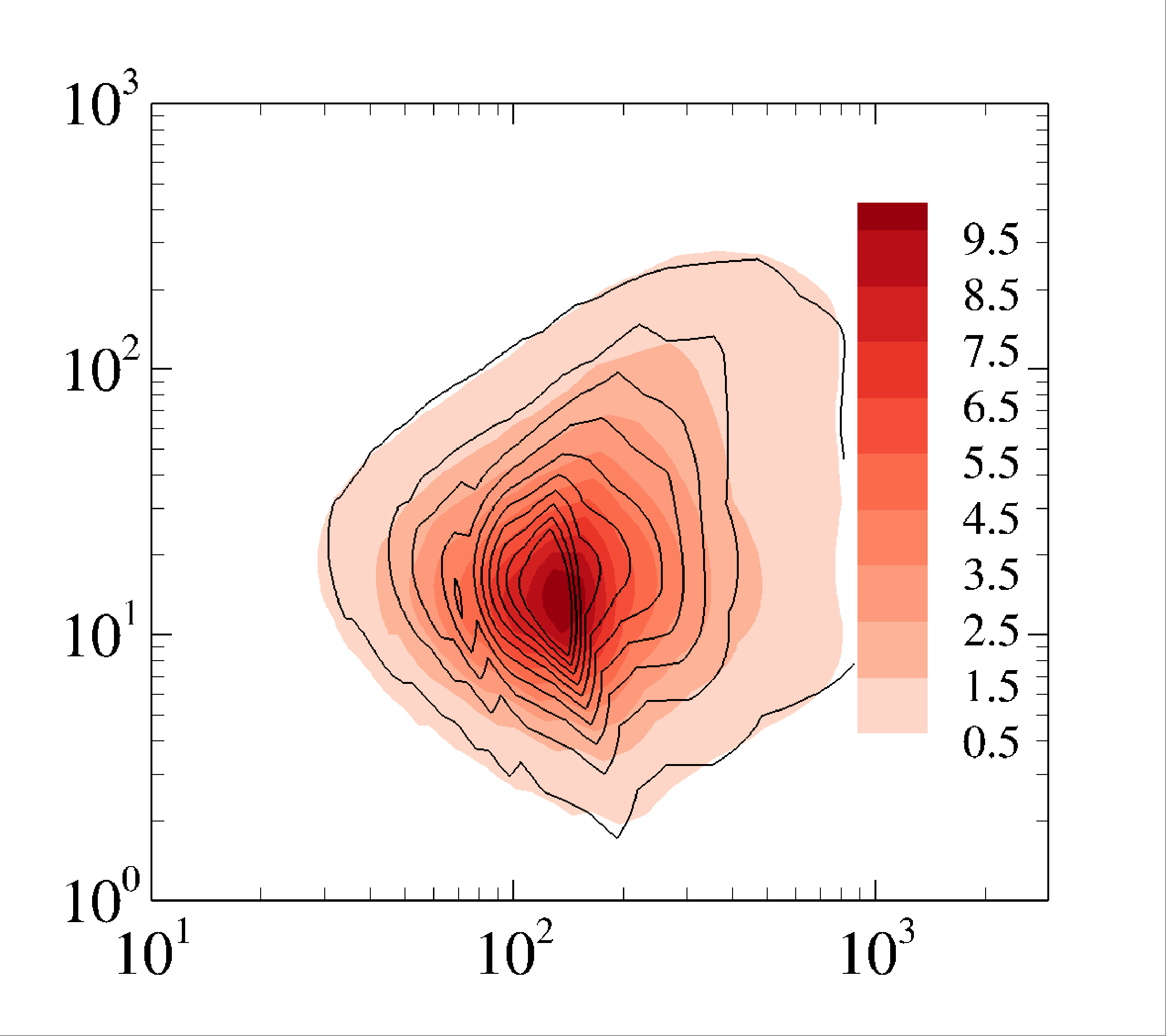}
\put(-5,80){(b)}
\put(48,0){$\lambda^+_{z,VP}$}
\put(-4,40){\rotatebox{90}{$y^+_{VP}$}}
\end{overpic}\\[1.5ex]
\begin{overpic}[width=0.5\textwidth,trim={0.2cm 0.2cm 0.2cm 0.2cm},clip]{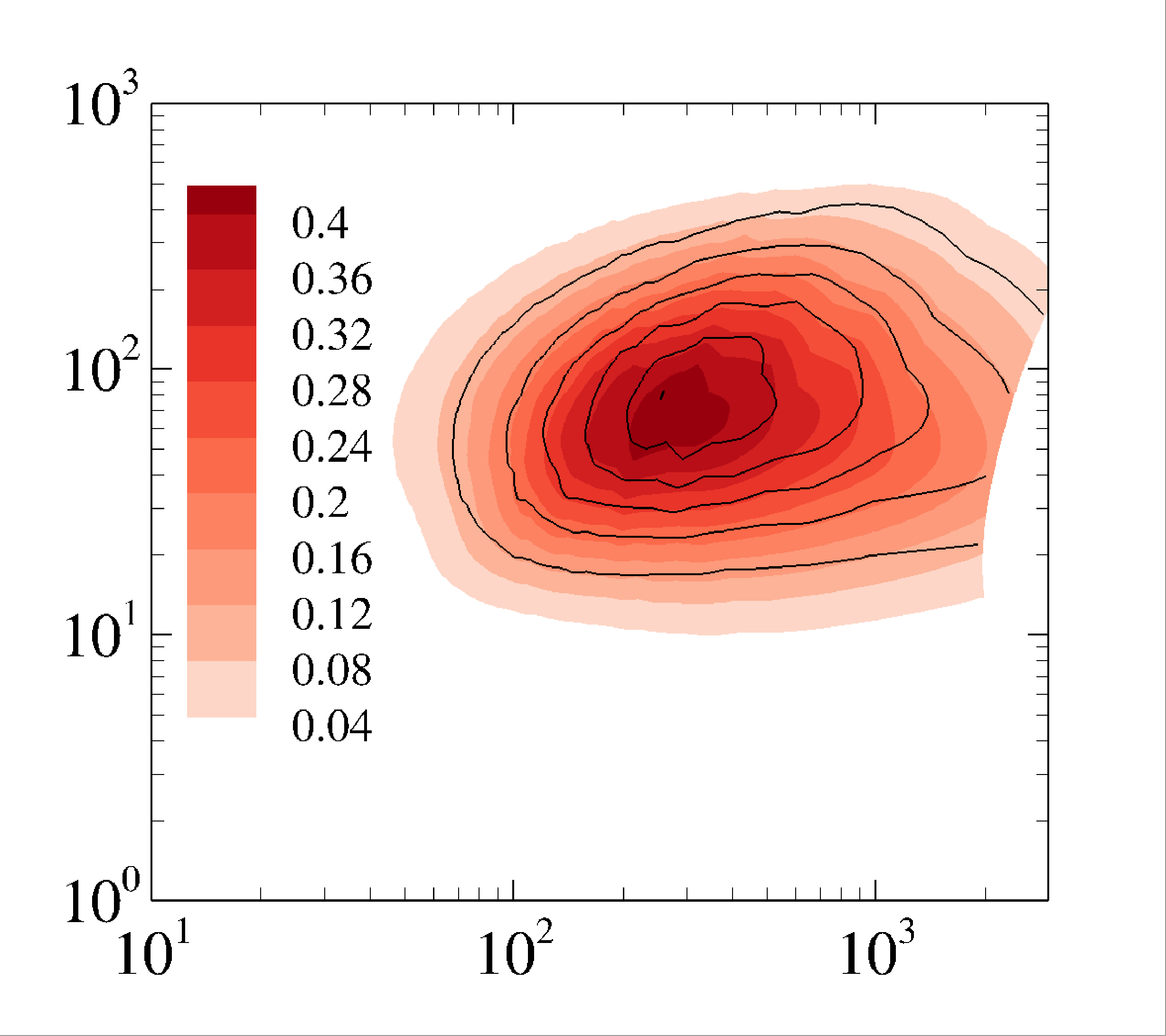}
\put(-5,80){(c)}
\put(48,0){$\lambda^+_{x,VP}$}
\put(-4,40){\rotatebox{90}{$y^+_{VP}$}}
\end{overpic}~
\begin{overpic}[width=0.5\textwidth,trim={0.2cm 0.2cm 0.2cm 0.2cm},clip]{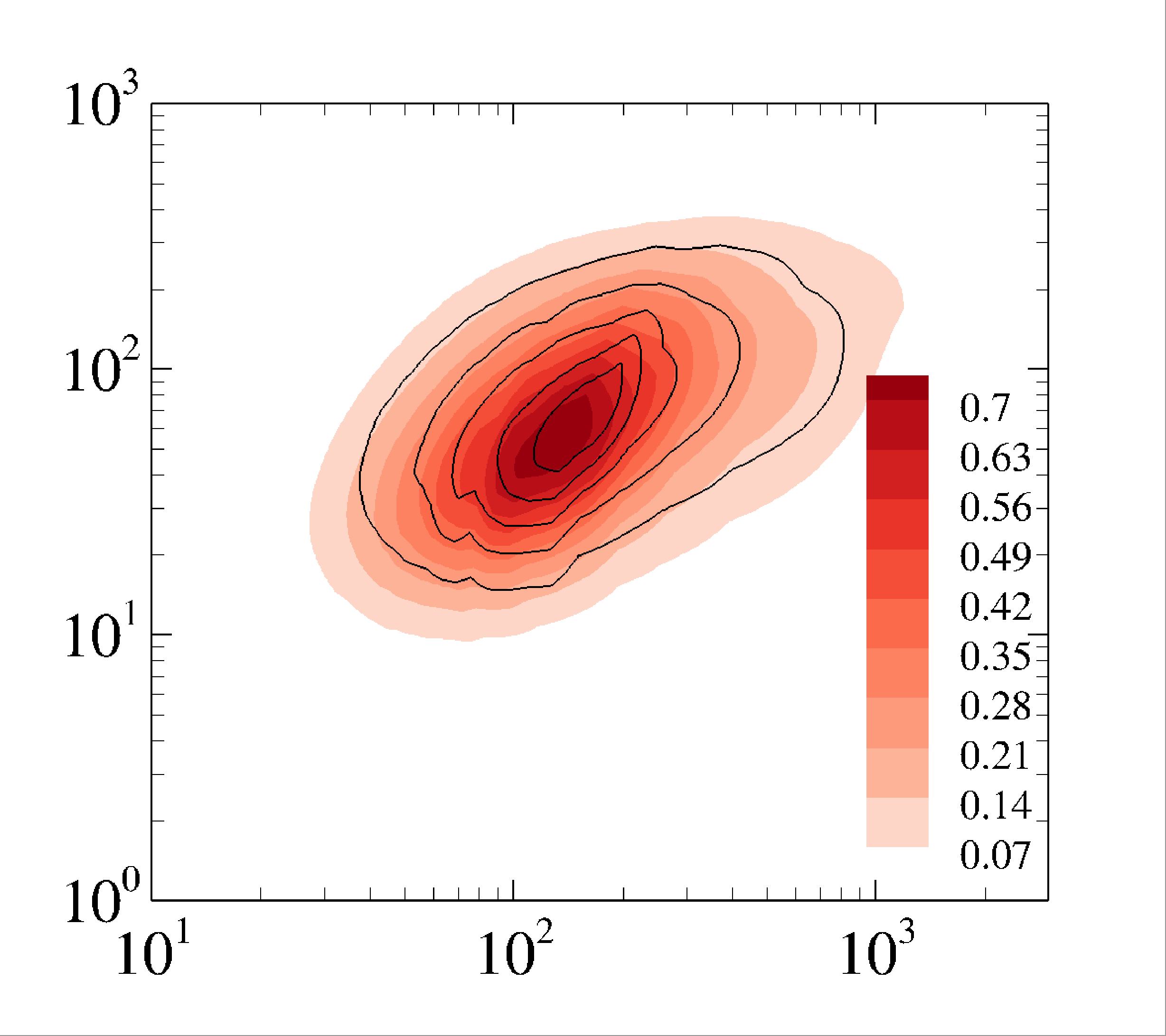}
\put(-5,80){(d)}
\put(48,0){$\lambda^+_{z,VP}$}
\put(-4,40){\rotatebox{90}{$y^+_{VP}$}}
\end{overpic}\\[1.0ex]
\caption{Pre-multiplied (a,b) streamwise and (c,d) spanwise spectra of solenoidal velocity 
(a,c) $u''^s_1$, (b,d) $u''^s_2$, flooded: case M6C-0, lines: case M6C-2.}
\label{fig:specs}
\end{figure}

\begin{figure}
\centering
\begin{overpic}[width=0.5\textwidth,trim={0.2cm 0.2cm 0.2cm 0.2cm},clip]{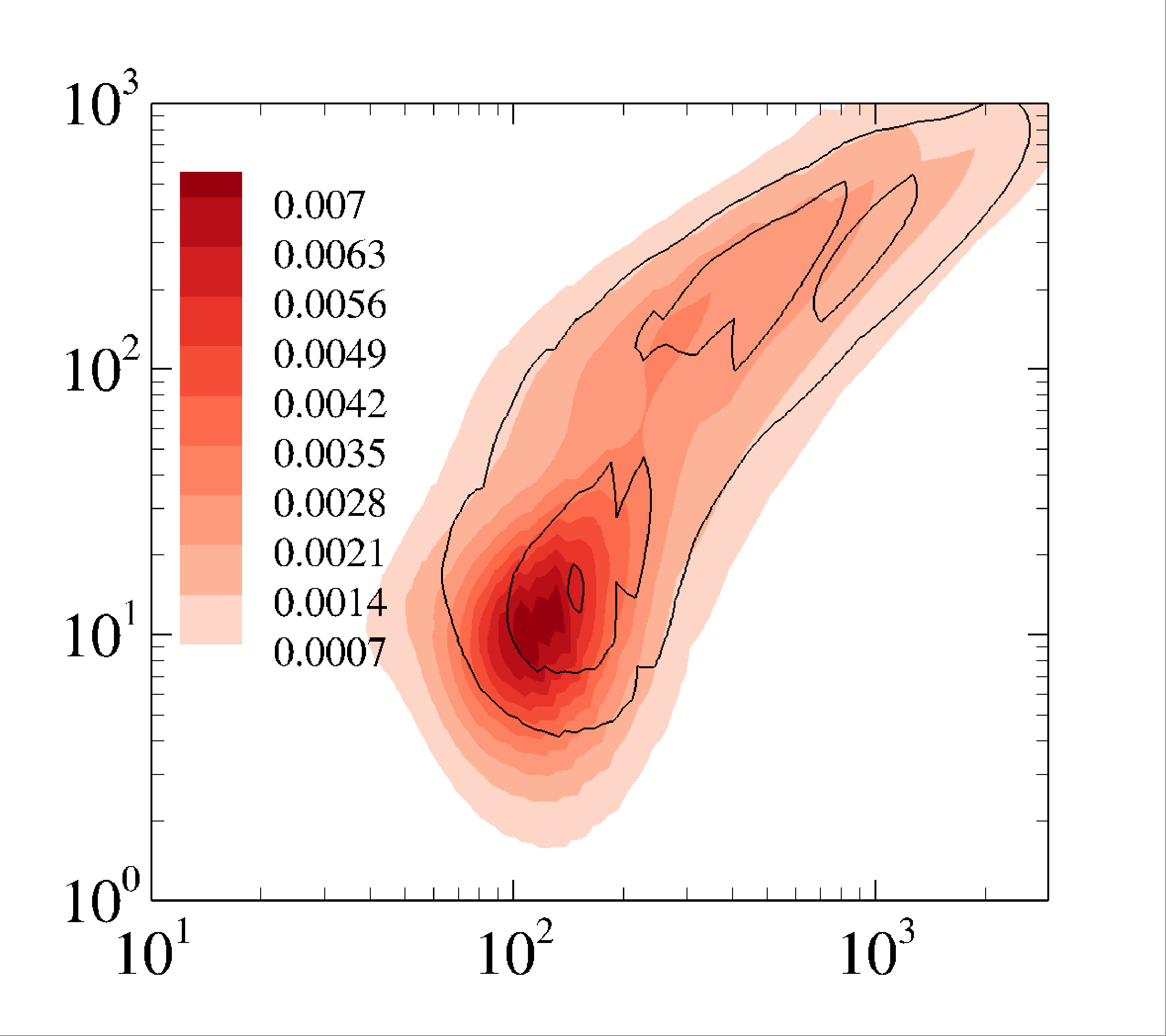}
\put(-5,80){(a)}
\put(48,0){$\lambda^+_{x,VP}$}
\put(-4,40){\rotatebox{90}{$y^+_{VP}$}}
\end{overpic}~
\begin{overpic}[width=0.5\textwidth,trim={0.2cm 0.2cm 0.2cm 0.2cm},clip]{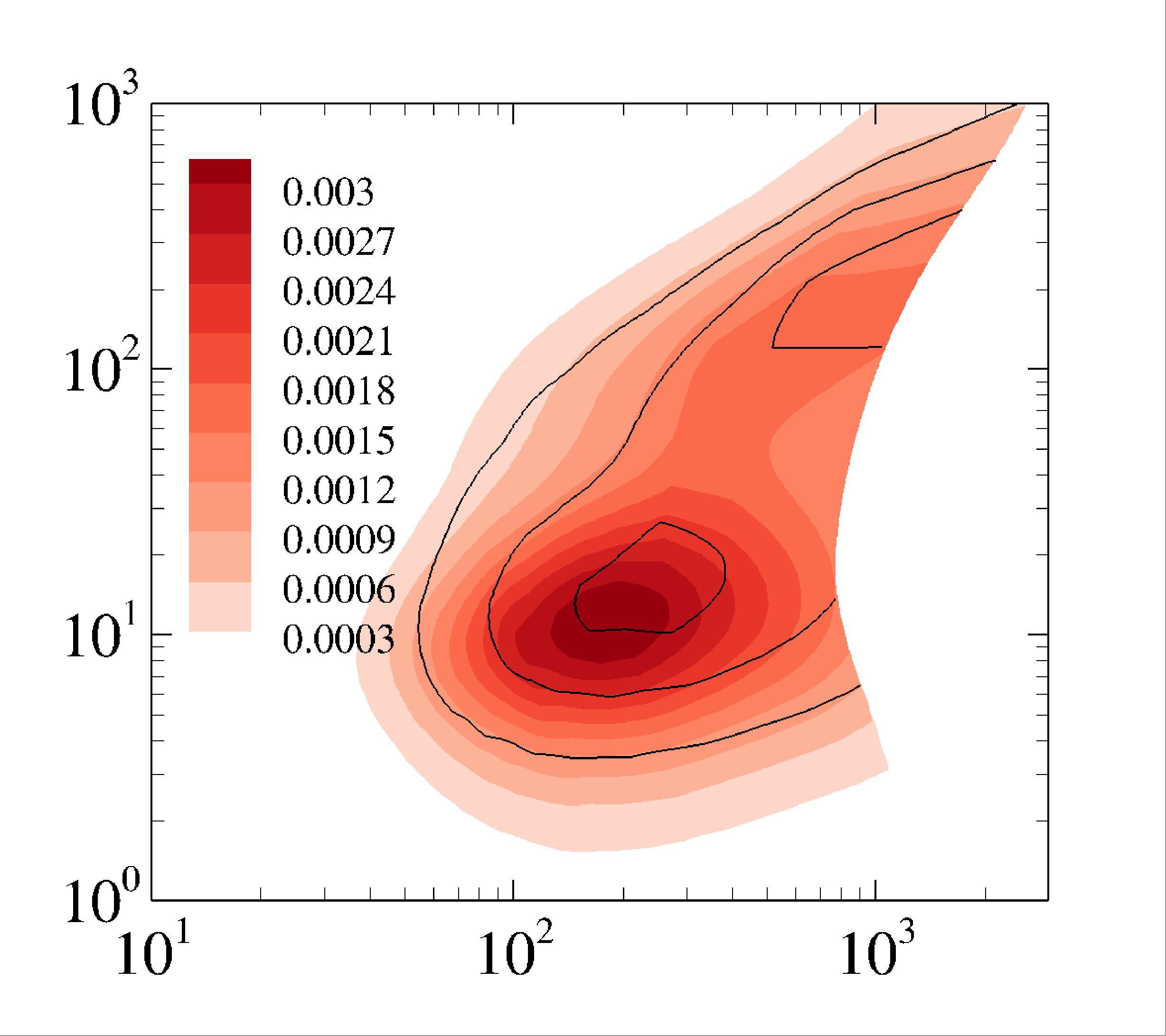}
\put(-5,80){(b)}
\put(48,0){$\lambda^+_{z,VP}$}
\put(-4,40){\rotatebox{90}{$y^+_{VP}$}}
\end{overpic}\\[1.5ex]
\begin{overpic}[width=0.5\textwidth,trim={0.2cm 0.2cm 0.2cm 0.2cm},clip]{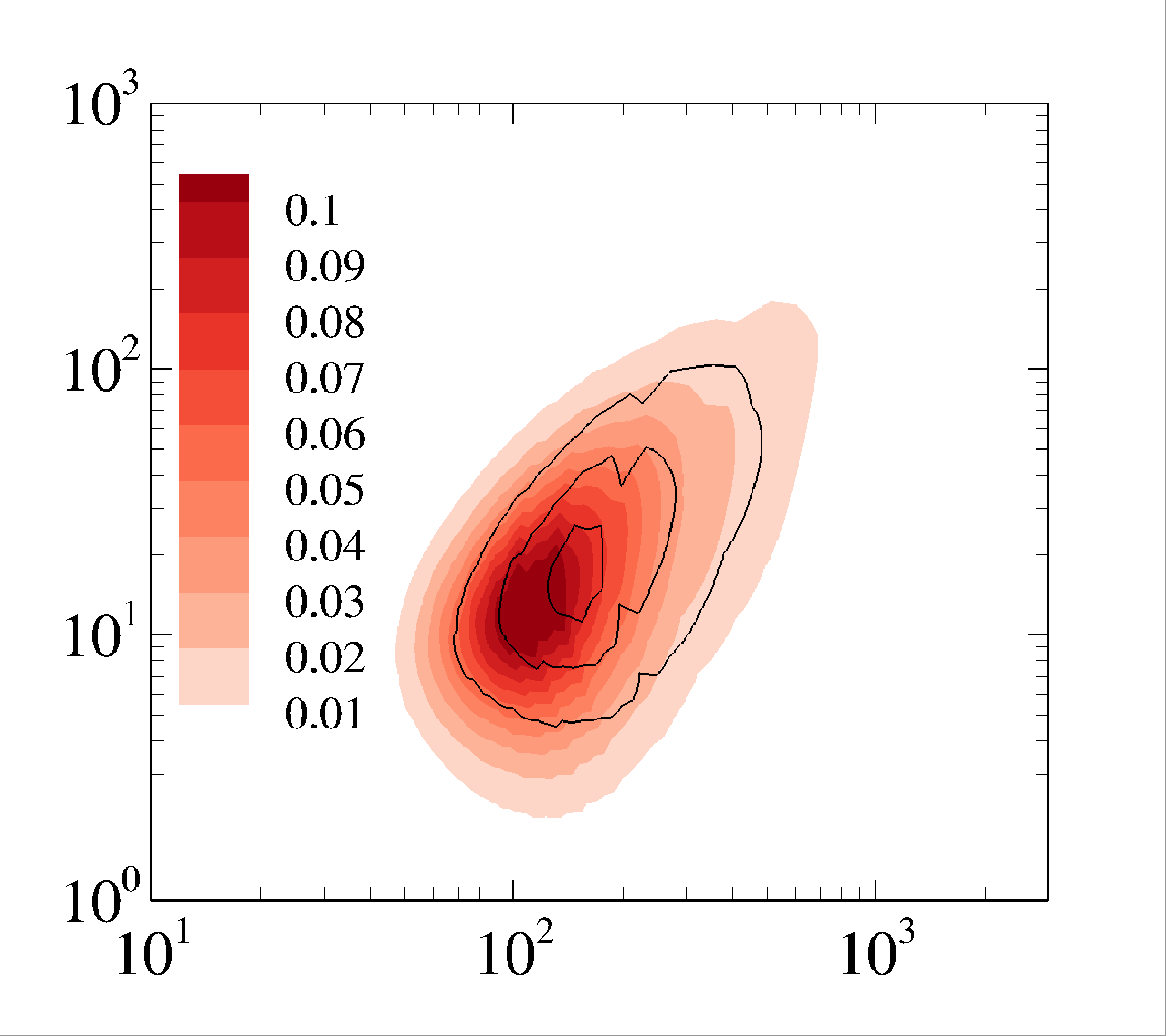}
\put(-5,80){(c)}
\put(48,0){$\lambda^+_{x,VP}$}
\put(-4,40){\rotatebox{90}{$y^+_{VP}$}}
\end{overpic}~
\begin{overpic}[width=0.5\textwidth,trim={0.2cm 0.2cm 0.2cm 0.2cm},clip]{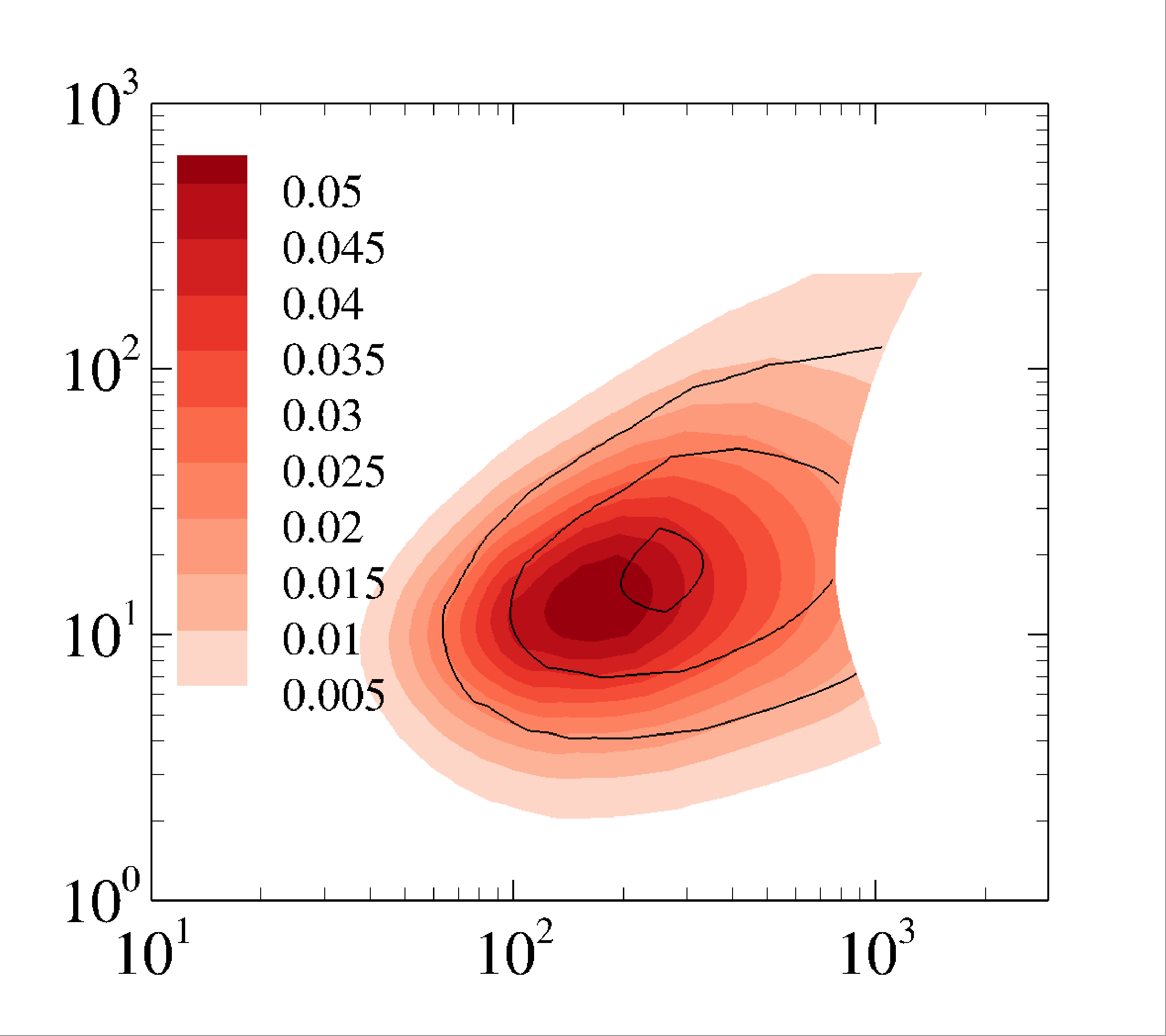}
\put(-5,80){(d)}
\put(48,0){$\lambda^+_{z,VP}$}
\put(-4,40){\rotatebox{90}{$y^+_{VP}$}}
\end{overpic}\\
\caption{Pre-multiplied (a,b) streamwise and (c,d) spanwise spectra of dilatational velocity
(a,c) $u''^d_1$, (b,d) $u''^d_2$, flooded: case M6C-0, lines: case M6C-2.}
\label{fig:specd}
\end{figure}

For a detailed analysis of the effects of particle feedback force on turbulent fluctuations 
at different length scales, in Figure~\ref{fig:specs} we present the pre-multiplied streamwise 
and spanwise spectra of solenoidal streamwise and wall-normal velocities in the cold wall 
cases M6C-0 and M6C-2. 
These spectra are plotted under the coordinates integrated by Equation~\eqref{eqn:vp}
which provides improved universal scalings for cold wall turbulence, as evidenced by 
the nearly identical off-wall locations of the Reynolds stress peaks (Figure~\ref{fig:rey}). 
In case M6C-0, the streamwise velocity spectra peaks are located at the streamwise 
length scale of $\lambda^+_{x,VP} \approx 1000$ and the spanwise scale of 
$\lambda^+_{z,VP} \approx 120$, with an off-wall location at $y^+_{VP} \approx 12$, 
corresponding to the characteristic length scales of velocity streaks. 
These length scales are similar to those observed in incompressible and compressible 
turbulence over adiabatic walls~\citep{lagha2011numerical,yao2020turbulence}, 
consistent with previous findings when analyzed under semi-local viscous scales
~\citep{huang2022direct}. 
The wall-normal velocity spectra peak around $y^+_{VP} \approx 70$, with streamwise 
and spanwise length scales of $\lambda^+_{x,VP} \approx 200 \sim 300$ and 
$\lambda^+_{z,VP} \approx 120$, respectively, 
corresponding to the characteristics of vertical motions induced by bursting events 
or streamwise vortices.

In the particle-laden case M6C-2, the streamwise spectra of $u''^s_1$ indicate a weakening of 
near-wall small-scale motions compared to case M6C-0. 
Although the peak values remain almost unchanged, their positions shift towards larger scales. 
Such a variation corresponds to the longer and more coherent velocity streaks manifested
in the instantaneous flow fields.
Conversely, the spanwise spectra are intensified near their peaks at 
$\lambda^+_{z,VP} \approx 120$ and $y^+_{VP}=12$, while
the spectra intensities at smaller and larger scales are comparatively less affected. 
Regarding the wall-normal velocity component $u''^s_2$, both the streamwise and spanwise spectra 
exhibit lower intensities across all scales,
with peak locations remaining almost consistent with those in case M6C-0.

Figure~\ref{fig:specd} displays the spectra of dilatational velocities $u''^d_1$ and $u''^d_2$. 
In general, the dilatational streamwise velocity is significantly weaker than 
the dilatational wall-normal velocity by approximately one order of magnitude. 
The streamwise spectra peak around $\lambda^+_{x,VP} \approx 100$, 
while the spanwise spectra peak at $\lambda^+_{z,VP} \approx 120$, 
corresponding to the characteristic length scales of near-wall dilatational motions 
in the form of travelling wave packets~\citep{yu2019genuine,yu2021compressibility}. 
The addition of particles in case M6C-2 suppresses dilatational motions at all scales
across the boundary layer, particularly in the near-wall region, 
with the peaks of the spectra shifted towards larger scales and higher locations.

The different trend of variation of the solenoidal and dilatational motions also provides
interesting insights into the dynamic relevance between them.
In our previous study~\citep{yu2024generation}, we have demonstrated via numerical experiments 
that the dilatational motions are generated in the process of the velocity streak breakdown
and the generation of the vortices, especially those associated with the strong bursting events.
The less meandering instantaneous near-wall velocity streaks and the lowering wall-normal
Reynolds stress in the particle-laden flows presented herein thereby imply 
the less effective generation of the streamwise vortices and the lower possibility of 
the occurrence of the bursting events~\citep{wang2019modulation},
consequently leading to a synchronous decrease in the dilatational motions.

From another perspective, we have also suggested in the previous study~\citep{yu2024generation} that
the near-wall dilatational perturbations in the shape of travelling wave packets cannot amplify, 
thereby excluding the possibility of their generating via instability or transient growth.
The results given herein confirm such an elucidation.
As we can infer from the spectra of $u''^s_1$, the strength of velocity streaks remains
almost the same as, and even stronger than, that in the particle-free case, 
but the dilatational motions are nonetheless weakened.
The concurrent weakening of the solenoidal wall-normal velocity $u''^s_2$ and 
the dilatational velocity components proves that the generation of the latter should 
be attributed to the former, which is associated with the bursting events and 
the streamwise vortices.
It also refutes the idea that these dilatational travelling wave packets are the evolving
perturbations lying beneath the velocity streaks.
Otherwise, these flow structures would be amplified by the disturbances imposed by 
the particle feedback forces.

\subsection{Near-wall particle feedback force} \label{subsec:force}

\begin{figure}
\centering
\begin{overpic}[width=0.5\textwidth]{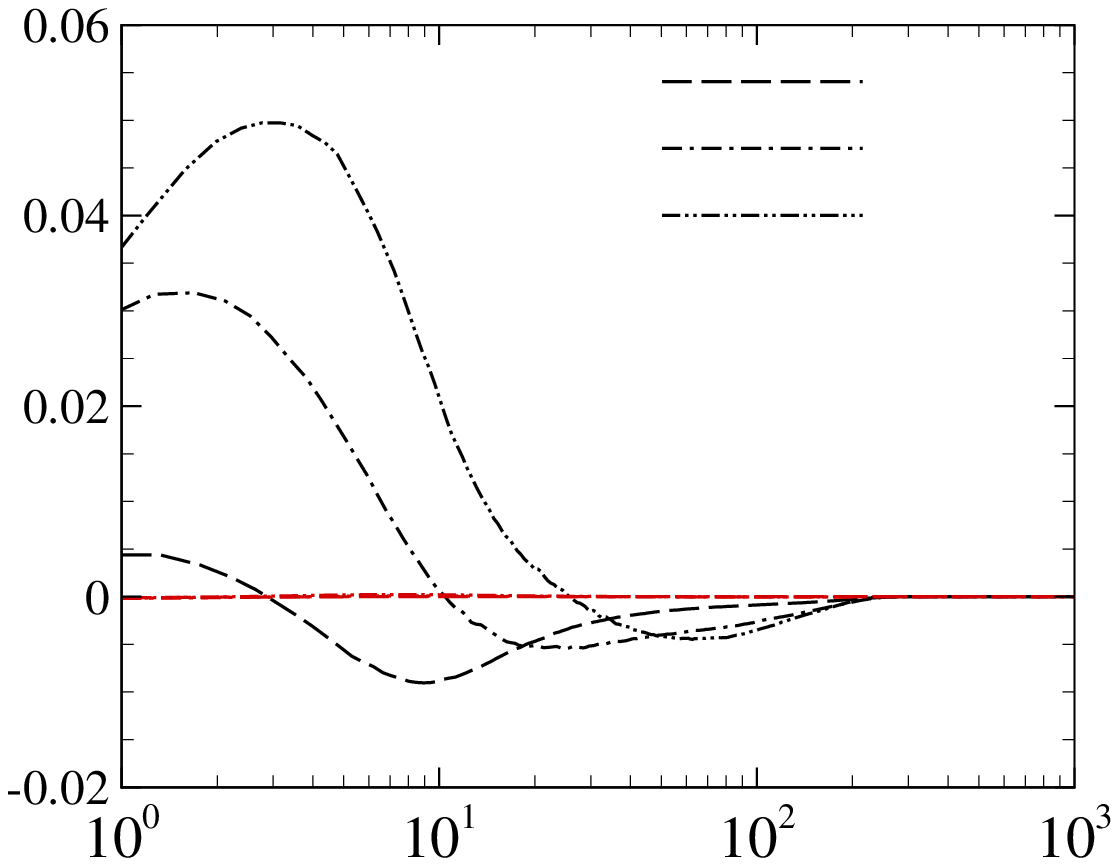}
\put(-2,70){(a)}
\put(48,0){$y^+$}
\put(-4,35){\rotatebox{90}{$\overline{F_{p1} u''_{1}}^+$}}
\put(73,66){\small M6-1}
\put(73,61){\small M6-2}
\put(73,55.5){\small M6-3}
\end{overpic}~
\begin{overpic}[width=0.5\textwidth]{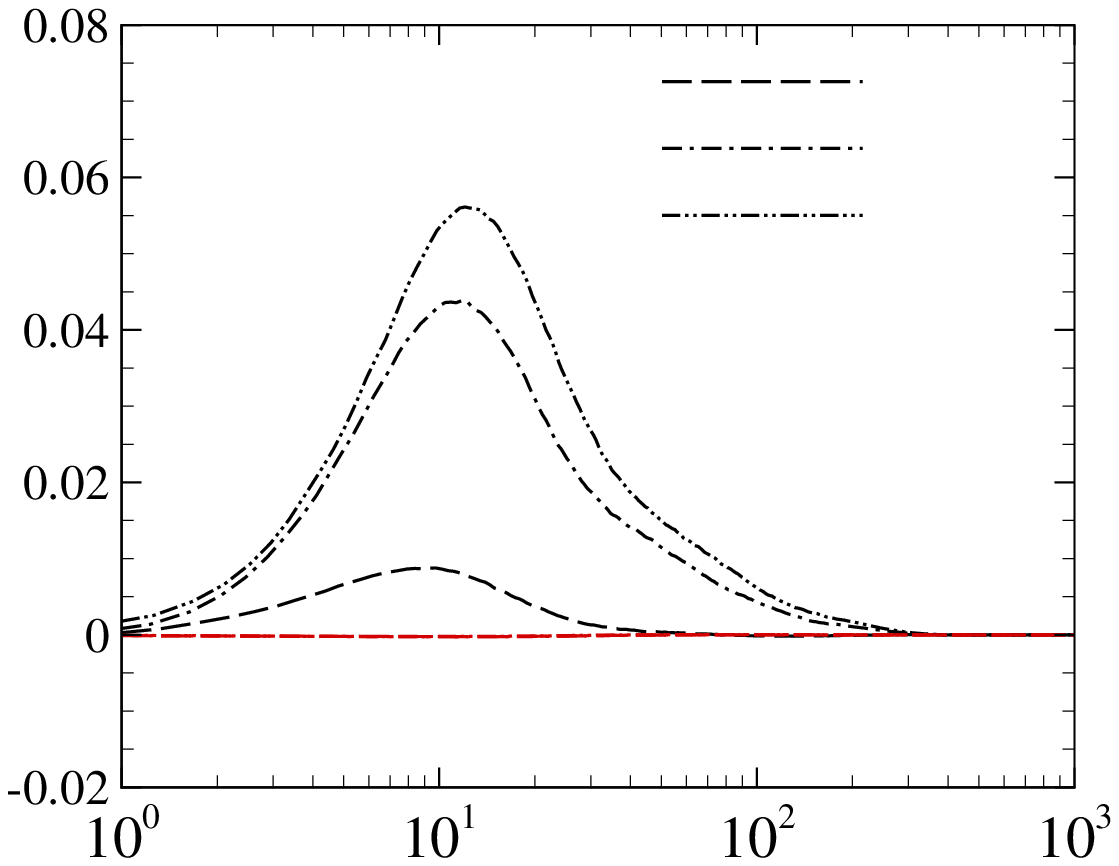}
\put(-2,70){(b)}
\put(48,0){$y^+$}
\put(-4,35){\rotatebox{90}{$\overline{F_{p1} u''_{1}}^+$}}
\put(73,66){\small M6C-1}
\put(73,61){\small M6C-2}
\put(73,55.5){\small M6C-3}
\end{overpic}\\[1.0ex]
\begin{overpic}[width=0.5\textwidth]{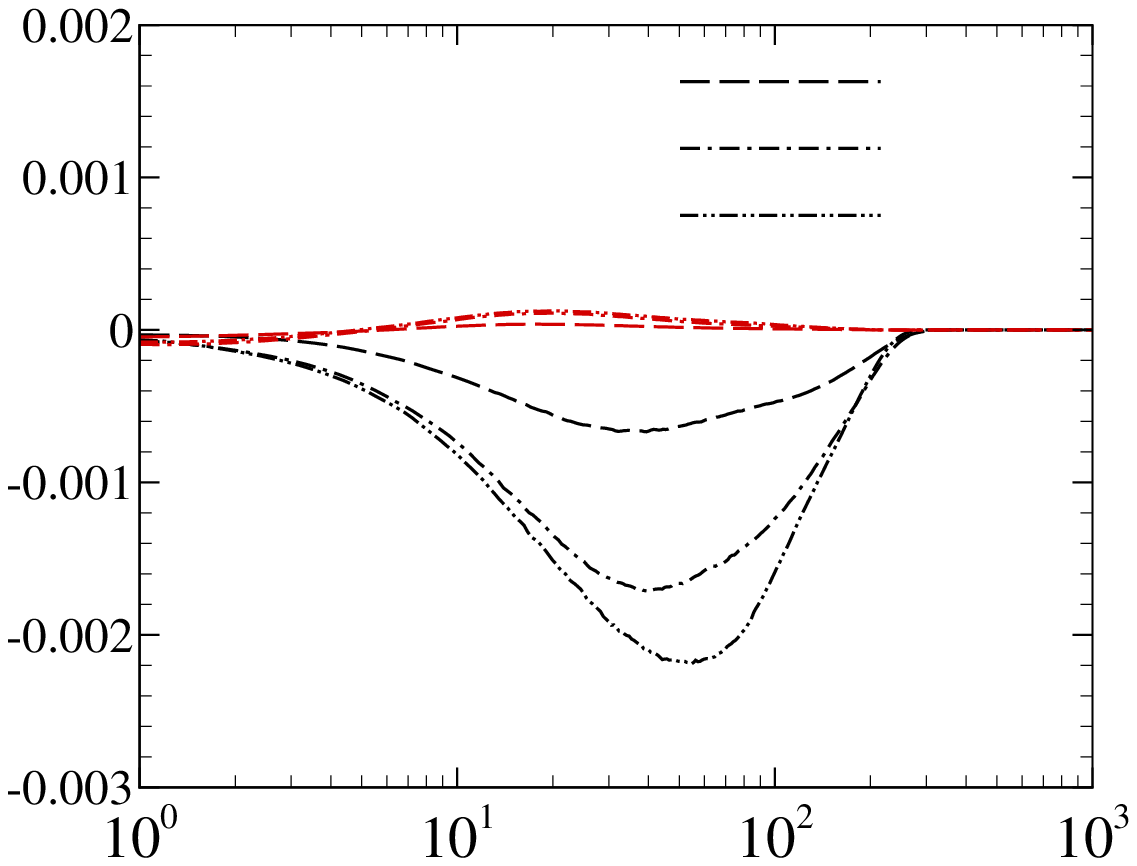}
\put(-2,70){(c)}
\put(48,0){$y^+$}
\put(-4,35){\rotatebox{90}{$\overline{F_{p2} u''_{2}}^+$}}
\put(73,66){\small M6-1}
\put(73,61){\small M6-2}
\put(73,55.5){\small M6-3}
\end{overpic}~
\begin{overpic}[width=0.5\textwidth]{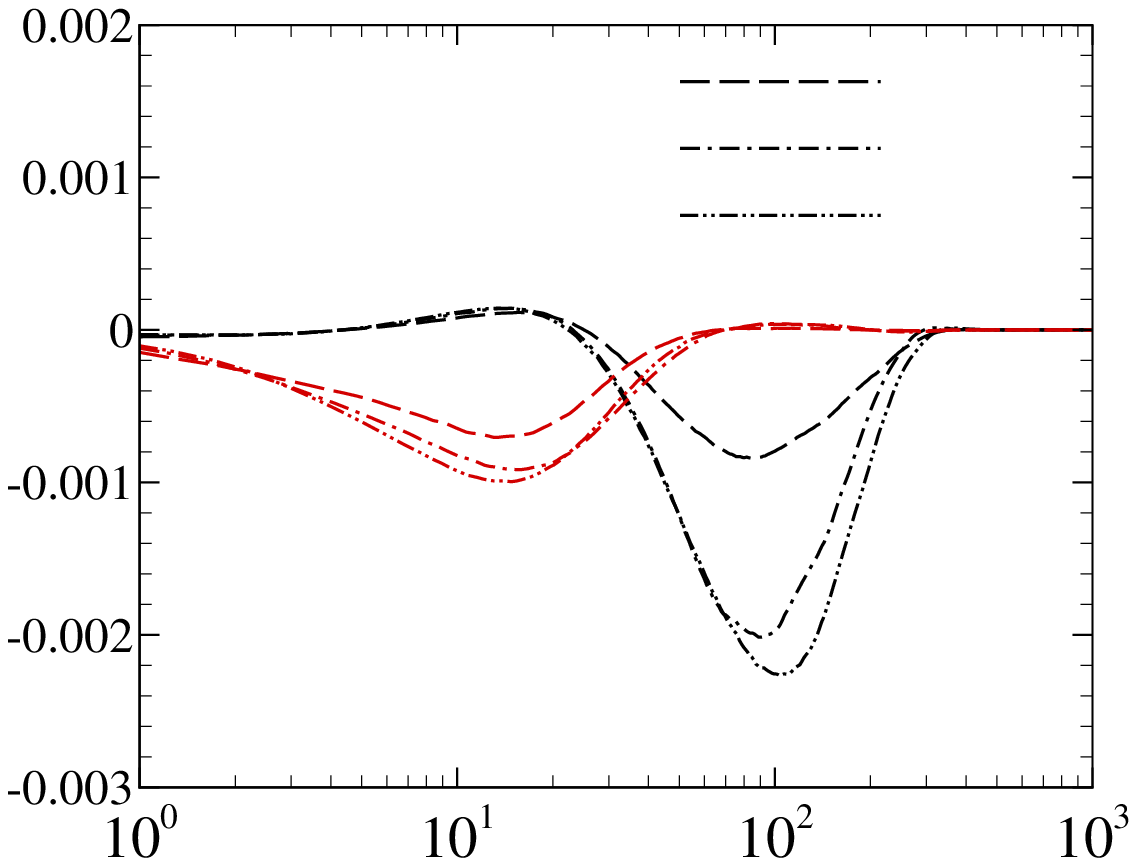}
\put(-2,70){(d)}
\put(48,0){$y^+$}
\put(-4,35){\rotatebox{90}{$\overline{F_{p2} u''_{2}}^+$}}
\put(73,66){\small M6C-1}
\put(73,61){\small M6C-2}
\put(73,55.5){\small M6C-3}
\end{overpic}\\
\caption{Work of particle force on the solenoidal (black lines) and dilatational (red lines) 
velocity, (a,b) $\overline{F_{p1} u''_{1}}^+$, (c,d) $\overline{F_{p2} u''_{2}}^+$, 
(a,c) cases M6, (b,d) cases M6C.}
\label{fig:pforce}
\end{figure}

Knowing that the dilatational motions are weakened by the presence of particles at all scales,
in this section, we explore the dynamic interactions between particles and the fluid 
by analyzing the particle feedback force and its correlation with dilatational motions. 
Figure~\ref{fig:pforce} displays the work of the particle feedback force on solenoidal and 
dilatational velocity fluctuations separately to assess their respect contributions. 
In general, the work done by particle forces is slightly lower than but of the same order as
the production and viscous dissipation in turbulent kinetic energy transport 
(not shown here for brevity), 
indicating their significant influence on near-wall turbulence, particularly for solenoidal motions. 
In both adiabatic and cold wall cases, the work of the particle force on streamwise solenoidal 
velocity $u''^s_1$ surpasses that on dilatational velocity by at least an order of magnitude. 
In adiabatic wall cases, the work $\overline{F_{p1} u''^s_{1}}^+$ varies from positive 
near the wall to negative values towards the outer region, with the critical point 
getting higher with mass loading. 
In low mass loading case M6-1, particles primarily extract turbulent kinetic energy 
from the fluid flow in the integral sense, but the energy transfer direction
is reversed in the moderate mass loading cases. 
This observation is consistent with the findings from instantaneous distributions of 
velocity streaks and particle feedback force (Figures~\ref{fig:inst-m6} and
\ref{fig:inst-m6c}), or equivalently the role played by particles in modulating
the near-wall dynamics, which is essentially the same as that in incompressible flows
~\citep{zhao2010turbulence,zhao2013interphasial}.
For the cold wall cases M6C, the work $\overline{F_{p1} u''^s_{1}}^+$ remains consistently 
positive and escalates with particle mass loading. 

As for the wall-normal component, 
the work of the particle force on its solenoidal component in adiabatic wall cases M6
is negative and independent of particle mass loadings, 
suggesting that the particle force suppresses the wall-normal velocity fluctuations. 
The work on the dilatational component $\overline{F_{p2} u''^d_{2}}^+$ remains relatively small. 
In cold wall cases M6C, the work on dilatational velocity $\overline{F_{p2} u''^d_{2}}^+$ is
of the same order as that of the work on the solenoidal velocity $\overline{F_{p2} u''^s_{2}}^+$. 
Although both terms are overall negative, the work on solenoidal velocity 
$\overline{F_{p2} u''^s_{2}}^+$ is significant within $y^+ \approx 30 \sim 200$, 
while that on dilatational velocity $\overline{F_{p2} u''^d_{2}}^+$ is prominent below 
$y^+ \lesssim 30$ in the buffer and viscous sublayer.

\begin{figure}
\centering
\small
\begin{overpic}[width=0.33\textwidth]{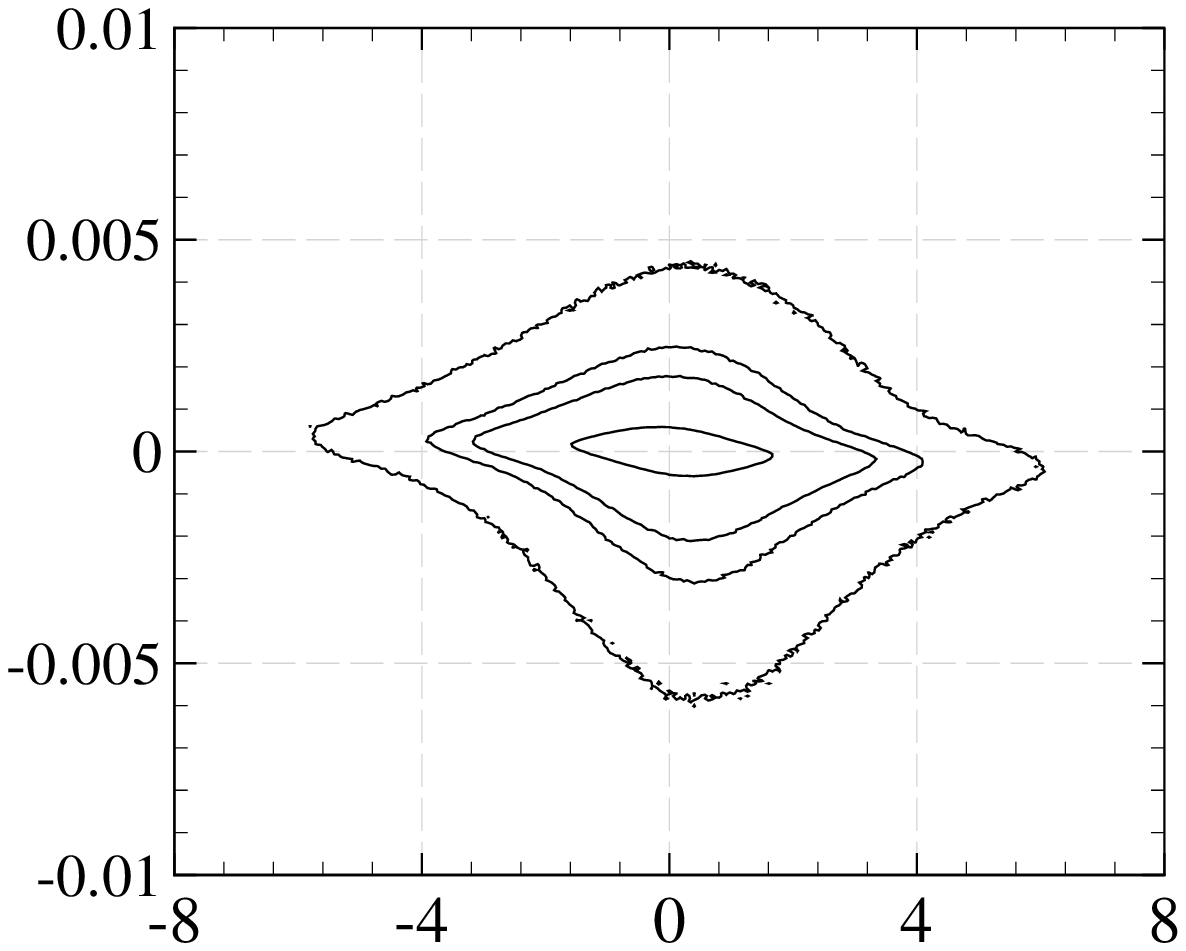}
\put(-5,78){(a)}
\put(-3,38){\rotatebox{90}{$F_{p2}$}}
\end{overpic}~
\begin{overpic}[width=0.33\textwidth]{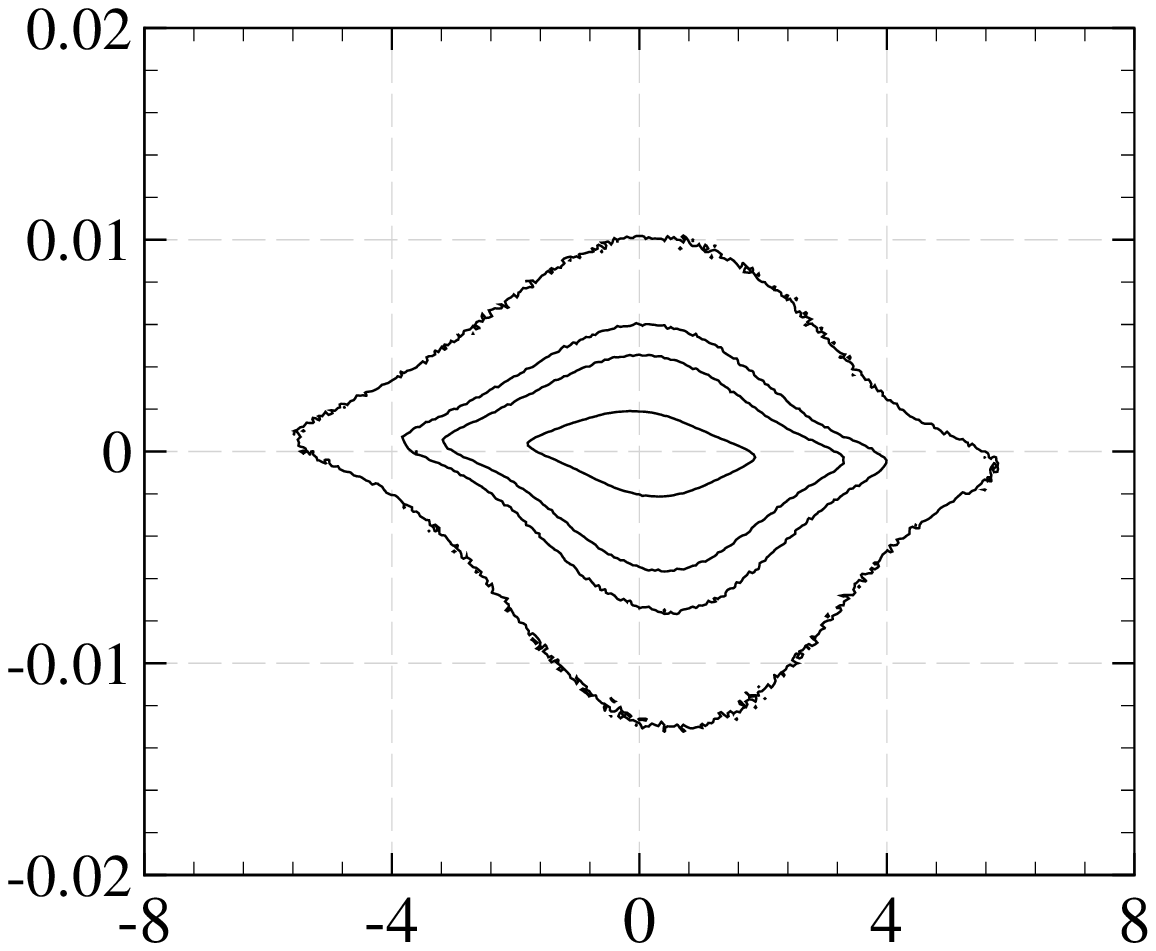}
\put(-5,78){(b)}
\end{overpic}~
\begin{overpic}[width=0.33\textwidth]{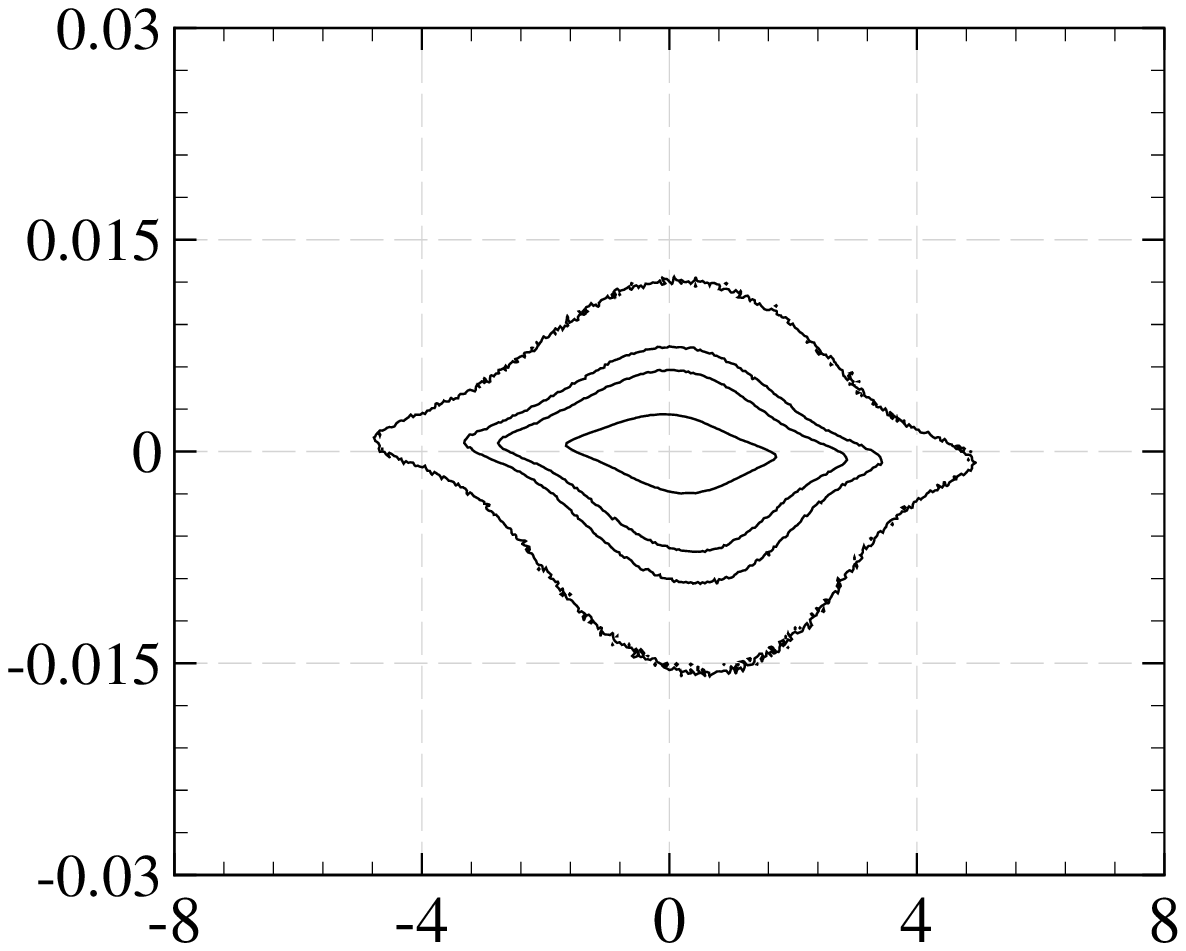}
\put(-5,78){(c)}
\end{overpic}\\
\begin{overpic}[width=0.33\textwidth]{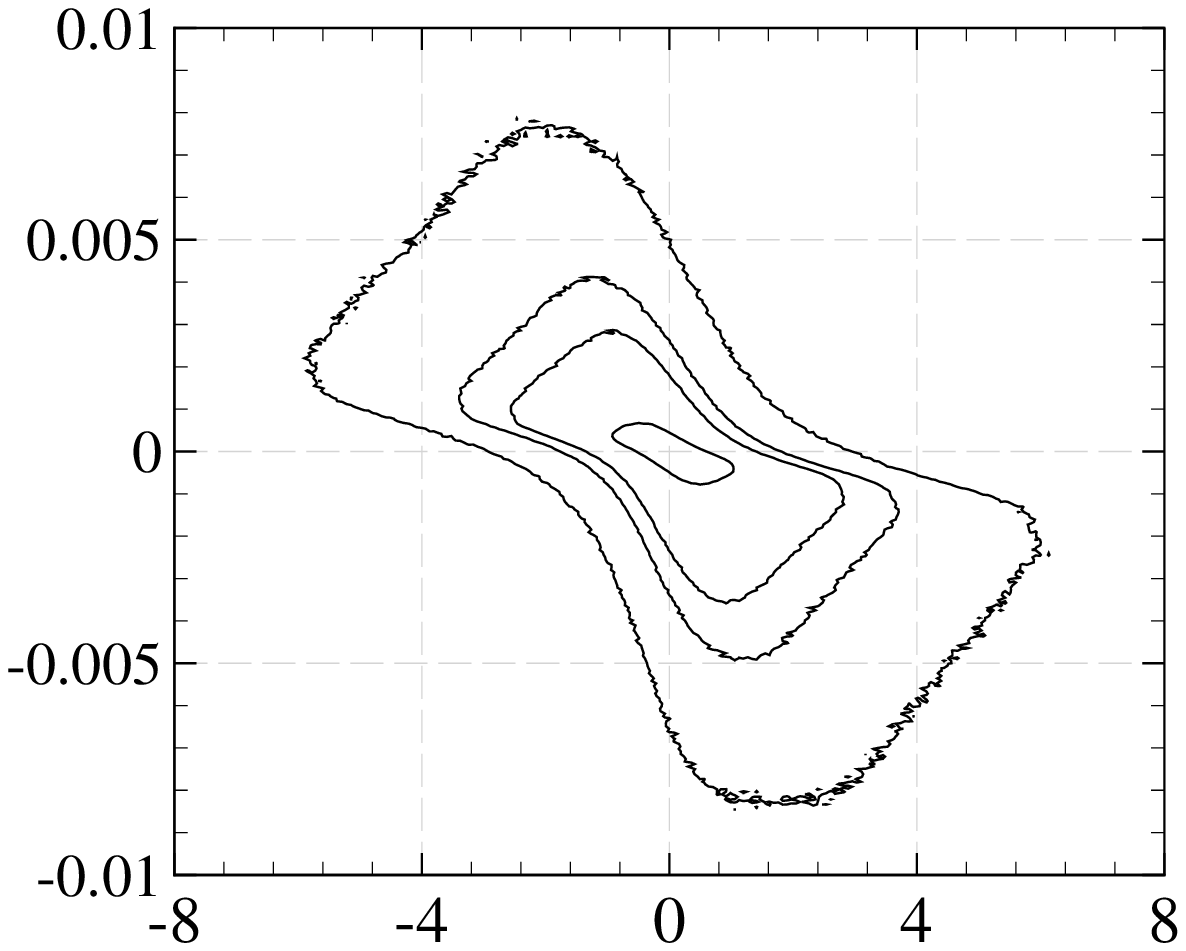}
\put(-5,78){(d)}
\put(-3,38){\rotatebox{90}{$F_{p2}$}}
\put(48,-1){$\theta'/\theta'_{w,rms}$}
\end{overpic}~
\begin{overpic}[width=0.33\textwidth]{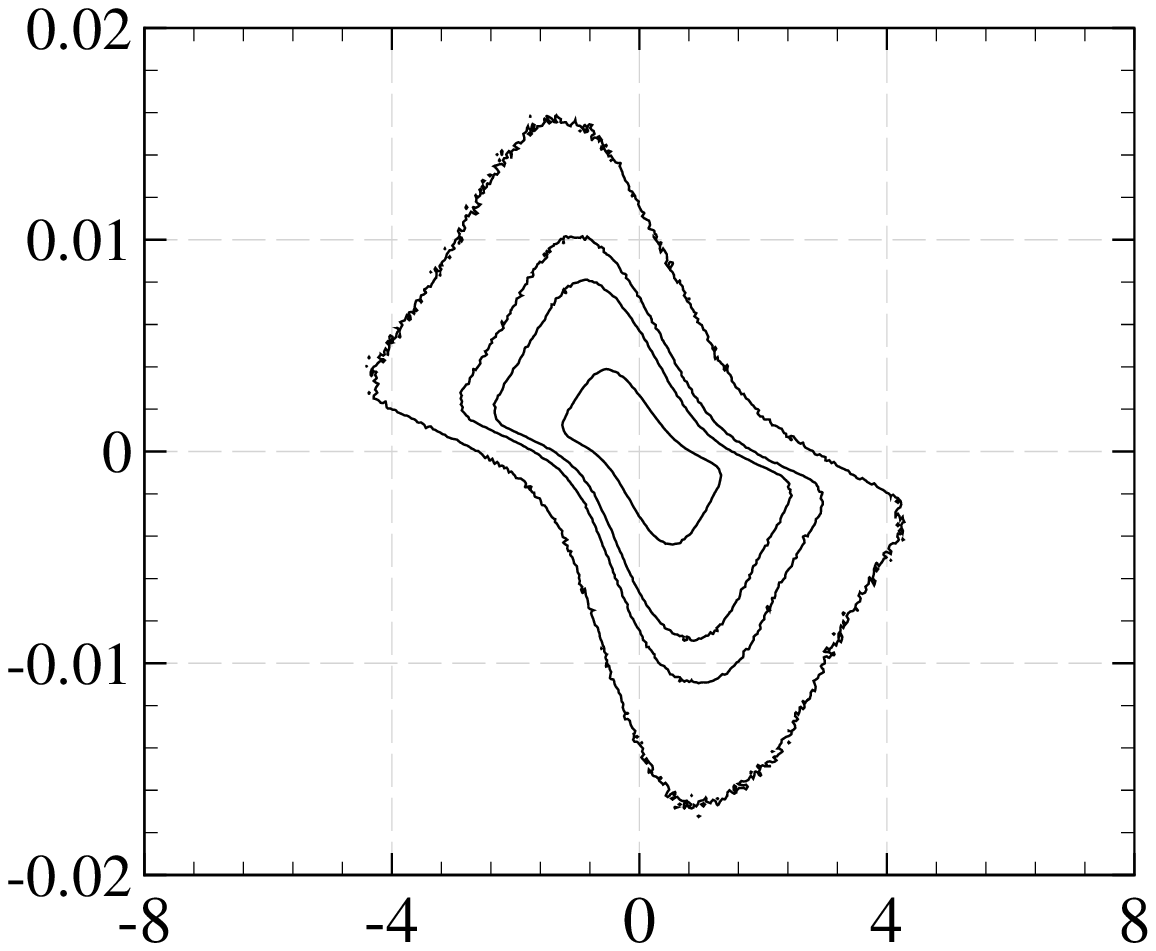}
\put(-5,78){(e)}
\put(48,-1){$\theta'/\theta'_{w,rms}$}
\end{overpic}~
\begin{overpic}[width=0.33\textwidth]{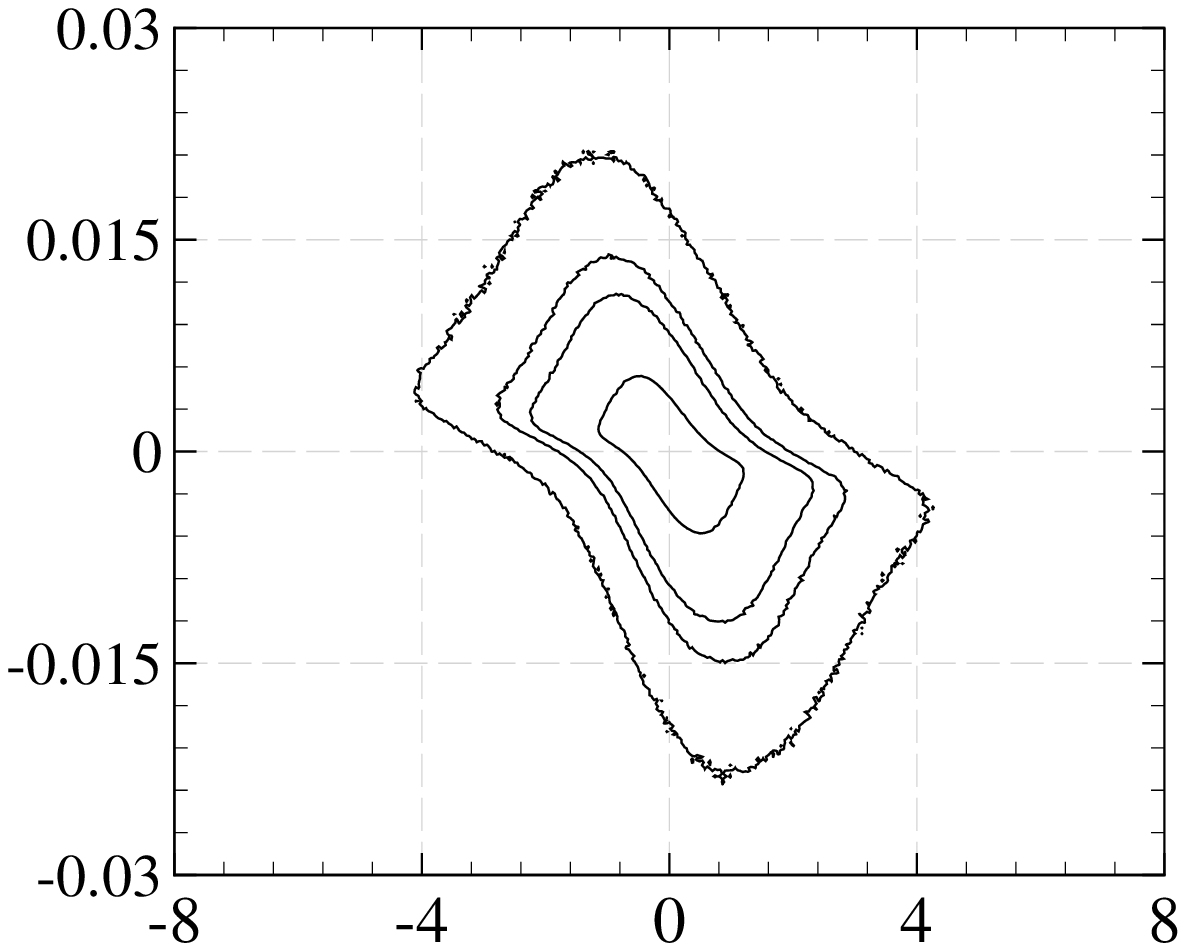}
\put(-5,78){(f)}
\put(48,-1){$\theta'/\theta'_{w,rms}$}
\end{overpic}\\
\caption{Joint PDF distribution $P(\theta',F_{p2})$ within $y^+=1 \sim 15$ in cases 
(a) M6-1, (b) M6-2, (c) M6-3, (d) M6C-1, (e) M6C-2, (f) M6C-3.
Contour levels: 0.001, 0.005, 0.01, 0.05.}
\label{fig:pdf}
\end{figure}

The comparatively high contribution of the particle force work on the dilatational velocity 
$\overline{F_{p2} u''^d_{2}}^+$ indicates the significant role of particles in 
influencing near-wall dilatational motions. 
To further elucidate their correlations, in Figure~\ref{fig:pdf} we display
the joint probability density function (PDF) between velocity divergence fluctuation 
$\theta'$ and wall-normal particle force $F_{p2}$. 
In the three adiabatic wall cases M6, the joint PDF distribution shows no preference 
in any quadrant. 
Conversely, in the cold wall cases M6C, the joint PDF predominantly occupies 
the second and fourth quadrants, indicating their negative correlations. 
With the increasing mass loadings, the extension of the joint PDF is smaller in the dimension of 
$\theta'$, suggesting that the probability of extreme compressive/expansive events is lower.
This correlation with $\theta'$ can be easily converted to that with the dilatational velocity.
Based on the definition of $\theta'$, the Taylor expansion near the wall yields the approximation
$u''^d_2 \approx \theta' (y + O(y^2))$, as long as the other components are relatively small due to 
the no-slip condition imposed by the wall, i.e. $\partial u''^d_1/\partial x =0$ and 
$\partial u''^d_3/\partial z=0$.
Consequently, the particle feedback force $F_{p2}$ exhibits a predominantly negative correlation 
with velocity component $u''^d_2$, consistent with observations from Figure~\ref{fig:pforce}. 
Joint PDF values in the first and third quadrants, on the other hand, are almost zero.

\begin{figure}
\centering
\begin{overpic}[width=0.5\textwidth,trim={0.2cm 0.2cm 0.2cm 0.2cm},clip]{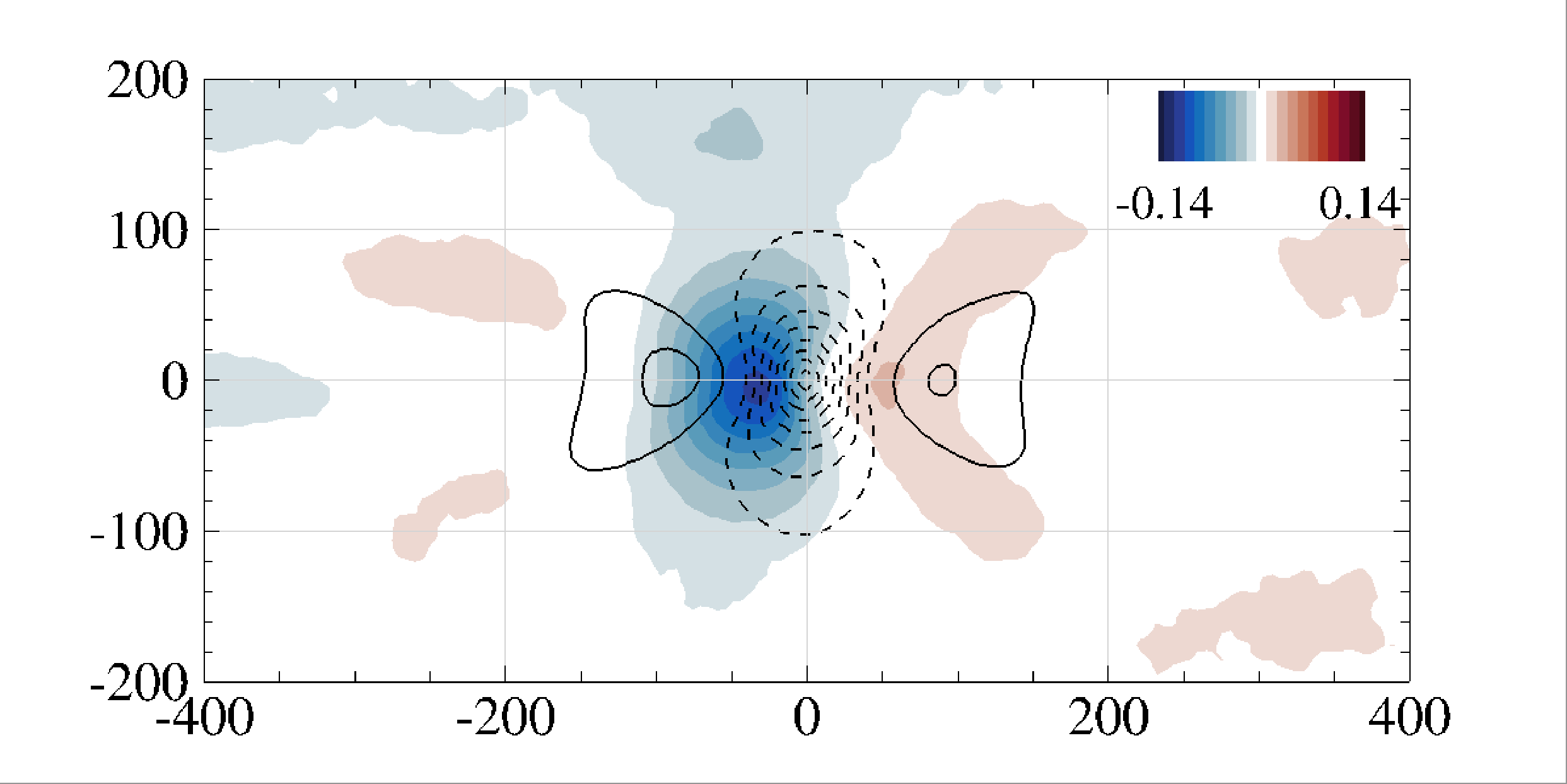}
\put(0,45){(a)}
\put(50,-2){$\Delta x^+$}
\put(0,24){\rotatebox{90}{$\Delta z^+$}}
\end{overpic}~
\begin{overpic}[width=0.5\textwidth,trim={0.2cm 0.2cm 0.2cm 0.2cm},clip]{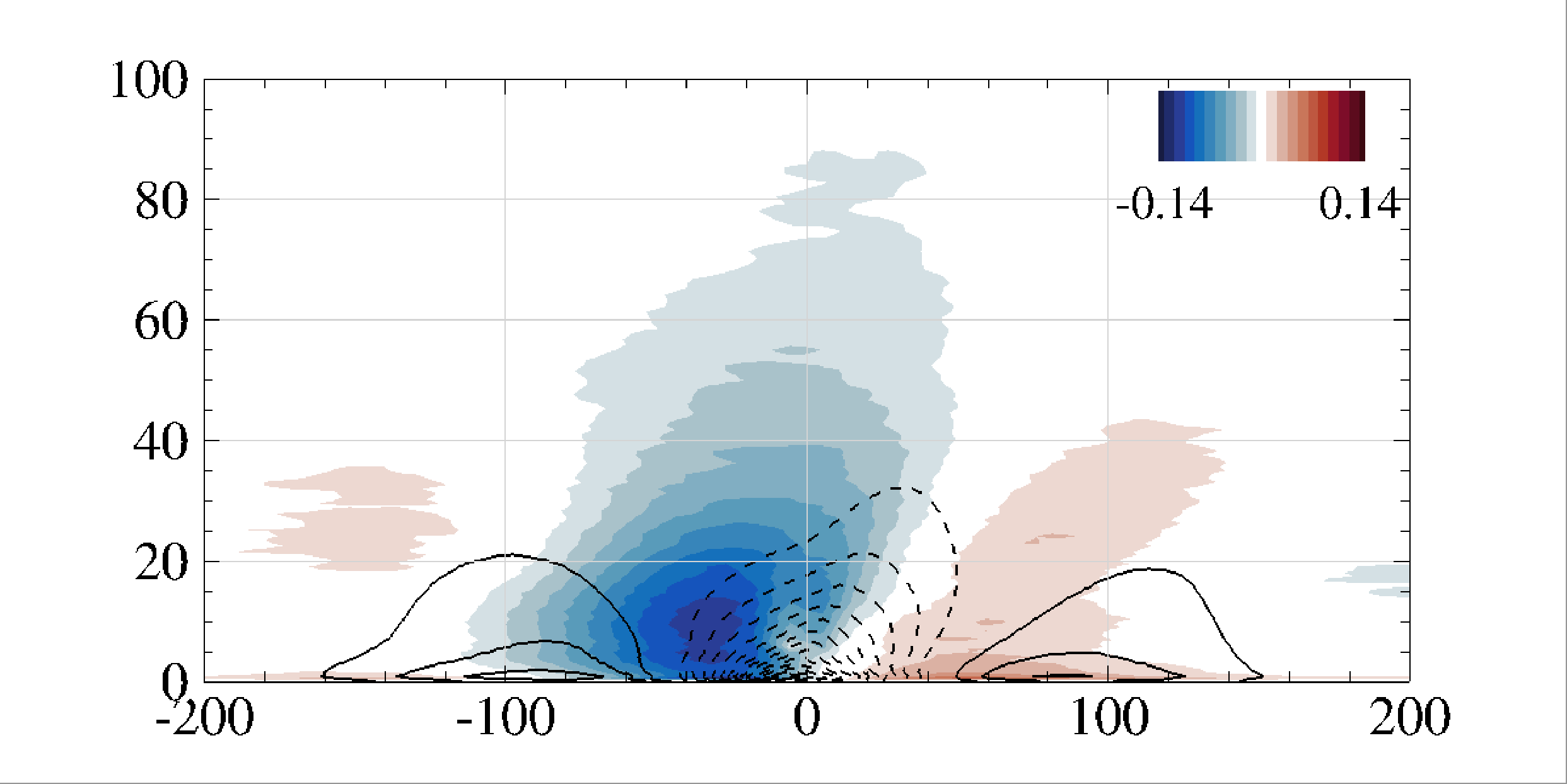}
\put(0,45){(b)}
\put(50,-2){$\Delta x^+$}
\put(0,24){\rotatebox{90}{$\Delta z^+$}}
\end{overpic}\\[1.0ex]
\begin{overpic}[width=0.5\textwidth,trim={0.2cm 0.2cm 0.2cm 0.2cm},clip]{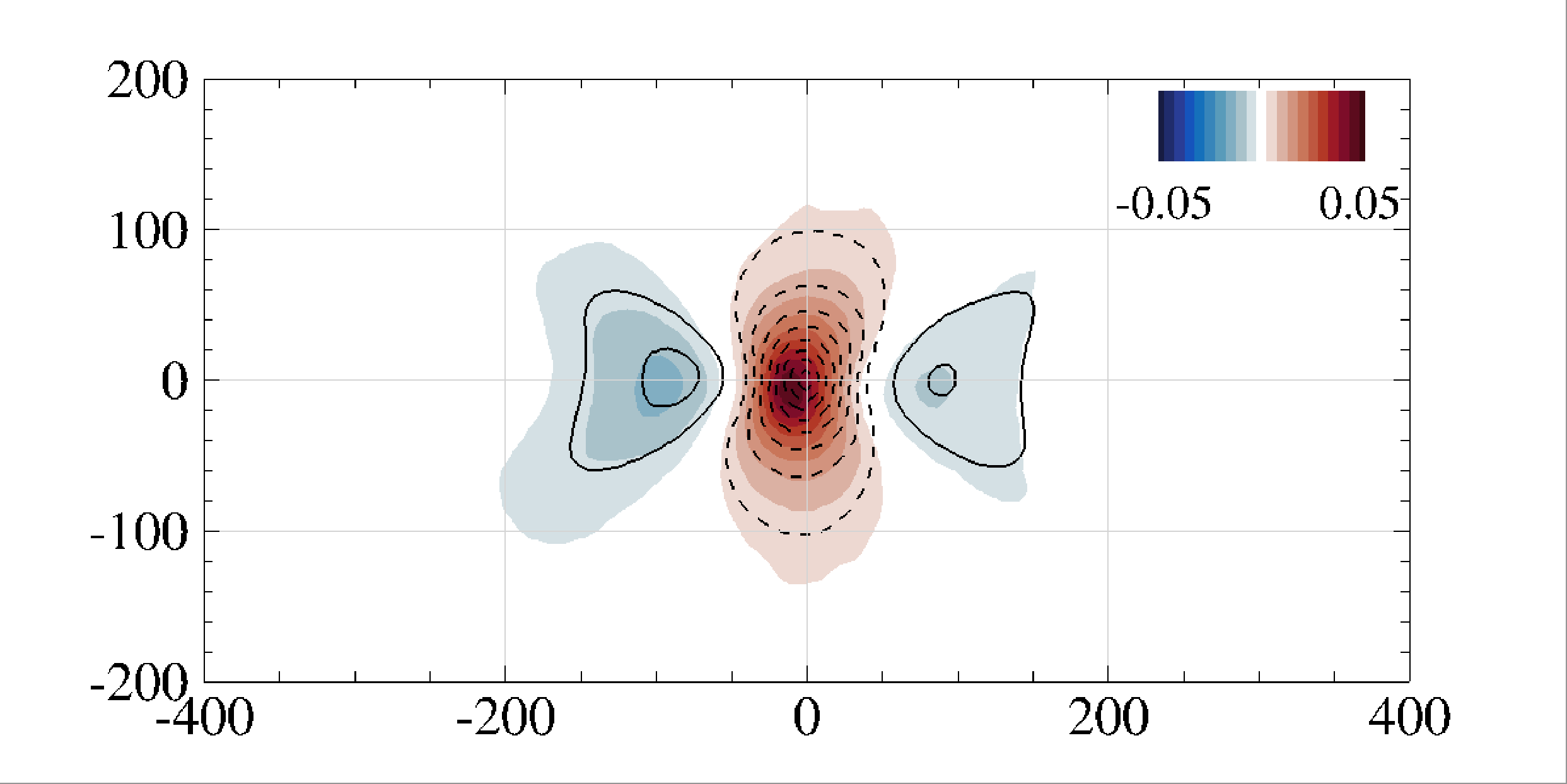}
\put(0,45){(c)}
\put(50,-2){$\Delta x^+$}
\put(0,24){\rotatebox{90}{$\Delta z^+$}}
\end{overpic}~
\begin{overpic}[width=0.5\textwidth,trim={0.2cm 0.2cm 0.2cm 0.2cm},clip]{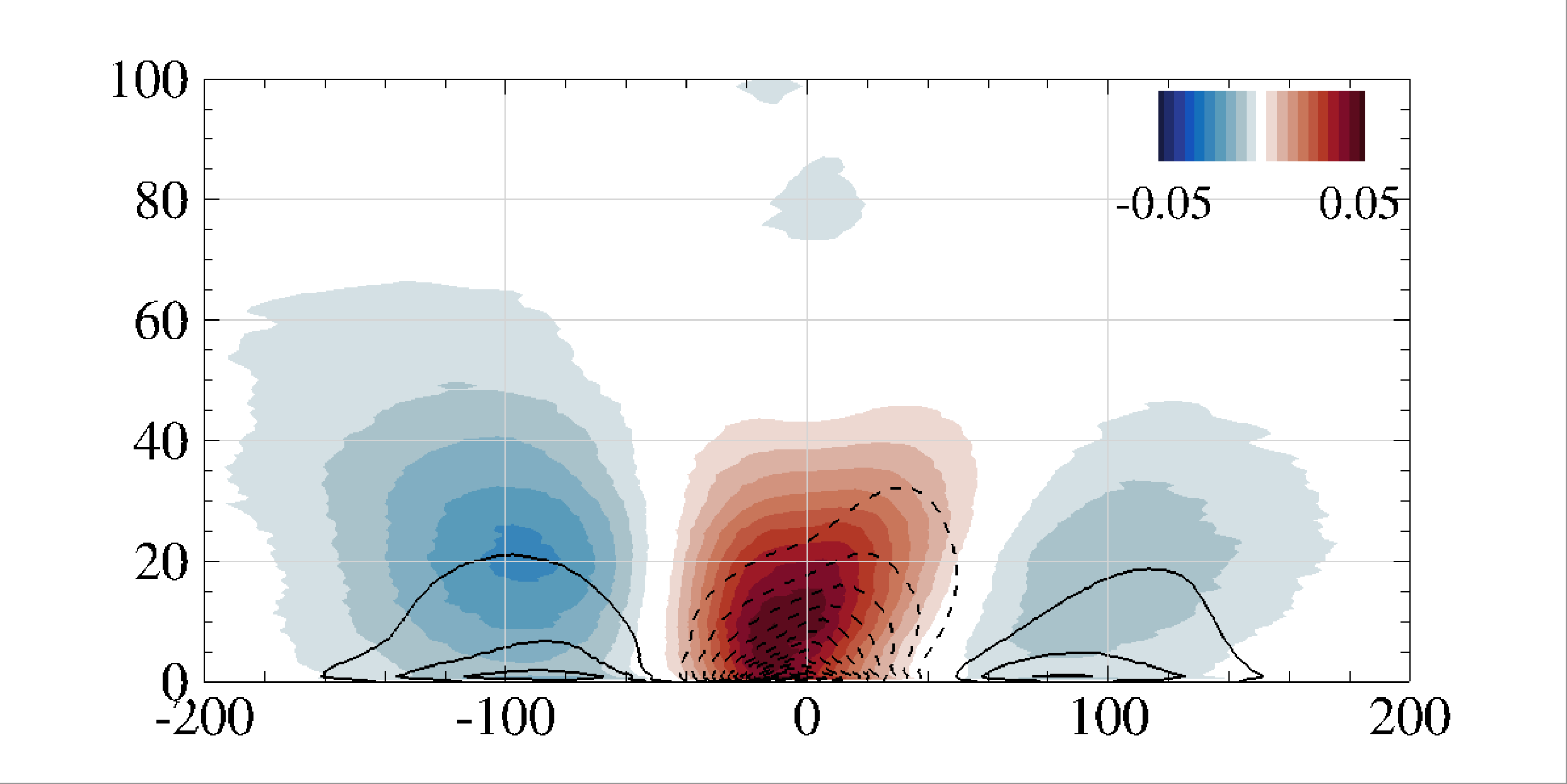}
\put(0,45){(d)}
\put(50,-2){$\Delta x^+$}
\put(0,24){\rotatebox{90}{$\Delta z^+$}}
\end{overpic}\\[1.0ex]
\begin{overpic}[width=0.5\textwidth,trim={0.2cm 0.2cm 0.2cm 0.2cm},clip]{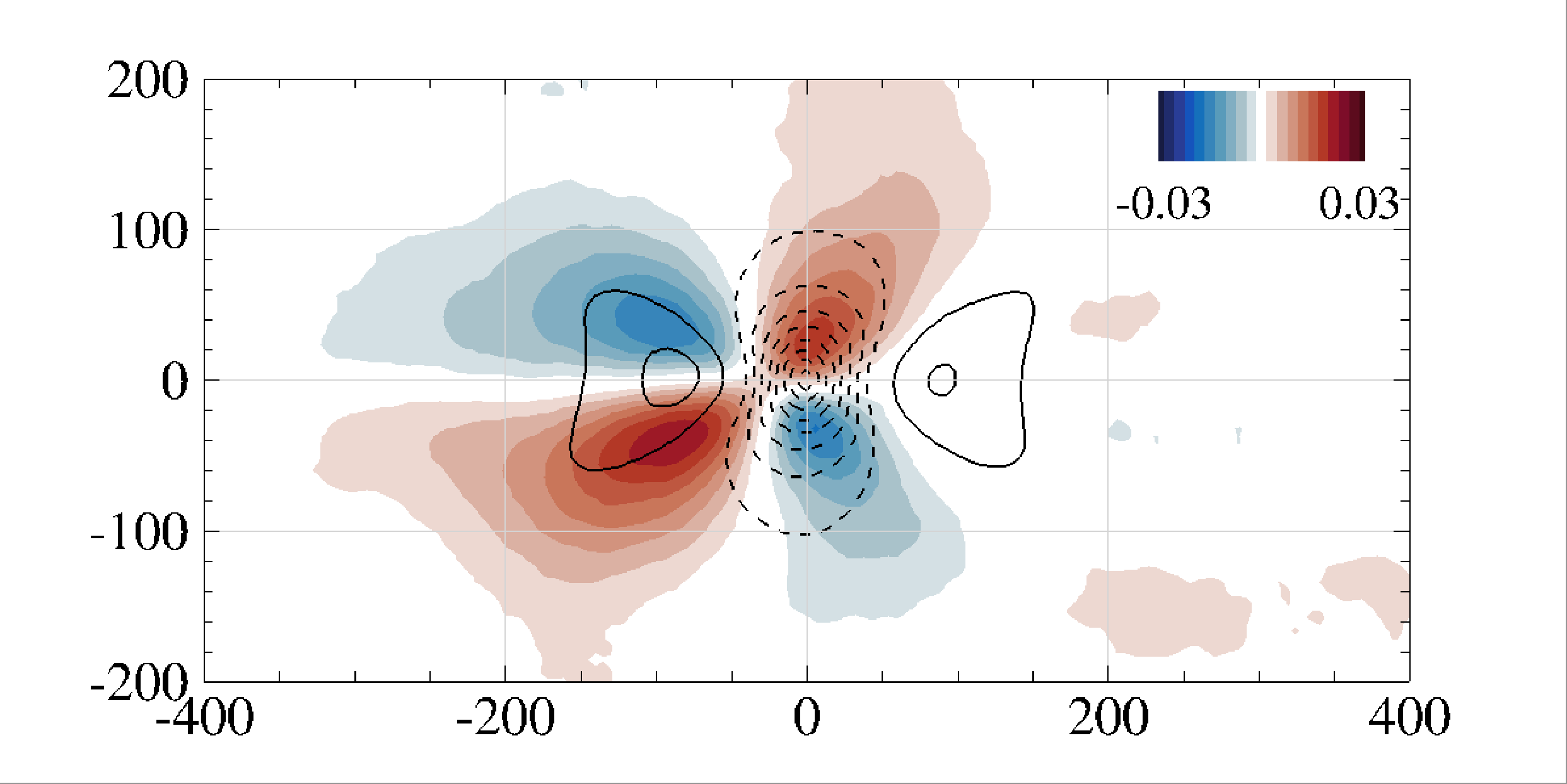}
\put(0,45){(e)}
\put(50,-2){$\Delta x^+$}
\put(0,24){\rotatebox{90}{$\Delta z^+$}}
\end{overpic}~
\begin{overpic}[width=0.5\textwidth,trim={0.2cm 0.2cm 0.2cm 0.2cm},clip]{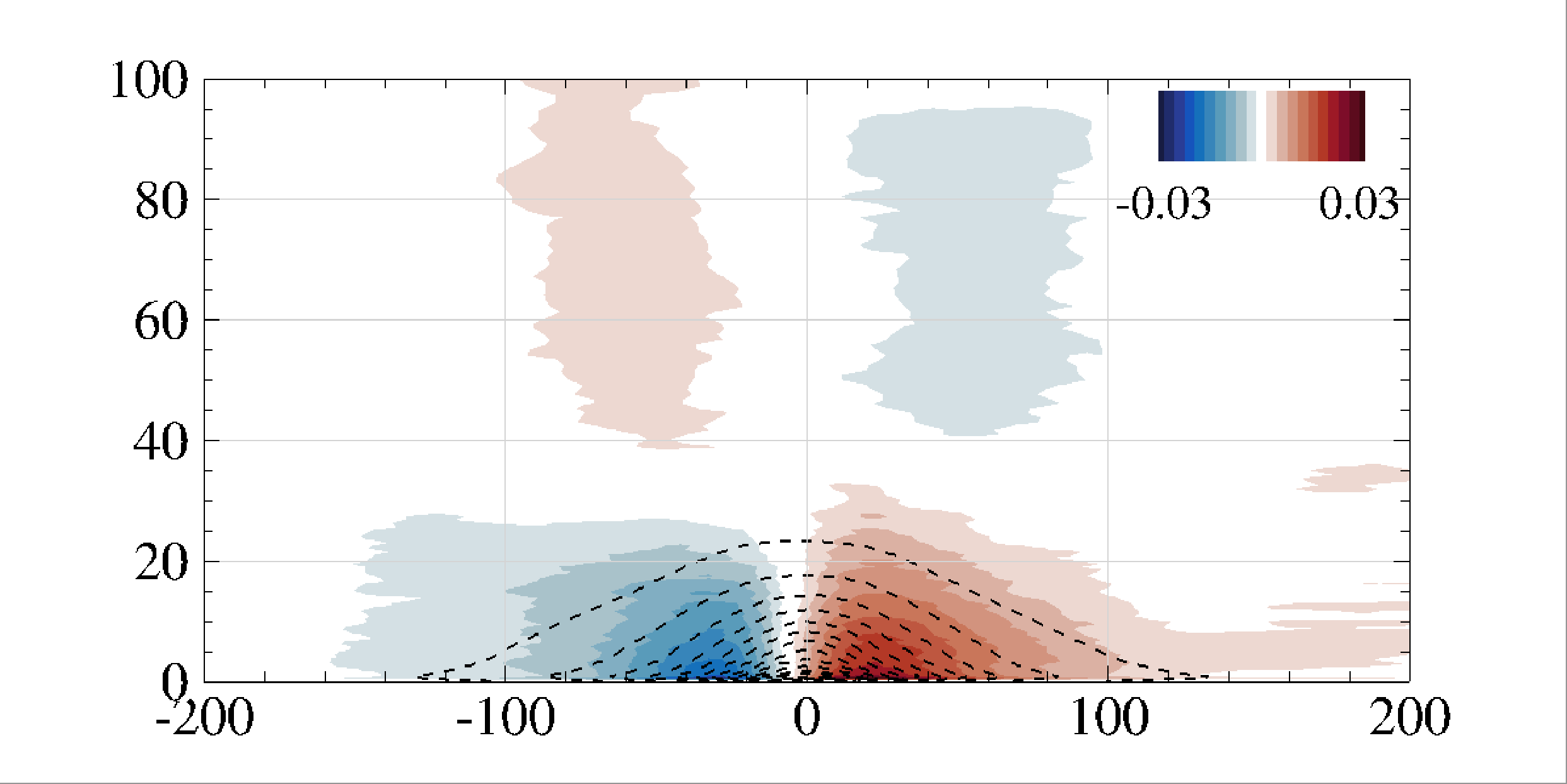}
\put(0,45){(f)}
\put(50,-2){$\Delta x^+$}
\put(0,24){\rotatebox{90}{$\Delta z^+$}}
\end{overpic}
\caption{Conditional average of (a,b) $F_{p1}$, (c,d) $F_{p2}$ and (e,f) $F_{p3}$
under the condition of $\theta' < -\theta'_{rms}$ at $y^+=15$ in case M6C-2.
Flooded: conditional averaged particle force, lines: velocity divergence (solid: positive,
dashed: negative).}
\label{fig:cond}
\end{figure}
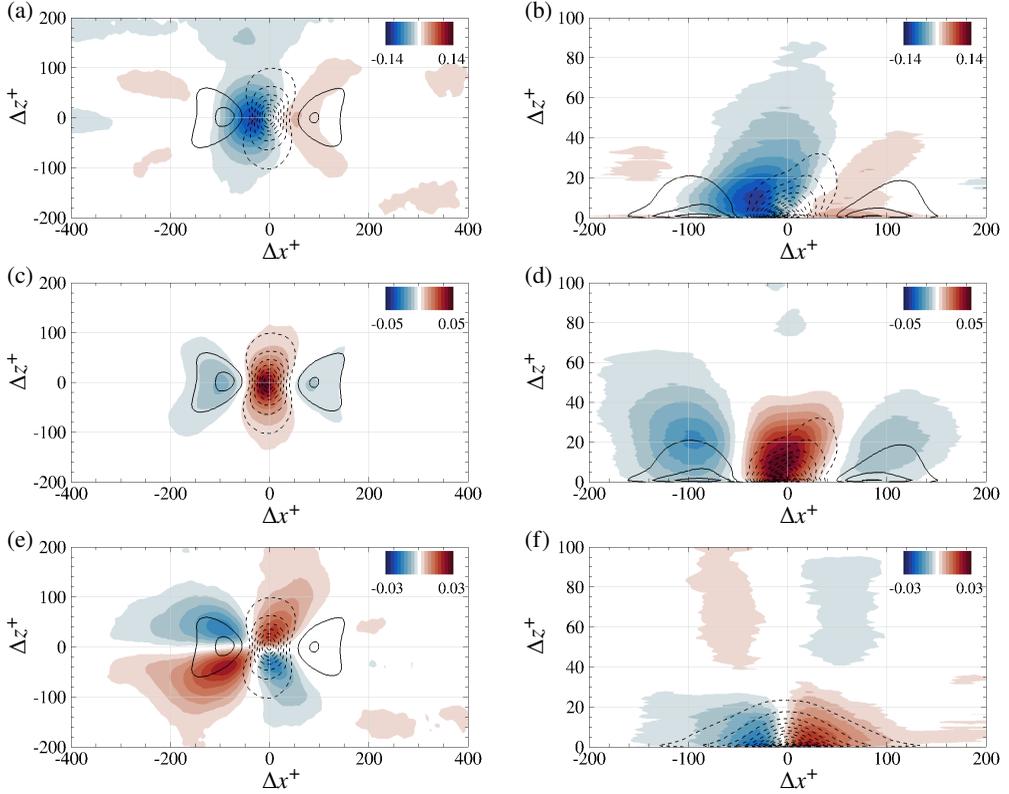

We further calculate the averaged particle force surrounding the events 
where $\theta' < -\theta'_{rms}$ at $y^+=15$ for case M6C-2, as shown in Figure~\ref{fig:cond}.
The results of the rest of the cases are similar so they will not be displayed
herein for brevity.
The conditionally averaged streamwise and wall-normal particle forces $F_{p1}$ and $F_{p2}$ 
are organized in a pattern similar to that of the velocity divergence, which is consistent
with the findings in the observation in the instantaneous flow field.
The difference between these components lies in their relative spatial position.
Based on the observation that the particle force and dilatational velocity exhibit 
characteristics of travelling waves,
the streamwise particle force $F_{p1}$ and the velocity divergence $\theta'$
are offset by an approximately $\pi/2$ phase angle.
This phase difference results in their trivial correlation and hence the work of
the particle force in this direction, even though the conditional averaged $F_{p1}$ surpasses
the particle force components in the other two directions.
This is reminiscent of the pressure-dilatational term in turbulent kinetic energy transfer 
as reported by~\citet{yu2020compressibility}, where the conditional average of pressure and 
velocity divergence are also offset by the phase angle of $\pi/2$.
The wall-normal component $F_{p2}$ is phase-shifted by $\pi$ relative to $\theta'$,
so they are negatively correlated.
From the perspective of particle motions, near-wall particles tend to experience 
vertical acceleration or deceleration as they approach the wall,
transported in the streamwise direction by compressive and expansive fluid motions.
These effects will probably alter the degree of the near-wall particle accumulation,
as inferred by~\citet{wang2024turbophoresis}.
Considering that the intensity of the dilatational motions is dependent on the wall temperature,
the near-wall particle concentration will probably change under different thermal boundary
conditions, which will be considered in our future work.
The conditionally averaged spanwise component $F_{p3}$ behaves as a saddle point,
driving motions opposing the conditional average of the local velocity divergence.
All three components of the particle force disrupt the near-wall dilatational motions,
in the similar way that they hinder the streamwise vortices as well~\citep{dritselis2008numerical}.

\subsection{Integral identity for skin friction}  \label{subsec:skin}

To reveal the particle modulations on the dynamics of the fluid, in this subsection,
we explore the variation of the mean momentum balance and its contribution to the skin friction
$C_f = 2\tau_w /(\rho_\infty U^2_\infty)$ utilizing the skin friction integral identity
proposed by \citet{wenzel2022influences}.
By performing two-fold integration along the wall-normal direction, the mean momentum equation
can be cast as the following formula that decomposes the skin friction into four terms
\begin{equation}
C_f = \frac{2}{\rho_\infty U^2_\infty \delta} \left( \int^\delta_0 \bar \tau_{xy} {\rm d} y
- \int^\delta_0 R^+_{12} {\rm d} y + \int^\delta_0 (y - \delta ) I_x {\rm d} y
+ \int^\delta_0 (y - \delta ) F_{p1} {\rm d} y \right).
\label{eqn:cfdec}
\end{equation}
The terms on the right-hand side represent the contribution to the skin friction of
the viscous shear stress (hereinafter referred to as $C_V$), Reynolds shear stress ($C_T$), 
mean convection ($C_G$) and the particle force ($C_{FP}$).
The $I_x$ in the integral of the mean convection term is formulated as
\begin{equation}
I_x = - \frac{\partial \bar \rho \tilde u_1 \tilde u_1}{\partial x}
- \frac{\partial \bar \rho \tilde u_1 \tilde u_2}{\partial y}
- \frac{\partial \overline{\rho u''_1 u''_1}}{\partial x}
- \frac{\partial \bar p}{\partial x}
+ \frac{\partial \bar \tau_{xx}}{\partial x}.
\end{equation}

\begin{figure}
\begin{overpic}[width=0.45\textwidth]{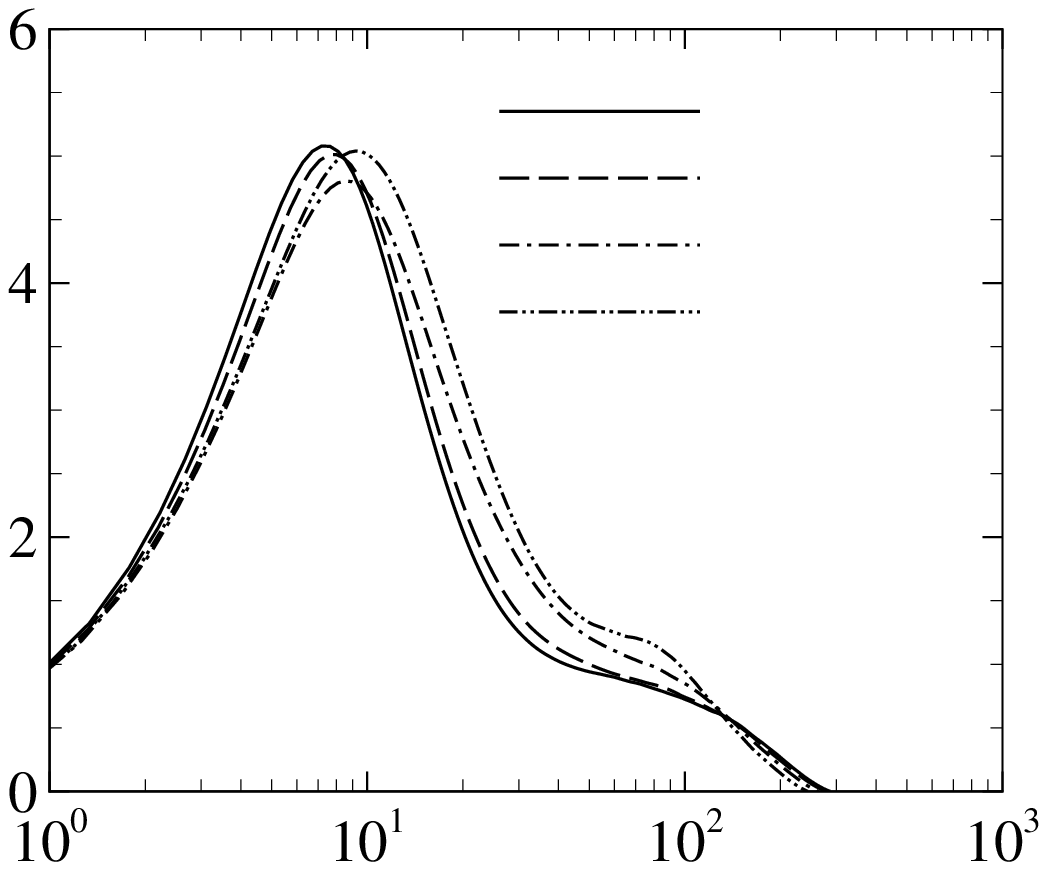}
\put(0,70){(a)}
\put(48,2){$y^+$}
\put(0,32){\rotatebox{90}{$y^+ \bar \tau_{xy}$}}
\put(67,64.5){\small M6-0}
\put(67,59.0){\small M6-1}
\put(67,53.5){\small M6-2}
\put(67,48.0){\small M6-3}
\end{overpic}~
\begin{overpic}[width=0.45\textwidth]{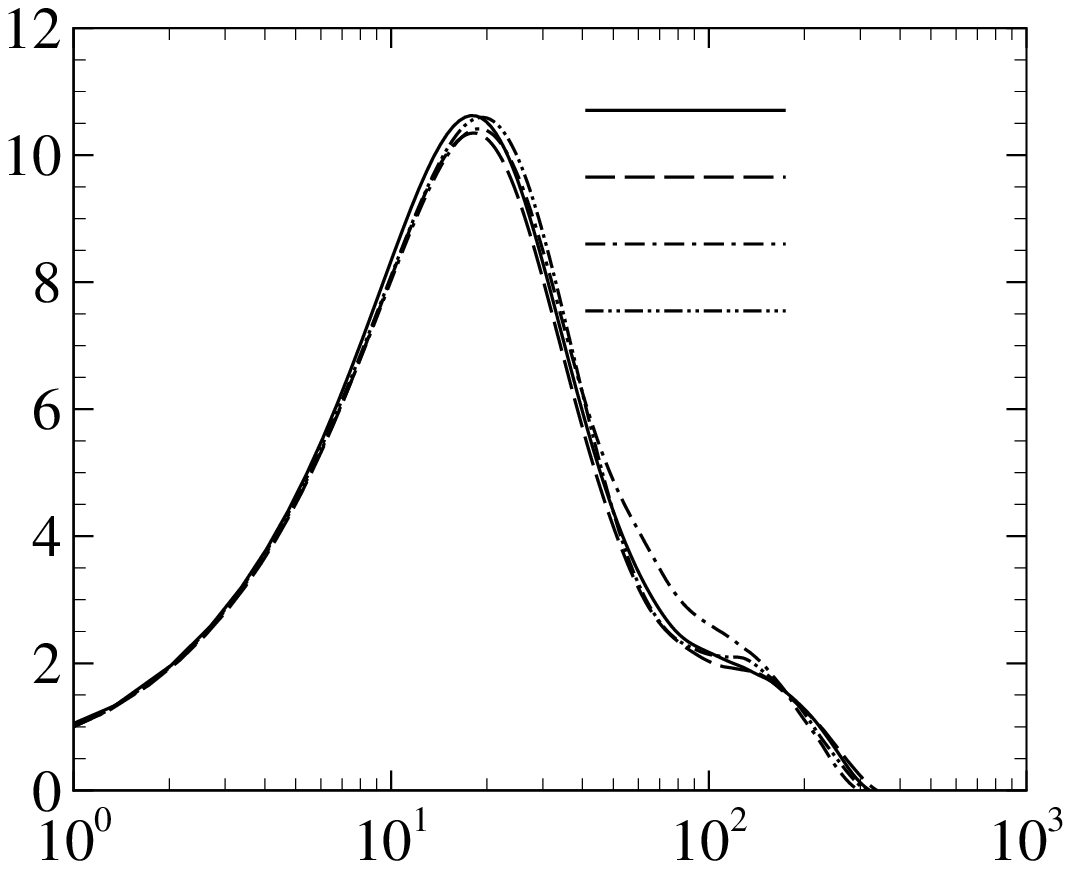}
\put(0,70){(b)}
\put(48,2){$y^+$}
\put(0,32){\rotatebox{90}{$y^+ \bar \tau_{xy}$}}
\put(72,64.5){\small M6C-0}
\put(72,59.0){\small M6C-1}
\put(72,53.5){\small M6C-2}
\put(72,48.0){\small M6C-3}
\end{overpic}\\[0.0ex]
\begin{overpic}[width=0.45\textwidth]{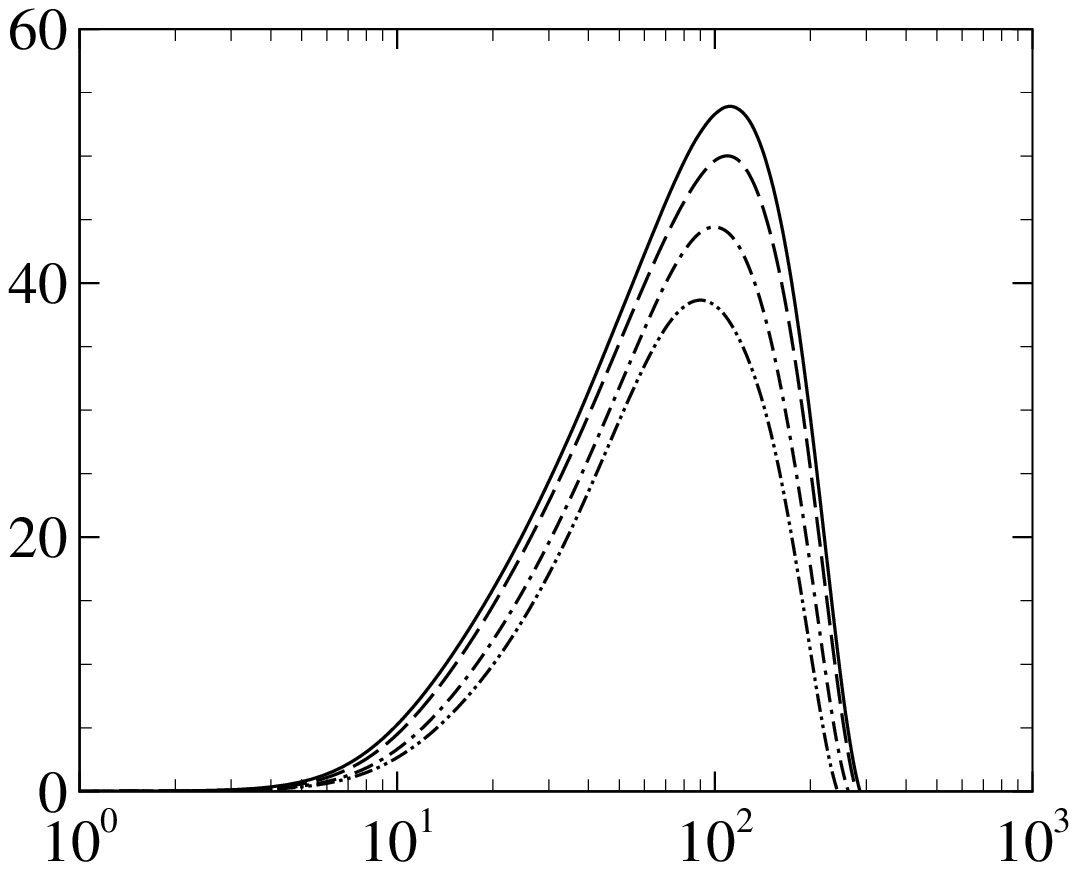}
\put(0,70){(c)}
\put(48,2){$y^+$}
\put(0,32){\rotatebox{90}{$y^+ R^+_{12}$}}
\end{overpic}~
\begin{overpic}[width=0.45\textwidth]{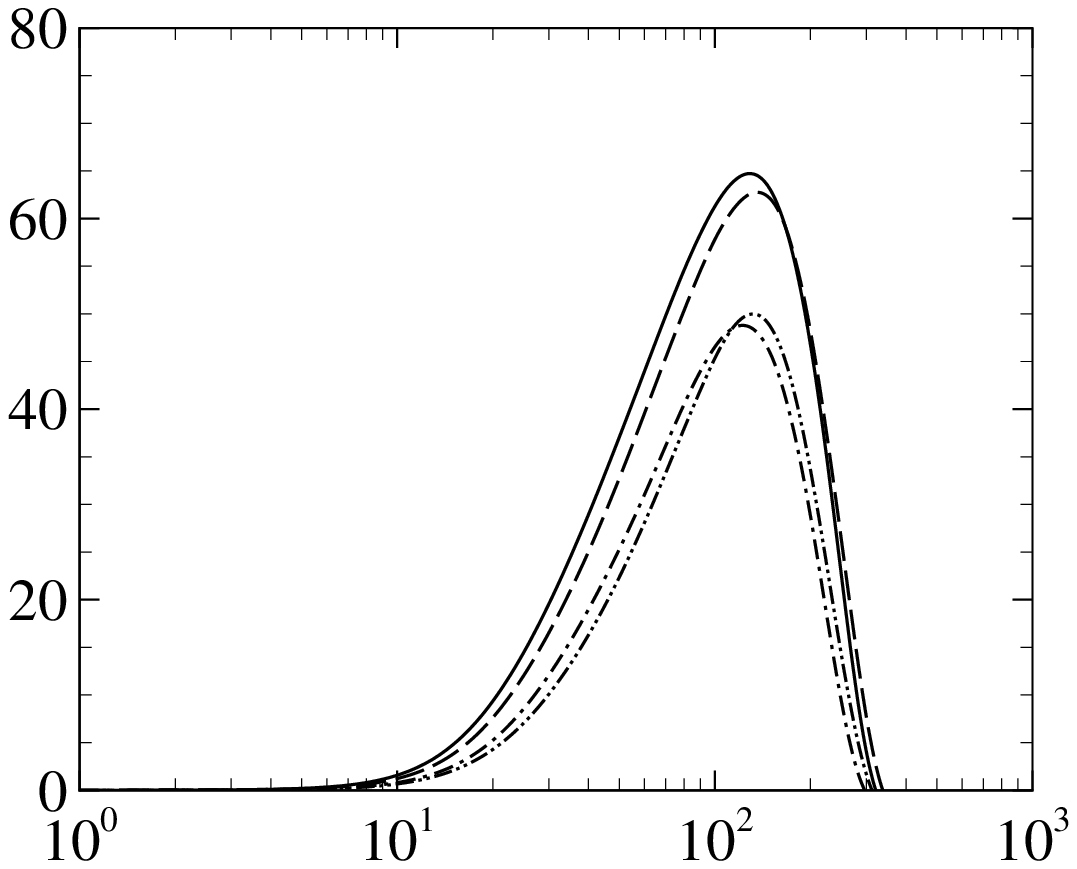}
\put(0,70){(d)}
\put(48,2){$y^+$}
\put(0,32){\rotatebox{90}{$y^+ R^+_{12}$}}
\end{overpic}\\[0.0ex]
\begin{overpic}[width=0.45\textwidth]{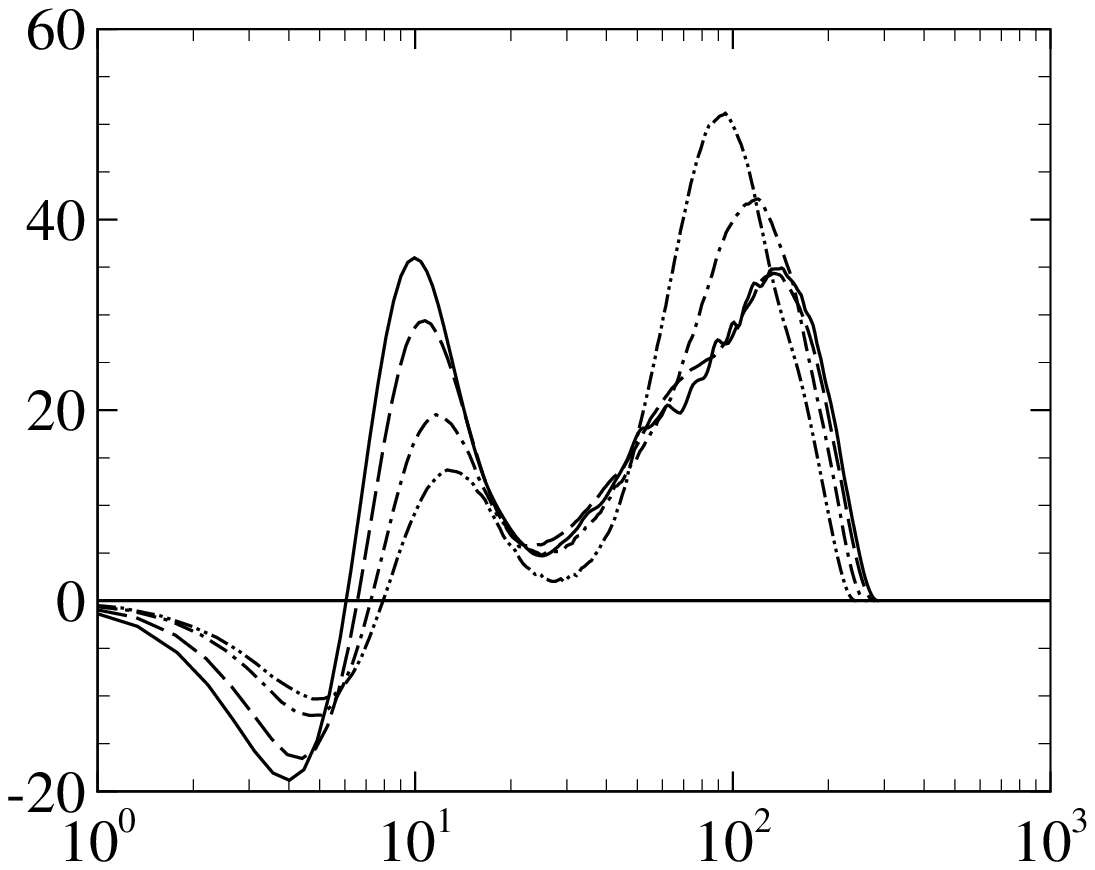}
\put(0,70){(e)}
\put(48,2){$y^+$}
\put(0,30){\rotatebox{90}{$y^+ (y - \delta ) I_x $}}
\end{overpic}~
\begin{overpic}[width=0.45\textwidth]{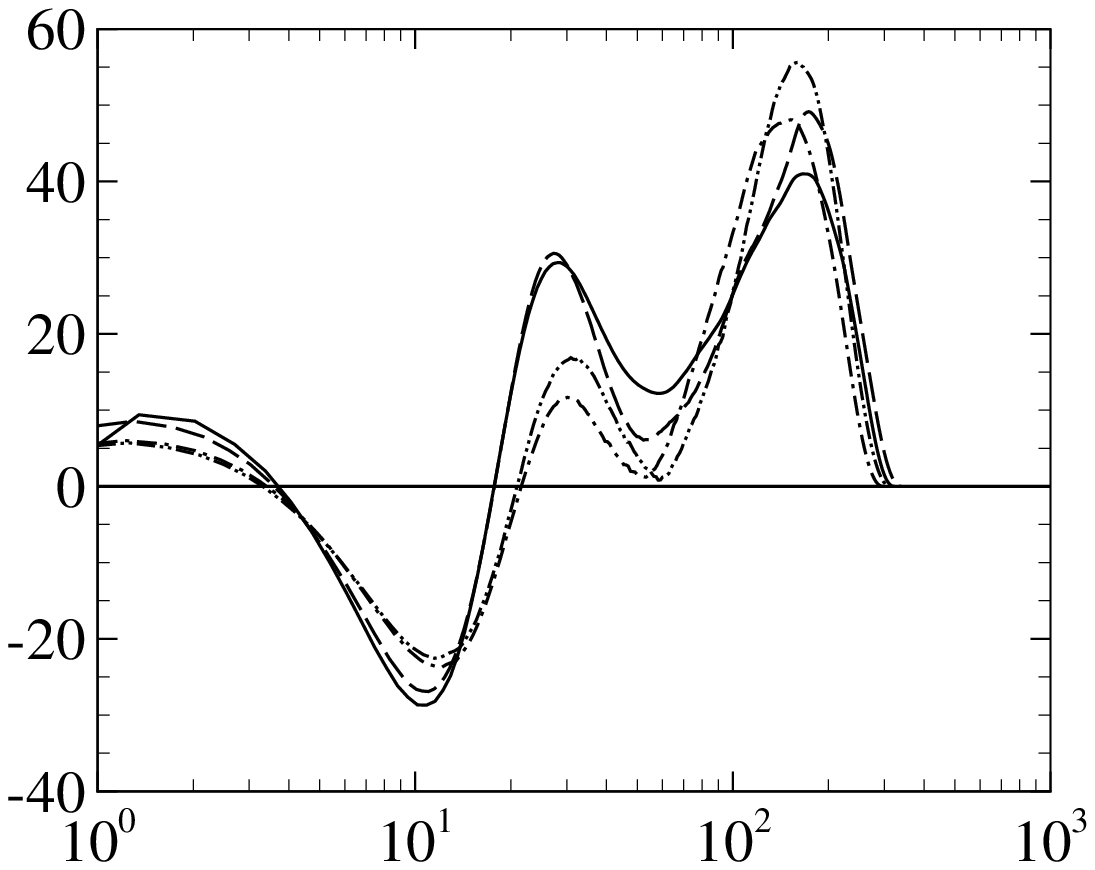}
\put(0,70){(f)}
\put(48,2){$y^+$}
\put(0,30){\rotatebox{90}{$y^+ (y - \delta ) I_x $}}
\end{overpic}\\[0.0ex]
\begin{overpic}[width=0.45\textwidth]{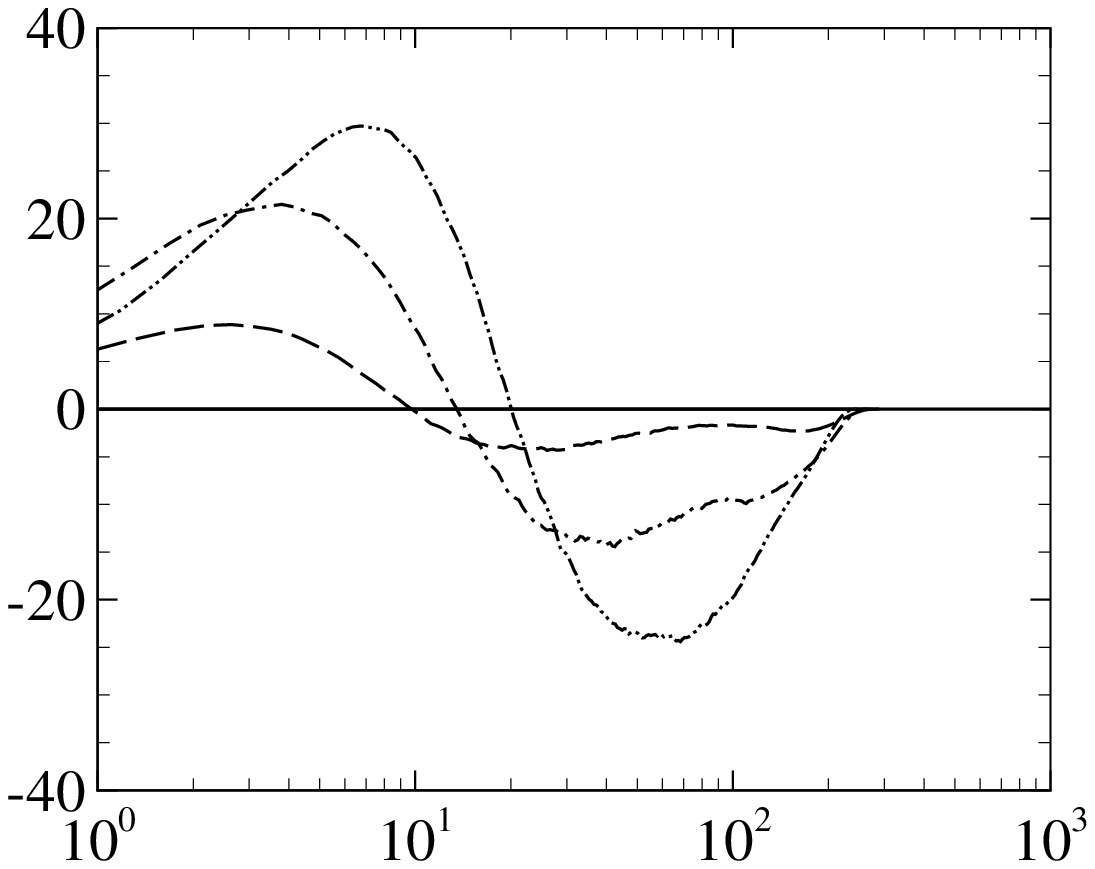}
\put(0,70){(g)}
\put(48,2){$y^+$}
\put(0,26){\rotatebox{90}{$y^+ (y - \delta ) F_{p1} $}}
\end{overpic}~
\begin{overpic}[width=0.45\textwidth]{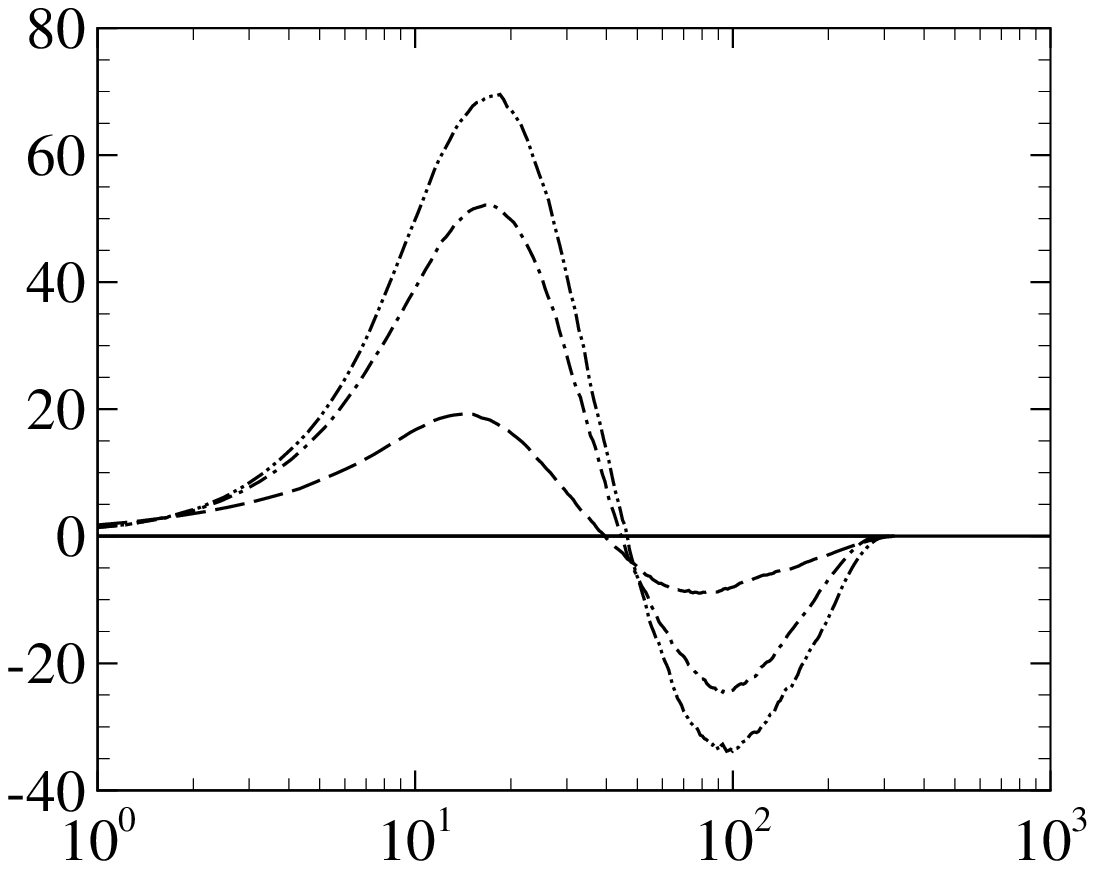}
\put(0,70){(h)}
\put(48,2){$y^+$}
\put(0,26){\rotatebox{90}{$y^+ (y - \delta ) F_{p1} $}}
\end{overpic}
\caption{Pre-multiplied integral of skin friction decomposition formula~\eqref{eqn:cfdec},
(a,b) viscous stress term $C_V$, (c,d) Reynolds stress term $C_T$, 
(e,f) particle force term $C_{FP}$, (g,h) mean convection term $C_G$, 
(a,c,e,g) cases M6, (b,d,f,h) cases M6C.}
\label{fig:cfdecp}
\end{figure}

In Figure~\ref{fig:cfdecp} we first present the integral of the four terms on the right-hand side
of Equation~\eqref{eqn:cfdec}, with the pre-multiplication of $y^+$ under the logarithmic coordinate.
The integral of the viscous term and Reynolds shear stress term constitute a greater portion
of the skin friction in the near-wall region and outer region, with the peaks reached at
$y^+ \approx 10$ and $y^+ \approx 100$, respectively.
The integral of the mean convection term has a complex profile, showing multiple peaks and valleys,
whose magnitudes below $y^+=30$ are decreasing and those above $y^+=30$ are increasing with
the higher particle mass loading cases.
The overall integration is positive, as indicated by the higher magnitude of the positive values.
Lastly, the pre-multiplied integral of the particle force term is positive close to the wall
and gradually varies to negative values above the buffer region.
Such a trend of variation can be ascribed to the fact that high inertia particles
travel faster than the fluid in the near-wall region, which can easily be envisaged to be caused by
the particles with higher streamwise velocity and deceleration gradually due to their high inertia 
and hence the slow response to the local fluid velocity.
In the outer region, the velocity difference between them is reversed.
The change of sign results in its comparatively low contribution to the skin friction.

The specific values of the contributions of these terms to the skin friction are listed in
Table~\ref{tab:cfdecp}.
For the adiabatic wall cases M6, the viscous term $C_V$ constitutes approximately 10\%,
and this percentage is increasing with the mass loading.
The contribution of the Reynolds stress term $C_T$, on the other hand, is decreasing monotonically
from approximately 70\% (M6-0) to 55\% (M6-3).
This is consistent with the decreasing Reynolds shear stress as reported in Figure~\ref{fig:rey}.
The mean convection term $C_G$ constitutes $20\%$ in these adiabatic wall cases, 
showing comparatively small variations in the presence of particles. 
This indicates that the mean convection and the spatial development are not much affected by 
the particle feedback force, at least in the sense of the integral mean momentum transport.
The particle feedback force constitutes a considerable portion by approximately $12\%$ 
in case M6-3, which originates from the work of particle force on the fluid in the near-wall region.
Nevertheless, such an effect is not sufficient to balance the decrement of the skin friction
from the perspective of their absolute values.
Similar conclusions can be given for the cold wall cases M6C, except that the contribution of
the viscous term $C_V$ is higher, that of the Reynolds stress term $C_T$ is lower,
the mean convection term $C_G$ constitutes a lower portion by less than $16\%$,
and the contribution of the particle force term $C_{FP}$ is higher than those in the adiabatic wall
cases M6, with the percentage up to $23\%$ in case M6C-3.
These results suggest the presence of particles influences the flow dynamics and 
the skin friction not only directly by accelerating the near-wall fluid velocity but also 
indirectly by modulating the mean and fluctuating velocity of the fluid.

\begin{table}
\centering
\begin{tabular}{ccccccccc}
Case    &   $C_f$   &   $C_V/C_f$   &   $C_T/C_f$   &   $C_G/C_f$   &   $C_{FP}/C_f$    &
$C^s_T/C_f$ &   $C^d_T/C_f$     &   $C^{sd}_T/C_f$ \\
M6-0  & 9.060E-4 & 0.0881 & 0.7132 & 0.1986 & 0.0000 & 0.7360 & 0.0018 & -0.0343 \\
M6-1  & 8.858E-4 & 0.0908 & 0.6728 & 0.1912 & 0.0451 & 0.6958 & 0.0015 & -0.0339 \\
M6-2  & 8.183E-4 & 0.0983 & 0.6041 & 0.1992 & 0.0984 & 0.6261 & 0.0012 & -0.0327 \\
M6-3  & 7.238E-4 & 0.1126 & 0.5541 & 0.2148 & 0.1185 & 0.5786 & 0.0008 & -0.0340 \\
M6C-0 & 1.656E-3 & 0.1663 & 0.6886 & 0.1451 & 0.0000 & 0.7241 & 0.0027 & -0.0458 \\
M6C-1 & 1.733E-3 & 0.1532 & 0.6193 & 0.1542 & 0.0733 & 0.6513 & 0.0029 & -0.0404 \\
M6C-2 & 1.548E-3 & 0.1815 & 0.5079 & 0.1118 & 0.1987 & 0.5401 & 0.0013 & -0.0384 \\
M6C-3 & 1.620E-3 & 0.1702 & 0.4770 & 0.1208 & 0.2319 & 0.5062 & 0.0016 & -0.0366 \\
\end{tabular}
\caption{Contribution of the right-hand-side terms in Equation~\eqref{eqn:cfdec} to
the skin friction}
\label{tab:cfdecp}
\end{table}

\begin{figure}
\begin{overpic}[width=0.45\textwidth]{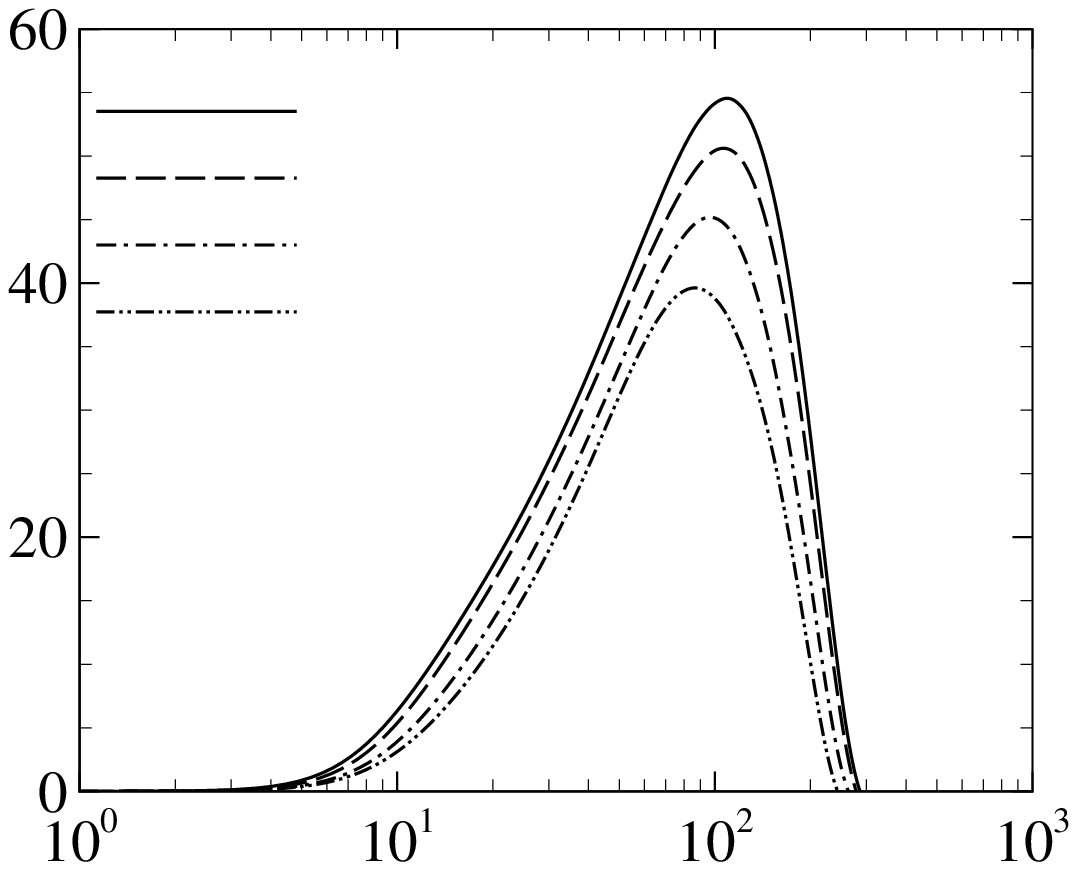}
\put(0,70){(a)}
\put(48,2){$y^+$}
\put(0,32){\rotatebox{90}{$y^+ R^{s+}_{12}$}}
\put(33,64.5){\small M6-0}
\put(33,59.0){\small M6-1}
\put(33,53.5){\small M6-2}
\put(33,48.0){\small M6-3}
\end{overpic}~
\begin{overpic}[width=0.45\textwidth]{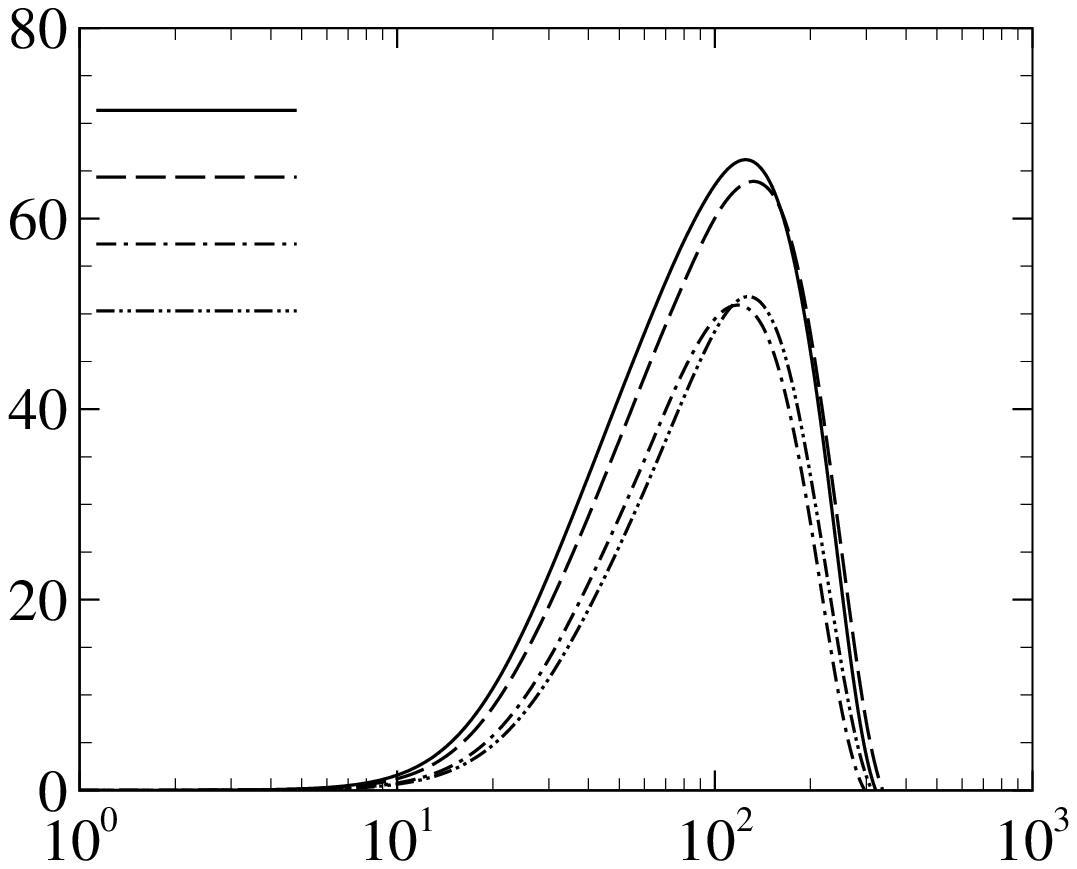}
\put(0,70){(b)}
\put(48,2){$y^+$}
\put(0,32){\rotatebox{90}{$y^+ R^{s+}_{12}$}}
\put(33,64.5){\small M6C-0}
\put(33,59.0){\small M6C-1}
\put(33,53.5){\small M6C-2}
\put(33,48.0){\small M6C-3}
\end{overpic}\\[0.0ex]
\begin{overpic}[width=0.45\textwidth]{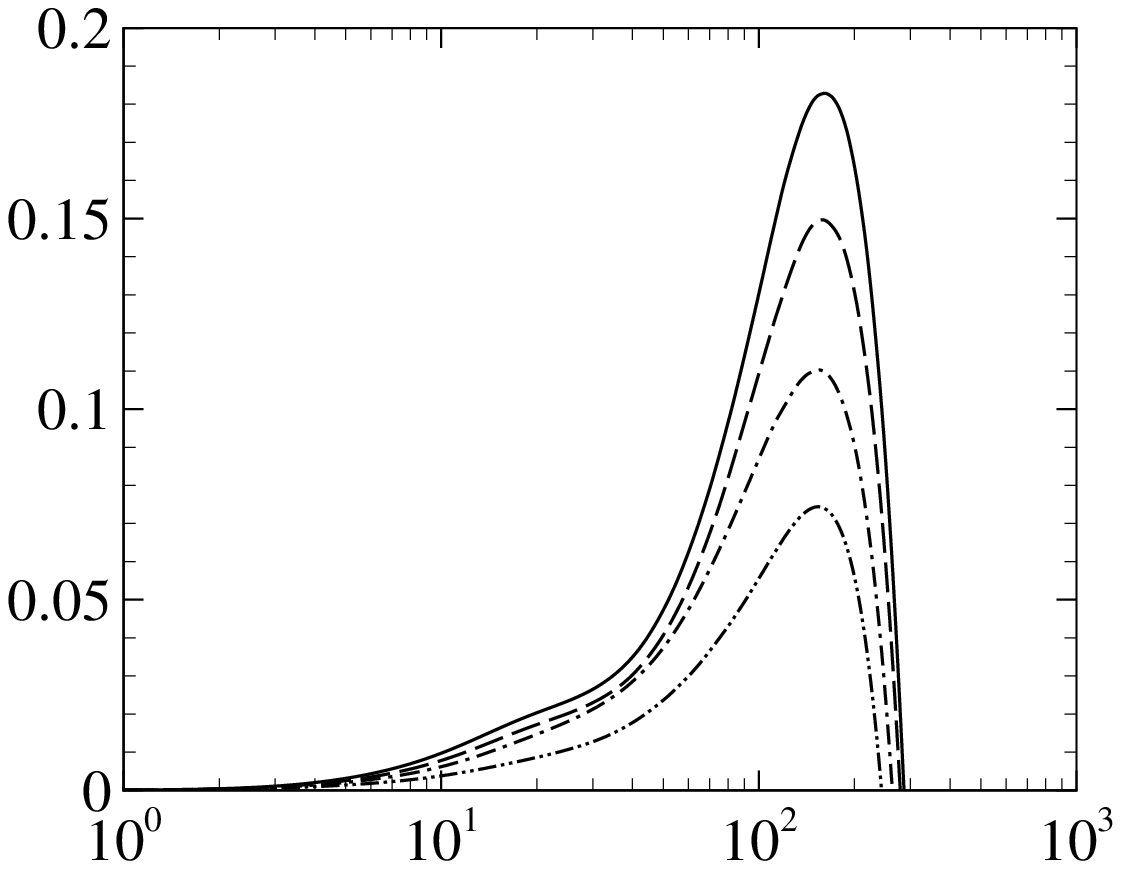}
\put(0,70){(c)}
\put(48,2){$y^+$}
\put(0,32){\rotatebox{90}{$y^+ R^{d+}_{12}$}}
\end{overpic}~
\begin{overpic}[width=0.45\textwidth]{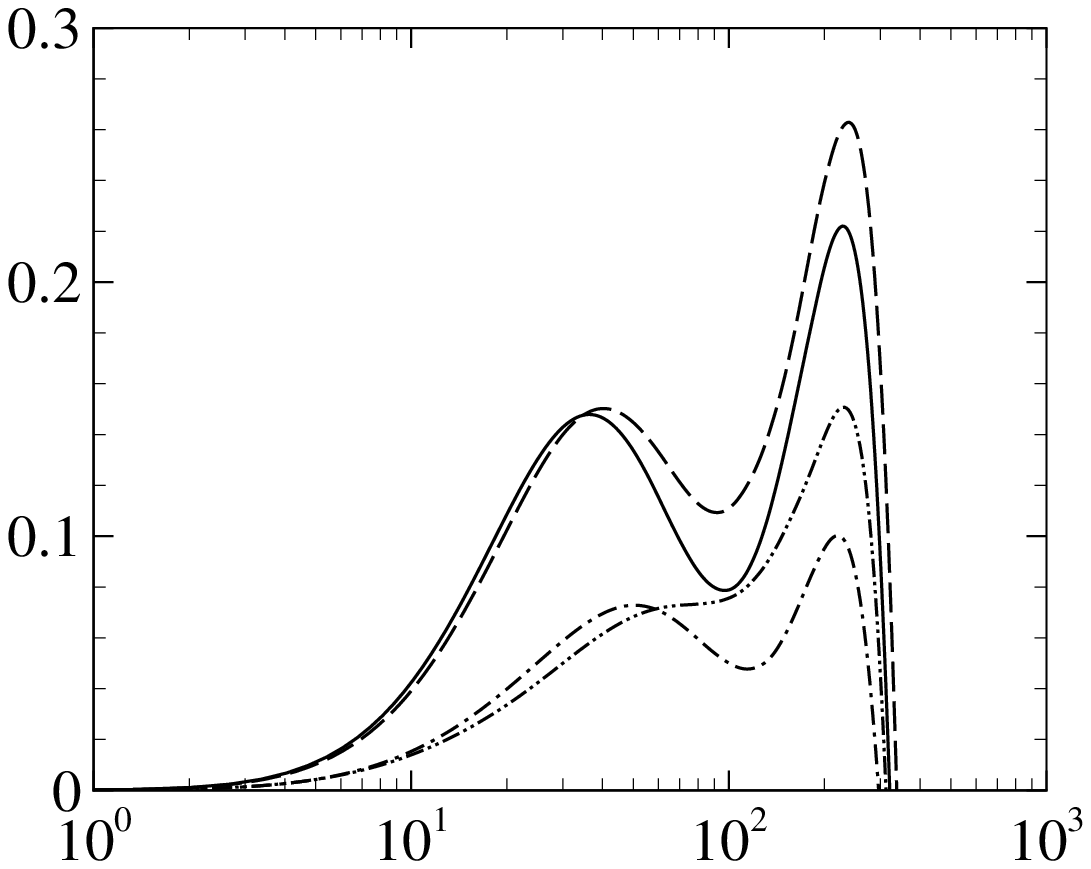}
\put(0,70){(d)}
\put(48,2){$y^+$}
\put(0,32){\rotatebox{90}{$y^+ R^{d+}_{12}$}}
\end{overpic}\\[0.0ex]
\begin{overpic}[width=0.45\textwidth]{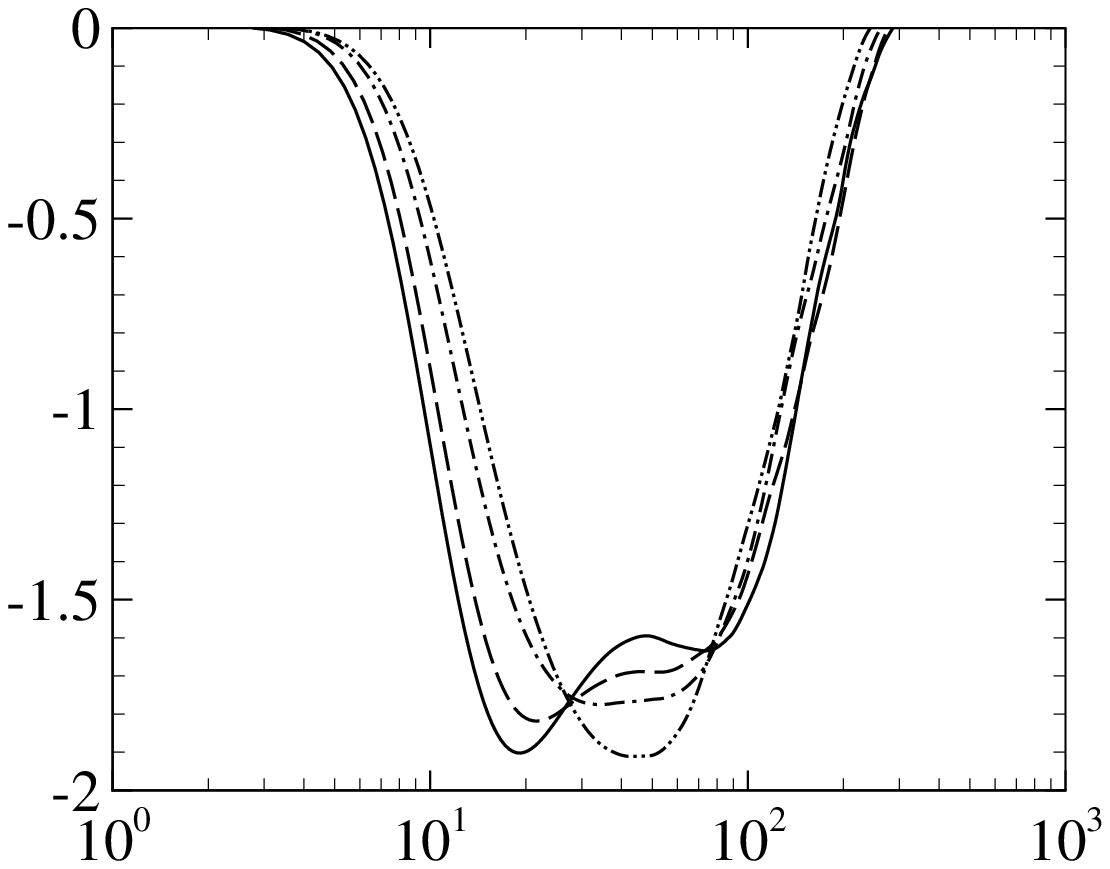}
\put(0,70){(e)}
\put(48,2){$y^+$}
\put(0,32){\rotatebox{90}{$y^+ R^{sd+}_{12}$}}
\end{overpic}~
\begin{overpic}[width=0.45\textwidth]{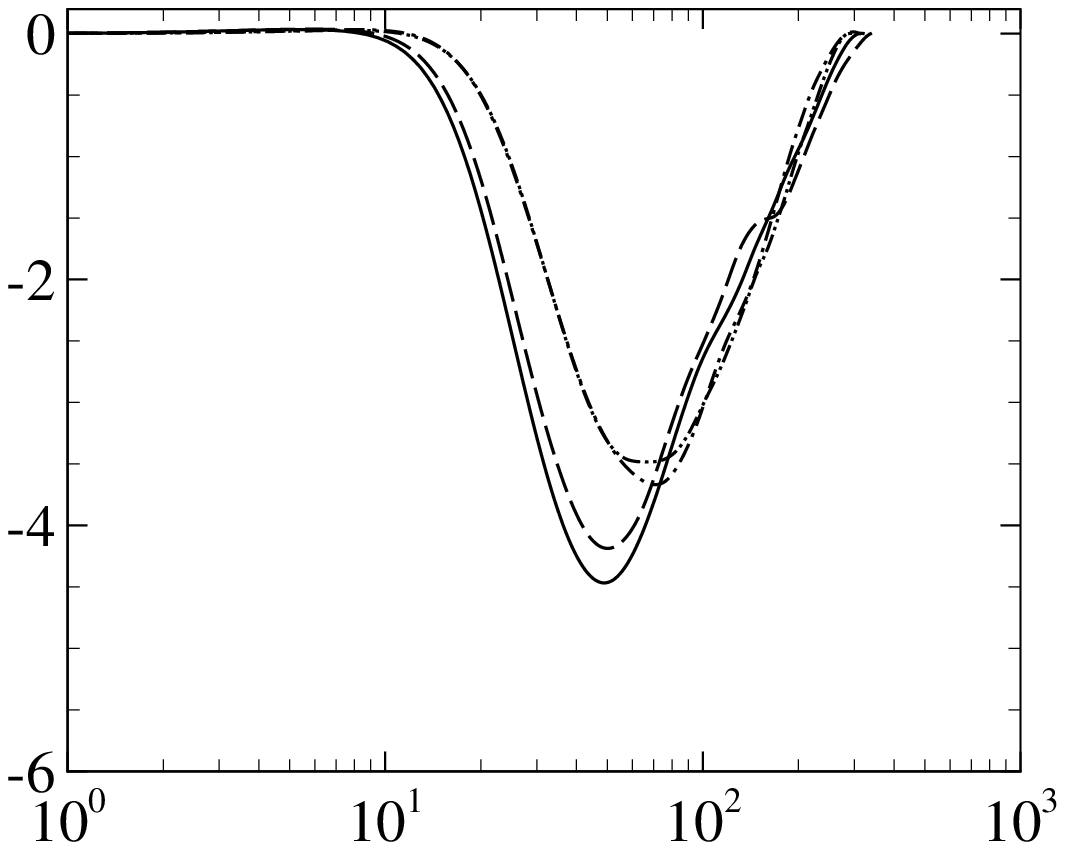}
\put(0,70){(f)}
\put(48,2){$y^+$}
\put(0,32){\rotatebox{90}{$y^+ R^{sd+}_{12}$}}
\end{overpic}
\caption{Pre-multiplied integral of the Reynolds shear stress term of
skin friction decomposition formula~\eqref{eqn:cfdec},
(a,b) solenoidal term $C^s_{T}$, (c,d) dilatational term $C^d_T$, 
(e,f) cross-correlation term $C^{sd}_T$, (a,c,e) cases M6, (b,d,f) cases M6C.}
\label{fig:cfdecpt}
\end{figure}

We further split the Reynolds shear stress into three components by substituting 
the Helmholtz decomposition of velocity fluctuations into its definition as follows,
\begin{equation}
R_{12} = \underbrace{\overline{\rho u''^s_1 u''^s_1}}_{R^s_{12}}
+ \underbrace{\overline{\rho u''^s_1 u''^s_1}}_{R^d_{12}}
+ \underbrace{\overline{\rho u''^s_1 u''^d_2} + \overline{\rho u''^d_1 u''^s_2}}_{R^{sd}_{12}}
\end{equation}
so their respect contribution to the skin frictions can be obtained by substituting
into Equation~\eqref{eqn:cfdec}, hereinafter referred to as the solenoidal term $C^s_{T}$, 
dilatational term $C^d_T$ and the cross-correlation term $C^{sd}_T$, respectively.
The wall-normal distributions of the integral of these terms are displayed in 
Figure~\ref{fig:cfdecpt}, and the specific values of their contributions to the skin friction
are also reported in Table~\ref{tab:cfdecp}.
Consistent with the previous findings in turbulent channel flows~\citep{yu2019genuine,
yu2021compressibility}, the Reynolds stress and its contribution to the skin friction
are primarily constituted by the solenoidal component.
The contribution from the dilatational component decreases monotonically with the mass loading
(except for case M6C-3), which is consistent with the trend of variation of the dilatational
motion in the discussions above.
However, this term remains trivial in all of the cases considered, further suggesting that 
the presence of particles does not alter the dynamical essence of wall-bounded turbulence being
populated by vortices.
The cross-correlation term is not much affected by the mass loading, contributing negatively
by approximately $-4\%$ to the skin friction.
By comparing these cases, it can be elucidated that the reduction of the contribution
of the Reynolds shear stress to the skin friction should be ascribed to the weaker
solenoidal motions associated with vortical and shear structures, instead of dilatational motions
or their correlation with the solenoidal motions.

\section{Concluding remarks} \label{sec:con}

This paper investigates the particle modulations of compressible turbulent boundary layers
at the Mach number of $6$ over quasi-adiabatic and cold walls by exploiting
the direct numerical simulation databases established
using the Eulerian-Lagragian point-particle method.
Four groups of cases with varying mass loadings ranging from 0 to 0.628 are analyzed
for each wall temperature, 
focusing on turbulent statistics variations, the influence of particle feedback forces 
on solenoidal and dilatational motions, and the modulation of the skin friction by particles.

The variation of the mean velocity and Reynolds stresses with increasing mass loading 
resemble those observed in incompressible wall-bounded turbulence. 
The transformed mean velocities under viscous scales are higher from the log layer till
the free-stream, leading to steeper slopes and intercepts in the log layer, irrelevant to
the specific transformation adopted.
The streamwise Reynolds stress remains high in the buffer layer, while the wall-normal 
and shear components decrease monotonically with the increasing particle mass loading. 
This decline also suggest the weaker turbulent heat flux,
resulting in variations in mean temperature distributions.

From the perspective of the coherent structures, we found that the velocity streaks are less 
meandering, while dilatational travelling wave structures remain evident in the presence of 
particles. 
We also identified the resemblance in instantaneous distributions between the velocity fluctuations 
and particle feedback force, implying significant modulation of coherent structures, 
particularly the presence of travelling wave packets in the wall-normal components of 
these flow quantities. 
By performing Helmholtz decomposition to split the velocity into solenoidal and 
dilatational portions, 
we observed that solenoidal streamwise velocity fluctuations, associated with velocity streaks, 
remain prominent in particle-laden flows. The solenoidal wall-normal velocity and 
dilatational velocity fluctuations decrease almost synchronously, 
supporting the previous elucidation that the dilatational motions
are generated by the streamwise vortices and/or strong bursting events,
instead of the evolving perturbations beneath the velocity streaks.
Additionally, we found a strong correlation between particle feedback force and 
dilatational motions.
This indicates that particles are accelerating/decelerating within 
travelling wave packets, leading to a weakening of dilatational motions during this process.
By analyzing the skin friction integral identity, we found that the Reynolds shear
stress contributes increasingly less portion to the skin friction with the increasing particle
mass loading due to the weakening of the vortical and shear motions,
which is partially compensated by the higher positive contribution of the particle feedback force.
The dilatational motions constitute merely a small portion of the skin friction 
in the presently considered cases with moderate particle mass loading,
suggesting the essentially unaltered nature of wall-bounded turbulence
populated by vortical and shear motions instead of gradually switching to the state dominated by
dilatational motions.

\bibliographystyle{jfm}
\bibliography{bibfile}

\end{document}